\documentclass[a4paper,11pt]{article}
\usepackage{jheppub}
\usepackage[table]{xcolor}
\usepackage{mathtools}
\usepackage{tensor}
\usepackage{multirow}
\usepackage{makecell}
\usepackage{comment}
\usepackage{nicematrix}
\usepackage{booktabs}
\usepackage[normalem]{ulem}
\usepackage{tikz}
\usepackage{mathrsfs}
\usetikzlibrary{patterns}

\allowdisplaybreaks

\newcommand{\scri}{\mathscr{I}}

\title{\boldmath Self-force framework for merger-ringdown waveforms}

\author[a]{Lorenzo K\"uchler}
\author[b]{Geoffrey Comp\`ere}
\author[a]{Adam Pound}
\affiliation[a]{School of Mathematical Sciences and STAG Research Centre, University of Southampton,\\Southampton SO17 1BJ, United Kingdom}
\affiliation[b]{Universit\'e Libre de Bruxelles, BLU-ULB Brussels Laboratory of the Universe and International Solvay Institutes, C.P. 231, B-1050 Bruxelles, Belgium}

\emailAdd{l.m.kuchler@soton.ac.uk, geoffrey.compere@ulb.be, A.Pound@soton.ac.uk}

\abstract{The prospect of observing asymmetric compact binaries with next-generation gravitational-wave detectors has motivated the development of fast and accurate waveform models in gravitational self-force theory. These models are based on a two-stage process: in a (slow) offline stage, waveform ingredients are pre-computed as functions on the orbital phase space; in a (fast) online stage, the waveform is generated by evolving through the phase space. While this framework has traditionally been restricted to the inspiral stage of a binary, we recently extended it across the transition to plunge, where the small companion crosses the innermost stable circular orbit around the primary black hole. In this paper, for the special case of quasicircular, nonspinning binaries, we show how the ``offline/online’’ phase-space paradigm also extends through the final plunge, which generates the binary’s merger-ringdown signal. We implement the method at leading, geodesic order in the plunge. The resulting plunge waveform agrees well with a stationary-phase approximation at early times and with a (self-consistently calculated) quasinormal mode sum at late times, but we highlight that neither of the two approximations reaches the peak of the full plunge waveform. Finally, we compare the plunge waveform to numerical relativity simulations. Our framework offers the prospect of fast, accurate inspiral-merger-ringdown waveform models for asymmetric binaries.}

\begin{document}
\maketitle
\flushbottom

\section{Introduction}

Since the first detection of a binary black hole merger in 2015~\cite{LIGOScientific:2016aoc}, three observing runs of the LIGO-Virgo-KAGRA Collaboration have seen close to a hundred events~\cite{LIGOScientific:2018mvr,LIGOScientific:2020ibl,KAGRA:2021vkt}. The majority of the observed signals originated from the merger of compact binaries with mass ratios typically close to 1, but several binaries have been detected with much more disparate masses. This includes one signal thought to have originated from the merger of a binary with  mass ratio $\approx$~1:26 (GW191219\_16312), outside the range in which current waveform models have been validated~\cite{KAGRA:2021vkt}. Future observing runs and next-generation ground-based detectors such as the Einstein Telescope~\cite{ET:2019dnz} and Cosmic Explorer~\cite{Evans:2021gyd} promise an increase in the number of detections~\cite{Baibhav:2019gxm,Kalogera:2021bya} and likely also in the number of events with smaller mass ratios. Space-based gravitational-wave (GW) detectors such as the Laser Interferometer Space Antenna (LISA)~\cite{amaroseoane2017laser} will be able to detect GWs in the mHz frequency band, allowing us to access a greater variety of sources such as intermediate- and extreme-mass-ratio inspirals (I/EMRIs)~\cite{Babak:2017tow,LISA:2022yao}. In addition, LISA will also detect the coalescence of massive black hole binaries, which could exhibit a long tail in their distribution of mass ratios extending to around 1:$10^3$~\cite{Barausse:2020mdt,Mangiagli:2022niy,Bellovary:2024akp}. Intermediate-mass-ratio binaries are also possible sources of multiband signals detectable with both space- and ground-based detectors~\cite{Jani:2019ffg}. The prospect of these future observations motivates the production of fast, accurate inspiral-merger-ringdown (IMR) models for asymmetric binaries, in which the secondary of mass $m_p$ (labeled with ``$p$'' because we treat it as a  particle) is significantly lighter than the primary of mass~$M$.

Gravitational self-force (GSF) theory is recognized as the primary method of modelling I/EMRIs~\cite{LISAConsortiumWaveformWorkingGroup:2023arg} and has been used to inform effective models that cover the full binary parameter space~\cite{Damour:2009sm,Nagar:2022fep,vandeMeent:2023ols,Leather:2025nhu}. Although originally designed for EMRIs, second-order GSF waveforms have proven to be very accurate even at mass ratios $\approx$1:10, with sub-radian dephasing compared to fully nonlinear numerical relativity (NR) simulations~\cite{Wardell:2021fyy,Albertini:2022rfe}. GSF waveforms can also now be generated on a timescale of milliseconds, fast enough for data analysis~\cite{Katz:2021yft}. 
However, the methods leading to these GSF waveforms are specialized to slow evolutions; they cannot capture the rapid final stages of a binary merger. Although these final stages contribute negligible signal-to-noise ratio for EMRIs~\cite{Amaro-Seoane:2012lgq}, they are critical for more moderate mass ratios and for ground-based detectors (which are most sensitive to the end stages of an asymmetric binary due to their low-frequency floor~\cite{Abac:2025saz}).

The basic anatomy of a binary evolution consists of three stages: the inspiral, which is well modeled by post-Newtonian (PN) theory~\cite{Blanchet:2013haa} (for comparable-mass binaries) or GSF theory~\cite{Barack:2018yvs,Pound:2021qin} (for asymmetric binaries); the merger, which has historically been the realm of fully nonlinear NR simulations~\cite{Duez:2018jaf}; and the post-merger ringdown, which is accurately described by vacuum black hole perturbation theory~\cite{London:2014cma,Berti:2025hly}.

Effective-one-body theory (EOB)~\cite{Buonanno:1998gg,Buonanno:2000ef} offered the simplest semi-analytical understanding of this evolution. Rather than a three-stage process, EOB proposed that there are only two stages: an extended inspiral that persists all the way to the light ring of the effective-one-body black-hole metric, followed by a ringdown. This description stems from the reduction of the two-body problem to a (reduced-mass) secondary object orbiting in an effective-one-body black hole spacetime. In that reduced problem, we can apply the following physical intuition. Before the secondary object crosses the effective spacetime's light ring, the waveform is dominated by radiation propagating directly from the orbiting secondary, and the amplitude of the radiation grows steadily as the secondary moves deeper into the strong-field, relativistic regime. After it crosses the light ring, any radiation the secondary emits falls into the black hole. The waveform is then dominated by radiation that was trapped on the light ring, which then leaks out to future null infinity as quasinormal modes.

To meet accuracy requirements for GW science, this simple description has had to be corrected by non-quasicircular corrections (NQCs)~\cite{Damour:2002vi,Damour:2007xr,Buonanno:2009qa,Taracchini:2014zpa,Riemenschneider:2021ppj} and other phenomenological adaptations calibrated to NR waveforms. However, the core idea proved remarkably accurate.

Complementary to EOB, the Phenom family of IMR models~\cite{Ajith:2007qp,Ajith:2007kx,Ajith:2009bn,Santamaria:2010yb} takes advantage of another feature of the merger. The amplitude and phase of the waveform through merger are surprisingly simple and can be very well approximated by elementary functions. Directly approximating the amplitude and phase observed in NR allows Phenom models to avoid stitching together an inspiral to a ringdown. Instead, an inspiral waveform based on PN and tuned to EOB is effectively stitched (in an NR-calibrated way) to an NR-informed merger-ringdown waveform.

Both these models leverage the fact that during the inspiral, the waveform can be written as a function of the two-body, mechanical phase-space variables---the bodies' relative positions and momenta, for example. GSF waveform models double down on this, using it to formulate a multiscale expansion of the Einstein field equations that puts the field equations in a practical form and simultaneously enables rapid waveform generation~\cite{Miller:2020bft,Pound:2021qin,Mathews:2025nyb}. In the multiscale expansion, waveform generation is divided into offline and online calculations. The offline calculations consist of solving the field equations to precompute waveform ingredients as functions on the binary's mechanical phase space. The online stage then consists of a rapid, inexpensive evolution through the phase space, together with a summation of waveform modes~\cite{Katz:2021yft}.

This GSF framework for modeling the inspiral has traditionally relied on a separation of time scales: the system's parameters (e.g., its fundamental frequencies) evolve on a radiation-reaction time scale that is large compared to the periods of orbital motion. During the inspiral, the radiation-reaction time scale is of order $M^2/m_p$, and the multiscale expansion takes the form of a post-adiabatic expansion~\cite{Hinderer:2008dm,Miller:2020bft,Pound:2021qin,Mathews:2025nyb}. This scaling breaks down as the secondary approaches the innermost stable circular orbit (ISCO), or more generally the separatrix between stable and plunging orbits~\cite{Stein:2019buj}, where the post-adiabatic expansion becomes singular. However, there is still a separation of time scales, as the ISCO-crossing time is long, of order $M(M/m_p)^{1/5}$, compared to the orbital period~\cite{Buonanno:2000ef,Ori:2000zn}. In Refs.~\cite{Compere:2021iwh,Compere:2021ERR,Compere:2021zfj,Kuchler:2024esj}, we exploited this fact to develop a multiscale expansion adapted to the transition across the ISCO. In Ref.~\cite{Kuchler:2024esj}, in particular, we showed how the phase-space paradigm persists in this regime and how it continues to facilitate an offline-online split of the field equations, ultimately enabling rapid waveform generation. (EOB models, in contrast, do not fully exploit the separation of scales, which allows them to smoothly evolve across the ISCO but prevents a complete offline-online split.)

In this paper, we continue this development by building a merger-ringdown GSF framework within the phase-space paradigm, restricting to the case of quasicircular, nonspinning binaries (as we also did for the transition to plunge). There is no separation of time scales during the final plunge, when the secondary falls from the ISCO down to the black hole horizon. Nevertheless, we show that one can continue to treat the waveform as a function on the binary's mechanical phase space, linking the waveform to the orbital dynamics all the way into the infinite future. Unlike in EOB, where the connection to the orbit is lost at the light ring, we obtain the entire merger-ringdown waveform without having to switch to a separate ringdown approximation. Like in the inspiral and transition to plunge, this enables an offline-online split in which we can pre-compute waveform ingredients as functions on the orbital phase space. 

Our phase-space formulation differs in two ways from previous approaches to modelling the merger-ringdown at leading order within GSF theory~\cite{Sundararajan:2007jg,Sundararajan:2008zm,Sundararajan:2010sr,Rifat:2019ltp,Islam:2022laz}: first, our approach maintains the rapid waveform generation framework of the inspiral; second, it can be systematically applied beyond leading order. 
We ultimately aim to combine our treatment of the plunge with the treatments of the inspiral and transition to plunge in order to construct a model that can seamlessly and rapidly generate complete IMR waveforms for asymmetric binaries. In the present paper, we limit our ambitions to three goals: (i) developing an appropriate ``post-geodesic'' expansion of the motion and field equations in the plunge regime; (ii) implementing the method to generate plunge waveforms at leading, geodesic order; (iii) exploring this leading-order waveform's features and accuracy.

For our formulation of the post-geodesic expansion in the plunge, we emphasize how the early-time behaviour of this expansion, when the secondary is near the ISCO, appropriately matches the late-time behaviour of our transition-to-plunge expansion. We then make use of this matching in our implementation of the leading, geodesic-order plunge. We compute plunge waveforms at that order using a Fourier transform adapted to the phase-space representation of the problem. Our calculations then closely follow earlier ones in Refs.~\cite{Hadar:2009ip,Folacci:2018cic}. However, we go beyond those calculations by placing them within a framework that can be applied at higher orders and by using the matching to the transition to plunge to more rigorously justify various steps.

With the geodesic plunge waveform in hand, we then explore how well it can be separated into two distinct segments: an extended inspiral and a ringdown, as in EOB. Our final waveform is given by an inverse Fourier transform, which presents two clear approximations. Before the waveform's peak amplitude, we can approximate the inverse Fourier transform by a stationary-phase approximation (SPA), in which the waveform frequency becomes equal to an integer multiple of the orbital frequency; this is in the spirit of EOB's waveform generation in its extended inspiral. The inverse Fourier transform can also be written in terms of a sum of quasinormal modes (QNMs), power-law tails, and prompt response, and we are able to internally compute the excitation coefficients of the QNMs. In line with EOB's basic description of merger, we find that the SPA works remarkably well until near the waveform's peak, and we find good agreement with the QNM sum after the peak. However, neither approximation is accurate at the waveform's peak, and stitching the two together does not reproduce the full plunge waveform. It is possible that inclusion of higher-order terms in the SPA will bring it closer to the peak. On the other hand, the QNM sum clearly breaks down near the peak and can only be sensibly used at times $\gtrsim 10M$ after the peak, consistent with studies of numerical fits to numerical merger-ringdown waveforms~\cite{Cheung:2023vki,Lim:2022veo,Carullo:2024smg,Mitman:2025hgy}.

The paper is structured as follows. Section~\ref{sec:merger-ringdown_phase-space} contains the phase-space description of our merger-ringdown model. In Sec.~\ref{sec:orbital_motion} we present the post-geodesic expansion of the orbital motion and asymptotically match it with the late-time transition-to-plunge solution of Ref.~\cite{Kuchler:2024esj}. We present the plunge field equations through second order in the mass ratio in Sec.~\ref{sec:field_equations}. We then proceed to solve the Regge-Wheeler-Zerilli equations to obtain first-order plunge waveforms in Sec.~\ref{sec:first-order_wf}. This section also explains the stationary-phase approximation to the plunge waveform and the construction of the QNM sum. In Sec.~\ref{sec:imlpementation_and_comparison} we present our numerical implementation, perform internal consistency checks and compare our first-order waveforms to NR simulations. Finally, we discuss our results and future directions in Sec.~\ref{sec:conclusions}.

We work in geometric units where the gravitational constant $G$ and the speed of light $c$ are set to unity, $G=c=1$. The small mass ratio is defined as $\varepsilon\coloneqq m_p/M$. It will also be useful to introduce the large mass ratio $q\coloneqq1/\varepsilon$.

\vspace{6pt} \noindent {\bf Data availability.} For the ease of reproducibility of our results, we provide several Mathematica notebooks in ancillary files~\cite{AncillaryFiles}.

\section{Merger-ringdown in a phase-space description}\label{sec:merger-ringdown_phase-space}

Before detailing our method, in this section we explain how the merger-ringdown regime can be described within the same phase-space paradigm as the inspiral and transition to plunge. For concreteness, we specialize immediately to the case of a nonspinning particle orbiting a slowly spinning primary black hole.

\subsection{Self-force primer}

We start from the equations of second-order self-force theory in the self-consistent approach~\cite{Pound:2012nt,Pound:2012dk,Pound:2017psq,Upton:2021oxf}, which are valid on all timescales. These equations will then be expanded on the relevant timescales of the inspiral, transition to plunge, and plunge.

The spacetime metric is split into a background plus a perturbation,  $g_{\alpha\beta}+h_{\alpha\beta}$, where $g_{\alpha\beta}$ is the Schwarzschild metric of the primary black hole as if it were isolated, with constant mass $M$. The perturbation $h_{\alpha\beta}\sim \varepsilon$, due to the presence of the orbiting secondary, encodes all corrections sourced by the secondary as well as all nonlinear effects of the two bodies' gravitational interaction, such as the primary's slow accumulation of mass and spin due to absorption of radiation. $h_{\alpha\beta}$ will be expanded in a series for small $\varepsilon$, but the form of that series will depend on the stage of the binary's evolution.

Without specifying the form of that expansion, we can write the Einstein equations perturbatively in $h_{\alpha\beta}$ as
\begin{align}\label{eq:EFE self-consistent}
    \delta G_{\alpha\beta}[h]+\delta^2G_{\alpha\beta}[h,h]+{\cal O}(\varepsilon^3) = 8\pi T_{\alpha\beta},
\end{align}
where $\delta G_{\alpha\beta}$ is the linearized Einstein tensor on the background $g_{\alpha\beta}$, $\delta^2G_{\alpha\beta}$ is quadratic in $h_{\alpha\beta}$,\footnote{Due to its strong singularity at the particle, $\delta^2G_{\alpha\beta}$ is not uniquely defined on a domain that includes the particle's worldline. We assign it the distributional definition of Ref.~\cite{Upton:2021oxf}, which is consistent with our use of the Detweiler stress-energy tensor.} and so on. On the right-hand side, the secondary is represented by the point-particle Detweiler stress-energy tensor~\cite{Detweiler:2011tt,Upton:2021oxf},
\begin{equation}\label{eq:TDet}
    T_{\alpha\beta} = m_p \int \tilde u_\alpha \tilde u_\beta \frac{\delta^4(x^\mu-z^\mu(\tilde\tau))}{\sqrt{-\tilde g}}d\tilde\tau,
\end{equation}
where $z^\mu(\tilde\tau)$ is the particle's orbital trajectory. $\tilde g_{\alpha\beta}=g_{\alpha\beta}+h^{\mathcal R}_{\alpha\beta}$ is a certain \emph{effective} metric, which is regular at the particle's position, and in which the particle moves as a test mass. Here we have split the metric perturbation into $h_{\alpha\beta} = h^{\cal P}_{\alpha\beta} + h^{\mathcal R}_{\alpha\beta}$, where $h^{\cal P}_{\alpha\beta}$ is an analytically known ``puncture'', which is singular at the particle's position, and $h^{\mathcal R}_{\alpha\beta}$ is the regular residual field, as defined in Ref.~\cite{Pound:2014xva}. The parameter $\tilde\tau$ is proper time in $\tilde g_{\alpha\beta}$, and the particle's 4-velocity is $\tilde u^\alpha \coloneqq dz^\alpha/d\tilde\tau$ (with $\tilde u_\alpha \coloneqq\tilde g_{\alpha\beta}\tilde u^\beta$).

While the particle obeys the geodesic equation in the effective metric $\tilde g_{\alpha\beta}$, its motion is accelerated in the background metric $g_{\alpha\beta}$, such that 
\begin{equation}\label{eom tau}
    \frac{D^2z^\mu}{d\tau^2} = f^\mu.
\end{equation}
Here $\tau$ is proper time in $g_{\alpha\beta}$ and $D/d\tau \coloneqq u^\alpha\nabla_\alpha$, with the 4-velocity $u^\alpha\coloneqq dz^\alpha/d\tau $ and covariant derivative $\nabla_\alpha$ compatible with $g_{\alpha\beta}$. Explicitly, the self-force (per unit mass) acting on the secondary is given by~\cite{Pound:2012nt,Pound:2017psq}
\begin{equation}\label{sfexpr}
    f^\mu = -\frac{1}{2}P^{\mu\nu}(\delta_\nu^\rho-h^{\mathcal R \rho}_\nu)(2 \nabla_\alpha h^{\mathcal R}_{\beta\rho} - \nabla_\rho h^{\mathcal R}_{\alpha\beta})u^\alpha u^\beta+{\cal O}(\varepsilon^3), \qquad P^{\mu\nu}\coloneqq g^{\mu\nu} + u^\mu u^\nu.
\end{equation}

Just as the particle's orbit evolves self-consistently in response to the metric perturbation, the primary's mass and spin evolve due to the GW fluxes of energy and angular momentum through its horizon. In order to build a consistent perturbative expansion, we need to take into account this dynamical change. We write the black hole's total mass as $M+\varepsilon\,\delta M$ and total spin as $\varepsilon\,\delta J$, where $M$ is the constant mass of the Schwarzschild background and $\delta M(v,\varepsilon)$ and $\delta J(v,\varepsilon)$ are the evolving corrections (normalized by $\varepsilon$), which are functions of advanced time $v$ along the horizon. These perturbations appear in $h_{\alpha\beta}$, rather than in the background, following the self-consistent prescription of Refs.~\cite{Miller:2020bft,Lewis:InPrep}. Adopting the conventions of Ref.~\cite{Kuchler:2024esj}, we define $\delta M\coloneqq \delta M^+$ and $\delta J\coloneqq \delta M^-$ and collectively denote them as $\delta M^\pm$.

\begin{figure}[t]
    \centering
    \includegraphics[width=0.475\textwidth]{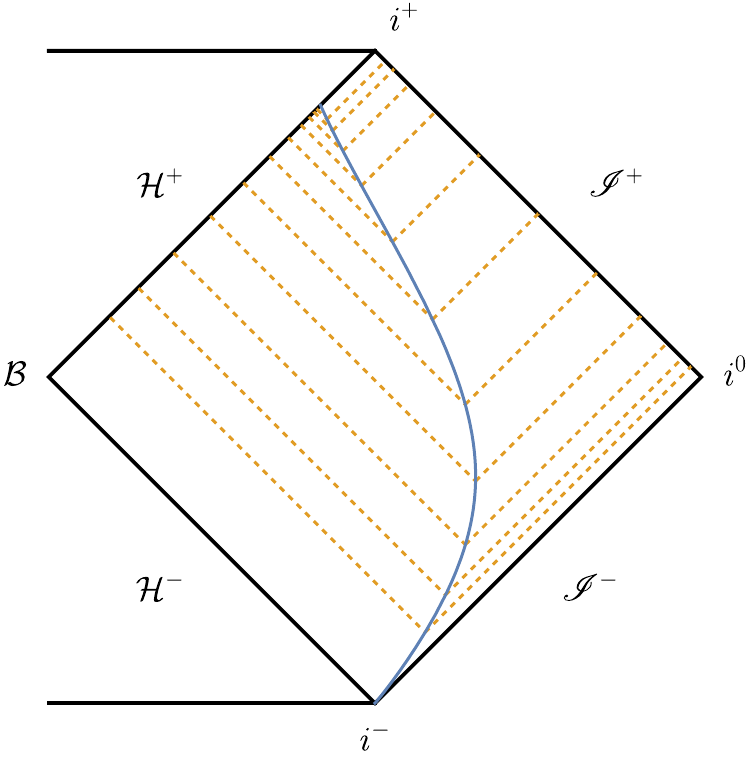}\hfill
    \includegraphics[width=0.475\textwidth]{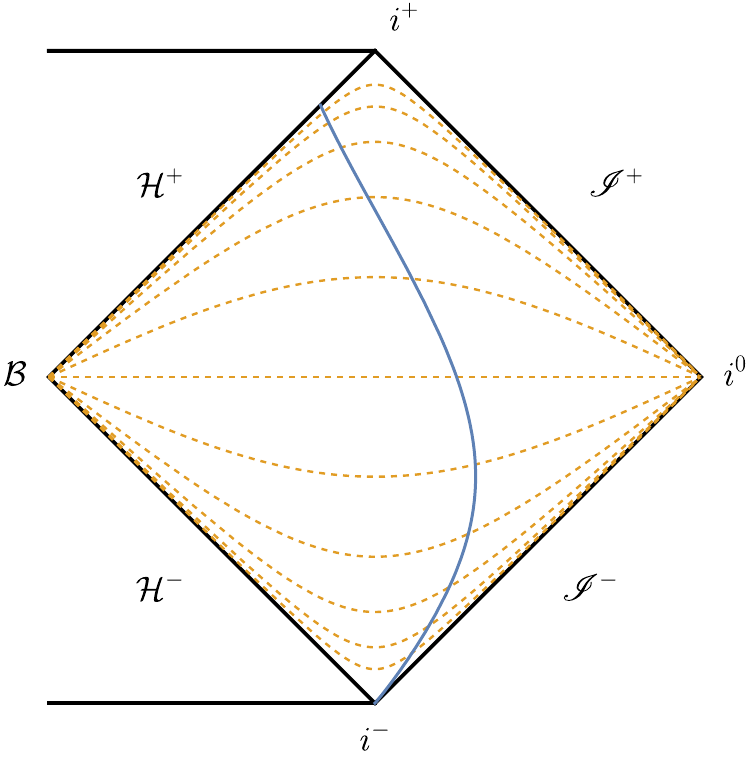}
    \caption{Plunging geodesic (blue curve) in the Schwarzschild exterior, using the standard $\arctan$ compactification~\cite{MTW}. Here the constant $t_0$ in Eq.~\eqref{trG} is chosen such that the geodesic passes through $r=3M$ at $t=0$. The waveform at $\scri^+$ is expressed as a function on the two-body phase space by foliating the spacetime with slices that connect $\scri^+$ to the particle, covering all of $\scri^+$ before the particle passes behind the horizon. Left: surfaces of constant $s$ (orange curves) with sharp $v$-$u$ slicing. Right: surfaces of constant $s$ in $t$ slicing.}
    \label{fig:Penrose}
\end{figure}

To describe the three stages of binary evolution, we now foliate the spacetime with surfaces of constant time $s$. As explained in detail in Refs.~\cite{Miller:2020bft,Miller:2023ers}, $s$ is most conveniently chosen to be a hyperboloidal time (or quasi-hyperboloidal, allowing null segments), such that each $s=\text{constant}$ slice extends from the black hole's future horizon $\mathscr{H}^+$ to future null infinity $\mathscr{I}^+$; this avoids some of the spurious divergences that arise at subleading orders in $\varepsilon$ if using Schwarzschild time $t$~\cite{Pound:2015wva,Cunningham:2024dog}. Here we leave the choice of time generic, only restricting it to the form $s=t-\kappa(x)$, where $t$ is Schwarzschild time and $x$ is the tortoise coordinate. In the present section, for simplicity we also assume $s=t$ at the particle's worldline, but in Secs.~\ref{sec:field_equations} and \ref{sec:RWZt} we will allow for more general choices. Figure~\ref{fig:Penrose} illustrates the two most extreme examples: ``sharp'' slicing in which $s=v$ to the left of the particle and $s=u$ to the right; and ``flat'' slicing $s=t$ everywhere. On each slice, we use Schwarzschild coordinates $x^i=(r,\theta,\phi)$. We can then conveniently use $t$ as a parameter along the particle's worldline, such that $z^\mu(t,\varepsilon) = (t,x^i_p(t,\varepsilon))$, with the spatial trajectory 
\begin{equation}\label{worldline}
    x^i_p(t,\varepsilon)=(r_p(t,\varepsilon),\pi/2,\phi_p(t,\varepsilon)).
\end{equation}

We refer to the space spanned by $(x^i_p,\dot x^i_p, \delta M^\pm)$ as the binary's phase space, where an overdot denotes $d/dt$. A particular binary system evolves through this space, and, via the Einstein field equations, the phase-space evolution is linked to the metric's evolution on spacetime. The particular form of the evolution, and the congruous form of the metric perturbation, depends on the regime: inspiral, transition to plunge, or plunge.

\subsection{Inspiral}

During the binary's quasicircular inspiral stage, the particle's orbit is characterized by a single slowly evolving orbital frequency, 
\begin{equation}\label{OmegaDef}
    \frac{d\phi_p}{dt}=\Omega,
\end{equation}
and a slowly evolving orbital radius. The mechanical variables describing the binary are then the constants $M$ and $m_p$, the orbital phase $\phi_p$, which varies on a ``fast'' orbital time scale $\sim 2\pi/\Omega\sim M$, and the  evolving parameters $J^a=(\Omega, \delta M^\pm)$, which evolve on the ``slow'' radiation-reaction time scale $\sim M/\varepsilon$. The orbital radius, rather than being independent, is expressed in terms of the other variables through an expansion of the form $r_p=r_{(0)}(\Omega)+\varepsilon\,r_{(1)}(J^a)+{\cal O}(\varepsilon^2)$,\footnote{Note that here and throughout this paper, we suppress functional dependence on the background mass~$M$.} where $r_{(0)}(\Omega)=M(M\Omega)^{-2/3}$ is the geodesic relationship, and the correction $r_{(1)}$ is due to the radial self-force.

We define the mechanical variables $(\phi_p,J^a)$ as functions on spacetime by making them constant on slices of constant global time $s$, recalling that $s=t$ on the particle's worldline. Their evolution from slice to slice is then governed by simple differential equations of the form~\cite{Miller:2020bft,Miller:2023ers}%
\begin{subequations}\label{eq:ODEs inspiral}
\begin{align}
    \frac{d\phi_p}{ds} &= \Omega ,\label{eq:phidot - inspiral}
    \\[1ex]
    \frac{d\Omega}{ds} &= \varepsilon \left[ F_{(0)}^\Omega(\Omega)+\varepsilon\,F_{(1)}^\Omega(J_a)+{\cal O}(\varepsilon^2)\right],\label{eq:Omegadot - inspiral}
    \\[1ex]
    \frac{d}{ds}\delta M^\pm &= \varepsilon\,F_{(1)}^\pm(\Omega)+{\cal O}(\varepsilon^2).\label{eq:Mdot - inspiral}
\end{align}
\end{subequations}
Here numeric labels $(n)$ denote the post-adiabatic order ($n$PA) at which a term contributes to the orbital phase. The forcing terms $F^{\Omega}_{(n)}$ are obtained from the self-force using the equation of motion~\eqref{eom tau}, while $F^{\pm}_{(n)}$ are determined from the fluxes of energy and angular momentum through the horizon.

The fundamental assumption in our phase-space approach is that the metric perturbation only depends on $s$ through a dependence on the evolving mechanical variables, such that 
\begin{equation}
    h_{\alpha\beta}(s, x^i,\varepsilon) = h_{\alpha\beta}(\phi_p(s,\varepsilon), J^a(s,\varepsilon), x^i,\varepsilon).   
\end{equation}
We can then treat the perturbation as a function on phase space, $h_{\alpha\beta}(\phi_p,J^a,x^i,\varepsilon)$. In the inspiral regime, we expand this function at fixed values of the phase-space coordinates, such that
\begin{equation}
    h_{\alpha\beta}(\phi_p,J^a,x^i,\varepsilon) = \sum_{n\geq1}\varepsilon^n h^{(n)}_{\alpha\beta}(\phi_p,J^a, x^i).
\end{equation}
Moreover, since $\phi_p$ is a $2\pi$-periodic variable, assuming continuity on phase space allows us to expand the metric perturbations in discrete Fourier series:
\begin{equation}\label{eq:h inspiral}
    h_{\alpha\beta}(\phi_p,J^a, x^i,\varepsilon) = \sum_{n\geq1}\varepsilon^n\sum_{m=-\infty}^\infty h^{(n),m}_{\alpha\beta}(J^a, x^i)e^{-im\phi_p}.
\end{equation}
For the metric perturbations, a label $(n)$ indicates their absolute order in $\varepsilon$.

Concrete calculations are performed by rewriting the Einstein equations as equations on phase space, as we review in Sec.~\ref{sec:field_equations} below. The leading-order field equations, for $h^{(1),m}_{\alpha\beta}$, are identical to traditional frequency-domain field equations for a particle on a precisely circular geodesic of frequency $\Omega$, even though $\phi_p$ and $J^a$ are never ascribed the time dependence they would have for a geodesic. The forcing functions $F^a_{(n)}=(F^\Omega_{(n)},F^\pm_{(n)})$ in Eqs.~\eqref{eq:Omegadot - inspiral} and \eqref{eq:Mdot - inspiral} are then calculated from the amplitudes $h^{(n),m}_{\alpha\beta}$.

In this approach, the asymptotic waveform is obtained as a function on phase space simply by taking the $r\to\infty$ limit of Eq.~\eqref{eq:h inspiral}. Rapid waveform generation is made possible by the fact that the waveform's inputs (the amplitudes $h^{(n),m}_{\alpha\beta}$ and forcing functions $F^{\Omega}_{(n)}$ and $F^\pm_{(n)}$) are all pre-computed in advance as functions on phase space, prior to specifying the value of $\varepsilon$ or any particular trajectory through phase space. Given the pre-computed inputs, waveforms are then generated by solving the ordinary differential equations~\eqref{eq:ODEs inspiral} to determine the waveform's time dependence.

\subsection{Transition to plunge}

When the particle approaches the ISCO, which lies at a radius $r_*\coloneqq 6M$ or frequency $\Omega_*\coloneqq 1/(6^{3/2}M)$,\footnote{Note that we chose to use $x$ rather than $r^*$ to denote the tortoise coordinate in order to avoid confusion with the numerical constant $r_*$.}  the frequency begins to evolve more rapidly, and the assumption that $d\Omega/dt\sim \varepsilon$ breaks down. The particle then transitions across the ISCO, on a time scale $\sim M/\varepsilon^{1/5}$ and over a frequency band of size $\sim \varepsilon^{2/5}$.

We describe this transition to plunge in essentially the same way as we did the inspiral. In place of $\Omega$, we adopt a scaled frequency  
\begin{equation}\label{eq:DOmega}
    \Delta\Omega \coloneqq \frac{\Omega-\Omega_*}{\varepsilon^{2/5}},
\end{equation}
which is of order unity in the transition regime. We then work with mechanical variables $\phi_p$ and $\Delta J^a = (\Delta\Omega,\delta M^\pm)$, and Eqs.~\eqref{eq:ODEs inspiral} are replaced by evolution equations of the form~\cite{Kuchler:2024esj}
\begin{subequations}\label{eq:ODEs transition}
\begin{align}
    \frac{d\phi_p}{ds} &= \Omega = \Omega_* + \varepsilon^{2/5}\Delta\Omega,\label{eq:phidot - transition}
    \\[1ex]
    \frac{d\Delta\Omega}{ds} &= \varepsilon^{1/5} \left[ F_{[0]}^{\Delta\Omega}(\Delta\Omega)+\varepsilon^{2/5} F_{[2]}^{\Delta\Omega}(\Delta\Omega)+\varepsilon^{3/5} F_{[3]}^{\Delta\Omega}(\Delta J^a)+{\cal O}(\varepsilon^{4/5})\right],\label{eq:DOmegadot - transition}
    \\[1ex]
    \frac{d}{ds}\delta M^\pm &= \varepsilon\,F_{[3]}^\pm(\Delta\Omega)+{\cal O}(\varepsilon^{6/5}). \label{eq:Mdot - transition}
\end{align}
\end{subequations}
We use numeric labels $[n]$, rather than $(n)$, to denote the post-leading-transition order ($n$PLT) at which a term contributes to the orbital phase\footnote{The quantity $F^{\delta M^\pm}_{[n]}$ defined in Ref.~\cite{Kuchler:2023jbu} is now denoted as $F^{\pm}_{[n+3]}$.}. Here, functions on phase space admit series expansions in powers of the small parameter $\varepsilon^{1/5}$ rather than in powers of~$\varepsilon$. This behaviour, as well as the $\varepsilon^{2/5}$ scaling of $\Delta\Omega$ in Eq.~\eqref{eq:DOmega}, is readily derived from the requirement that the transition-to-plunge expansion asymptotically matches the expansions in the inspiral and plunge. 

Although the frequency $\Omega$ now evolves at a rate $d\Omega/ds =\varepsilon^{2/5}(d\Delta\Omega/ds) \sim \varepsilon^{3/5}$, the rate of change of $\delta M^\pm$ remains ${\cal O}(\varepsilon)$. This is true in all regimes, including even the plunge, because the fluxes of energy and angular momentum across the horizon are always ${\cal O}(\varepsilon^2)$, as they are proportional to the square of the time derivative of the metric perturbation (and we recall that $\delta M^\pm$ are  normalized by $\varepsilon$).

Just as for the inspiral, we expand the metric perturbation for small $\varepsilon$ at fixed values of our phase-space variables:
\begin{equation}\label{eq:h transition}    
    h_{\alpha\beta} = \sum_{m=-\infty}^\infty\Bigl[\varepsilon\, h^{[5],m}_{\alpha\beta}(\Delta J^a,x^i) + \varepsilon^{7/5} h^{[7],m}_{\alpha\beta}(\Delta J^a,x^i)+ \varepsilon^{8/5} h^{[8],m}_{\alpha\beta}(\Delta J^a,x^i)+{\cal O}(\varepsilon^{9/5}) \Bigr]e^{-im\phi_p}.
\end{equation}
Similarly to the inspiral, for the metric perturbations a label $[n]$ indicates the absolute order of $\varepsilon^{1/5}$. Here the dependence on $\varepsilon^{1/5}$ is inherited from the orbital evolution through the field equations. Like in the inspiral, we express the field equations as equations on phase space, allowing us to directly compute the metric amplitudes  $h^{[n],m}_{\alpha\beta}(\Delta J^a,x^i)$ and forcing functions $F^a_{[n]}(\Delta J^b)$ as functions of $\Delta J^a$. Again, we review this method in Sec.~\ref{sec:field_equations}; and again, the forcing functions $F^{a}_{[n]}(\Delta J^b)$ can be computed from the amplitudes $h^{[n],m}_{\alpha\beta}$.

Since we are able to precompute all the waveform ingredients, we maintain the rapid waveform-generation framework of the inspiral. Given the waveform amplitudes and forcing functions, we rapidly generate waveforms by solving the orbital evolution equations~\eqref{eq:ODEs transition}.

\subsection{Plunge}\label{sec:plunge summary}

Once the particle has fallen sufficiently far below the ISCO, after a time $\gtrsim M/\varepsilon^{1/5}$, the transition-to-plunge approximation breaks down and the orbital frequency begins to evolve on a time scale $\sim M$. This occurs as the steepness of the radial potential begins to drive the inward motion, dominating over radiation-reaction effects. The secondary then plunges following a nearly geodesic orbit in the primary's geometry with almost constant orbital energy and angular momentum~\cite{Ori:2000zn}.

One might intuitively doubt that the phase-space picture can be extended to this final plunge regime. That intuition stems from thinking that the ringdown portion of the waveform occurs in some sense ``after'' the particle has fallen behind the black hole horizon. However, the meaning of ``after'' entirely depends on one's choice of spacetime foliation. From the perspective of an asymptotic observer, the particle \emph{never} falls behind the horizon. Using either our sharp or flat slicings, we can foliate the entire black hole exterior with constant-$s$ slices, and the particle's trajectory intersects every such slice, never crossing the horizon at any time $s<\infty$. See again Fig.~\ref{fig:Penrose} for an illustration of this. As a consequence, the entire merger and ringdown can be parameterized by the binary's trajectory through phase space. 

In the bulk of the paper, we show how to apply this idea to develop an appropriate expansion for the plunge. The fundamental difference between the plunge and the other regimes is that the orbital frequency now evolves on the same time scale $\sim M$ as the orbital phase. Hence, instead of equations of the form~\eqref{eq:ODEs inspiral} or \eqref{eq:ODEs transition}, in the plunge we have equations of the form
\begin{subequations}\label{eq:ODEs plunge}
\begin{align}
    \frac{d\phi_p}{ds} &= \Omega ,\label{eq:phidot - plunge}
    \\[1ex]
    \frac{d\Omega}{ds} &= F_{\{0\}}^{\Omega}(\Omega)+\varepsilon\, F_{\{1\}}^{\Omega}(J^a)+{\cal O}(\varepsilon^{2}),\label{eq:Omegadot - plunge}
    \\[1ex] 
    \frac{d}{ds}\delta M^\pm &= \varepsilon\,F_{\{1\}}^\pm(\Omega)+{\cal O}(\varepsilon^2),\label{eq:Mdot - plunge}
\end{align}
\end{subequations}
where we have reverted to phase-space coordinates $J^a=(\Omega,\delta M^\pm)$, and we use labels $\{n\}$ to denote post-geodesic orders ($n$PG). The metric perturbation, still treated as a function on phase space, has a corresponding expansion
\begin{equation}\label{eq:h plunge}    
    h_{\alpha\beta} = \sum_{m=-\infty}^\infty\Bigl[\varepsilon\,h^{\{1\},m}_{\alpha\beta}(J^a,x^i) + \varepsilon^{2} h^{\{2\},m}_{\alpha\beta}(J^a,x^i) +{\cal O}(\varepsilon^{3}) \Bigr]e^{-im\phi_p}.
\end{equation}
The label $\{n\}$ indicates the metric perturbation's absolute order in $\varepsilon$. Just as for the inspiral and transition to plunge, we will rewrite the field equations as equations for the functions $h^{\{n\},m}_{\alpha\beta}(J^a,x^i)$. 

This formulation of merger and ringdown markedly contrasts with EOB's. As alluded to in the Introduction, in the EOB description, the waveform's link to the binary phase space is broken when the effective particle passes the light ring of the effective-one-body metric; after that point, a phenomenological ringdown waveform is attached. By maintaining the waveform's link to the phase-space trajectory, we maintain the structure that enables fast waveform generation: all waveform ingredients are pre-computed by solving field equations on phase space, and waveforms are then rapidly generated by solving the evolution equations~\eqref{eq:ODEs plunge}.

We also emphasize that our treatment of the plunge fundamentally differs from an expansion around a geodesic solution, even though we refer to our expansion as ``post-geodesic'' for lack of a better name. An expansion around a geodesic solution would involve an expansion of the phase-space trajectory itself, as in $\phi_p(t,\varepsilon) = \phi_0(t) + \varepsilon\,\phi_1(t) + {\cal O}(\varepsilon^2)$, $\Omega(t,\varepsilon) = \Omega_0(t) + \varepsilon\,\Omega_1(t) + {\cal O}(\varepsilon^2)$, and $\delta M^\pm(t,\varepsilon) = \delta M^\pm_0(t)+{\cal O}(\varepsilon)$, where a subscript 0 denotes a geodesic solution. Correspondingly, the metric perturbation would be expanded as $h_{\alpha\beta} = \varepsilon\,h^{\{1\}}_{\alpha\beta}(s,x^i) + \varepsilon^2 h^{\{2\}}_{\alpha\beta}(s,x^i)+{\cal O}(\varepsilon^3)$, where $h^{\{1\}}_{\alpha\beta}$ would be sourced by the particle on the geodesic trajectory. This would be a ``Gralla-Wald'' formulation~\cite{Gralla:2008fg,Pound:2015fma}, which is an expansion in powers of $\varepsilon$ at fixed spacetime coordinates $(s,x^i)$; in our approach, we instead expand all quantities, including $\dot\Omega$ and $\dot{\delta M}{}^\pm$, at fixed values of $(\phi_p,J^a,x^i)$. Our leading perturbation $h^{\{1\}}_{\mu\nu}$ is a function of $(\phi_p,\Omega,\delta M^\pm)$, which becomes a function of time when evaluated along phase-space trajectories obtained by solving Eqs.~\eqref{eq:ODEs plunge}. Those phase-space trajectories are only geodesic trajectories when $F^a_{\{1\}}$ and higher terms are omitted. 

Historically, a Gralla-Wald expansion has not been used in the inspiral because it is only valid on time scales much shorter than the dephasing time $\sim M/\sqrt{\varepsilon}$ over which the accelerated trajectory dephases significantly from the geodesic trajectory~\cite{Pound:2015wva}. This restriction would not in itself be problematic in the transition or plunge, which occur on the time scales $\sim M/\varepsilon^{1/5}$ and $\sim M$, respectively. However, a Gralla-Wald expansion would not allow us to assemble the inspiral, transition to plunge and plunge in an efficient asymptotically matched expansion scheme.

By treating all three regimes in the same manner, we put the problem in a uniform offline/online format. 
Ultimately, one can hybridize the waveforms between the inspiral, transition-to-plunge, and plunge regimes at the level of the precomputed forcing functions $F_{(n)}^a$, $F_{[n]}^a$, and $F_{\{n\}}^a$ and the amplitudes $h^{(n),m}_{\alpha\beta}$, $h^{[n],m}_{\alpha\beta}$, $h^{\{n\},m}_{\alpha\beta}$ (with $r\to\infty$), such that the full IMR waveform can be generated from pre-computed ingredients by solving a single set of three ordinary differential equations for $\phi_p(s,\varepsilon)$, $\Omega(s,\varepsilon)$, and $\delta M^\pm(s,\varepsilon)$. 

In this paper, we defer the presentation of such a hybridization, focusing on the plunge regime and its asymptotic matching to the transition-to-plunge regime.

\section{Plunge: orbital motion}\label{sec:orbital_motion}

We now develop the equations~\eqref{eq:phidot - plunge} and \eqref{eq:Omegadot - plunge} describing the plunging orbit. We work in Schwarzschild coordinates $(t,r,\theta,\phi)$, such that the background metric reads
\begin{align}
    ds^2=-f(r)dt^2+\frac{dr^2}{f(r)}+r^2 \left(d\theta^2 + \sin^2\theta \, d\phi^2\right), \qquad f(r)\coloneqq1-\frac{2M}{r},
\end{align}
and the particle's spatial trajectory $x^i_p(t,\varepsilon)$ is as in Eq.~\eqref{worldline}.

We present two different formulations of the plunge expansion, based on two different choices of phase-space coordinates $(\phi_p,J^a)$. In Sec.~\ref{sec:expfixedOmega}, we choose $J^a=(\Omega, \delta M^\pm)$ as in Sec.~\ref{sec:plunge summary}. This choice exhibits potentially problematic features because $\Omega$ does not provide a single coordinate chart over the whole plunge; it increases to a maximum at the light ring and then decreases to zero at the horizon, such that $(\phi_p,\Omega,\delta M^\pm)$ actually represents two distinct coordinate patches (completely analogous to the ``isofrequency'' orbits described in Ref.~\cite{Warburton:2013yj}). In Sec.~\ref{sec:expfixedrp} instead we make a choice that avoids this issue: $J^a=(r_p, \delta M^\pm)$, where $r_p$ is the orbital radius. Since the two formulations only differ in their choice of phase-space coordinates, they are formally equivalent. In Appendix~\ref{app:equivalence} we derive the transformation between them. 

In either approach, the variables $J^a$ satisfy evolution equations of the form
\begin{equation}\label{F(Delta)Ja}
    \frac{dJ^a}{dt} = F^{a}(J^b,\varepsilon) = F^a_{\{0\}}(J^b) + \varepsilon\,F^a_{\{1\}}(J^b) + {\cal O}(\varepsilon^2).
\end{equation}
If $J^a=(\Omega,\delta M^\pm)$, then the functions $F^a_{\{n\}}$ are the only needed orbital input in the online waveform generation described in the previous section. If $J^a = (r_p,\delta M^\pm)$, then $\Omega$ in Eq.~\eqref{eq:phidot - plunge} becomes $\Omega(r_p,\varepsilon)=\Omega_{\{0\}}(r_p)+\varepsilon\,\Omega_{\{1\}}(J^a)+{\cal O}(\varepsilon^2)$, and the functions $\Omega_{\{n\}}$ become necessary inputs in the waveform generation.

To derive the forcing functions $F^a_{\{n\}}$ and the frequency corrections $\Omega_{\{n\}}$, we rewrite the orbital equation of motion~\eqref{eom tau} in terms of $t$. Defining the redshift $U\coloneqq dt/d\tau$, we can write the four-velocity $u^\mu\coloneqq dz^\mu/d\tau$ as
\begin{equation}\label{4vel}
    u^\mu(J^a,\varepsilon) = U\left(1, \frac{dr_p}{dt}, 0, \Omega\right).
\end{equation}
If we choose $J^a=(\Omega, \delta M^\pm)$, then $dr_p/dt$ here is expressed in terms of $F^a_{\{n\}}$ by applying the chain rule to $r_p(t,\varepsilon)=r_p(J^a(t,\varepsilon),\varepsilon)$. The normalization of the four-velocity for timelike curves, ${g_{\mu\nu}u^\mu u^\nu=-1}$, leads to an equation for the redshift,
\begin{equation}\label{4velnorm}
    U^{-2}=-g_{\mu\nu}\frac{dz^\mu}{dt}\frac{dz^\nu}{dt}.
\end{equation}
Rewritten in terms of the parameter $t$, the equation of motion~\eqref{eom tau} becomes
\begin{equation}\label{eom}
    \frac{d^2z^\mu}{dt^2} + U^{-1}\frac{dU}{dt}\frac{dz^\mu}{dt} + \Gamma^{\mu}_{\nu\sigma}\frac{dz^\nu}{dt}\frac{dz^\sigma}{dt} = U^{-2}f^{\mu},
\end{equation}
where $\Gamma^\mu_{\nu\sigma}$ are the background Schwarzschild Christoffel symbols. The self-force $f^\mu$ has only two independent components because $f^\theta=0$ on equatorial orbits and because the normalization ${g_{\mu\nu}u^\mu u^\nu=-1}$ implies $u^\mu f_\mu = 0$.

Obtaining the forcing functions $F^\Omega_{\{n\}}$ will additionally require information from the transition-to-plunge regime. This is because the PG expansion is not self-contained; without information from the transition to plunge, we would have no way of determining \emph{which} geodesic is the correct one to expand around, for example. Picking out the correct PG dynamics requires asymptotic matching conditions between the PG and PLT expansions. These conditions come from the requirement that the PG and PLT expansions commute. If we expand a function for small $\varepsilon$ at fixed $\Omega$ (a PG expansion) and then re-expand for small $\varepsilon$ at fixed $\Delta\Omega$ (a PLT expansion), we must obtain the same result as we do by first expanding for small $\varepsilon$ at fixed $\Delta\Omega$ and then re-expanding for small $\varepsilon$ at fixed $\Omega$. In both cases, we obtain a double series in $\varepsilon$ and $(\Omega-\Omega_*)$, and the two double series must agree term by term. Equivalently, we can say that a late-time expansion (for $M\Delta\Omega\gg1$) of the PLT dynamics must match, term by term, an early-time expansion ($M(\Omega-\Omega_*)\ll1$) of the PG dynamics.

\subsection{Expansion at fixed orbital frequency}\label{sec:expfixedOmega}

We start by expanding the orbital radius $r_p$ and the redshift $U$ in powers of the mass ratio $\varepsilon$ at fixed phase-space coordinates $\Omega$ and $\delta M^\pm$:
\begin{align}
    r_p(\Omega,\delta M^\pm,\varepsilon) &= r_{\{0\}}(\Omega) + \sum_{n=1}^\infty \varepsilon^n r_{\{ n\}}(\Omega,\delta M^\pm),
    \\[1ex]
    U(\Omega,\delta M^\pm,\varepsilon) &= U_{\{0\}}(\Omega) + \sum_{n=1}^\infty \varepsilon^n U_{\{ n\}}(\Omega,\delta M^\pm).
\end{align}
Note that these quantities are independent of $\phi_p$: as a consequence of the background's axisymmetry, the metric perturbation only ever depends on $\phi_p$ in the combination $(\phi-\phi_p)$~\cite{Pound:2021qin}, such that it is independent of $\phi_p$ when evaluated on the particle at $\phi=\phi_p$; this implies the self-force and other dynamical quantities derived from it (such as $r_{\{n\}}$) are likewise independent of $\phi_p$.

Since \textit{all} functions are expanded at fixed phase-space coordinates, we also expand the coordinates' rates of change (with respect to $t$) as
\begin{align}\label{dOmegafixedOmega}
    \dot\Omega(\Omega,\delta M^\pm,\varepsilon) &= F^\Omega_{\{0\}}(\Omega) + \sum_{n =1}^\infty \varepsilon^n F^\Omega_{\{ n\} }(\Omega,\delta M^\pm),
    \\[1ex]\label{ddeltaMpmfixedOmega}
    \delta\dot M{}^\pm( \Omega,\delta M^\pm,\varepsilon) &= \sum_{n =1}^\infty \varepsilon^{n} F^{\pm}_{\{ n\} }(\Omega,\delta M^\pm). 
\end{align}
Finally, the self-force can be similarly expanded as
\begin{subequations}\label{fmufixedOmega}
\begin{align}
    f^\mu(\Omega,\delta M^\pm,\varepsilon) &= \sum_{n=1}^\infty \varepsilon^n f^\mu_{\{n\}}(\Omega,\delta M^\pm).
\end{align}
\end{subequations}

We now substitute the post-geodesic expansion of the worldline~\eqref{worldline} and the four-velocity~\eqref{4vel} into the equation of motion~\eqref{eom} and the normalization condition~\eqref{4velnorm}. This leads straightforwardly to a sequence of equations for $r_{\{n\}}$, $U_{\{n\}}$, and $F^a_{\{n\}}$, describing the plunging motion at each $n$PG order. Those equations are then completed using information from the transition-to-plunge regime.

\subsubsection{Leading-order match with the late-time transition-to-plunge solution}\label{sec:leading-order match}

Our derivation of the 0PG dynamics in the next section will crucially rely on information from the asymptotic match with the transition to plunge close to the ISCO. Here we give a brief overview of how this matching works, focusing on the 0PG information it implies; we defer a more detailed analysis to Sec.~\ref{sec:match}.

The transition-to-plunge expansions to 2PLT order for the orbital radius and the rate of change of the orbital frequency read~\cite{Kuchler:2024esj}
\begin{align}\label{rp_transition}
    r_p &= 6M + \varepsilon^{2/5}\left[r_{[0]}(\Delta\Omega) + \varepsilon^{2/5}r_{[2]}(\Delta\Omega) + {\cal O}(\varepsilon^{3/5})\right],
    \\[1ex]\label{FDO_transition}
    \frac{d\Omega}{dt} &= \varepsilon^{3/5}\left[F^{\Delta\Omega}_{[0]}(\Delta\Omega) + \varepsilon^{2/5}F^{\Delta\Omega}_{[2]}(\Delta\Omega) + {\cal O}(\varepsilon^{3/5})\right].
\end{align}
The terms $r_{[n]}$ are algebraically determined, while $F^{\Delta\Omega}_{[n]}$ solve sourced linearized Painlev\'e transcendental equations of the first kind~\cite{Kuchler:2024esj}. In the late-time limit $\Delta\Omega\to+\infty$ (where the transition to plunge asymptotically matches with the plunge) we have
\begin{equation}
    r_{[0]} = r_{[0]}^{(2,1)} \Delta\Omega, \qquad r_{[2]} = r_{[2]}^{(4,2)}\Delta\Omega^2 + {\cal O}(\Delta\Omega^{-1}),
\end{equation}
and
\begin{equation}
    F^{\Delta\Omega}_{[0]} = F_{[0]}^{(3,3/2)}\Delta\Omega^{3/2} + {\cal O}(\Delta\Omega^{-1}), \qquad F^{\Delta\Omega}_{[2]} = F_{[2]}^{(5,5/2)}\Delta\Omega^{5/2} + {\cal O}(\Delta\Omega^{0}),
\end{equation}
where $r_{[0]}^{(2,1)}$, $r_{[2]}^{(4,2)}$, $F_{[0]}^{(3,3/2)}$ and $F_{[2]}^{(5,5/2)}$ are numerical constants given in Appendix~\ref{app:earlylate_transition}. Plugging these solutions into Eqs.~\eqref{rp_transition} and \eqref{FDO_transition} and re-expressing $\Delta\Omega$ as $(\Omega-\Omega_*)/\varepsilon^{2/5}$ yields
\begin{align}\label{rp_transition_late}
    r_p &= \left[6M + r_{[0]}^{(2,1)} (\Omega - \Omega_*) + r_{[2]}^{(4,2)} (\Omega - \Omega_*)^2 + {\cal O}\left[(\Omega - \Omega_*)^3\right] \right] + {\cal O}(\varepsilon),
    \\[1ex]\label{FDO_transition_late}
    \frac{d\Omega}{dt} &= \left[F_{[0]}^{(3,3/2)}(\Omega - \Omega_*)^{3/2} + F_{[2]}^{(5,5/2)}(\Omega - \Omega_*)^{5/2} + {\cal O}\left[(\Omega - \Omega_*)^{7/2}\right] \right] + {\cal O}(\varepsilon).
\end{align}
The ${\cal O}(\varepsilon)$ terms come from the subleading terms in the late-time expansion of the PLT orders, e.g., the ${\cal O}(\Delta\Omega^{-1})$ term in $F^{\Delta\Omega}_{[0]}$. Subleading terms in the square brackets originate from the leading-order terms in the late-time expansion of higher PLT orders. 

As explained above, the $\varepsilon^0$ terms in Eqs.~\eqref{rp_transition_late} and \eqref{FDO_transition} must match, term by term, with the near-ISCO expansion of 0PG quantities in the plunge. Hence, we can infer that, in the near-ISCO limit $\Omega\to\Omega_*$,
\begin{align}\label{rG_plunge_early}
    r_{\{0\}} &= 6M + r_{[0]}^{(2,1)} (\Omega - \Omega_*) + r_{[2]}^{(4,2)} (\Omega - \Omega_*)^2 + {\cal O}\left[(\Omega - \Omega_*)^3\right],
    \\[1ex]\label{FG_plunge_early}
    F^\Omega_{\{0\}} &= F_{[0]}^{(3,3/2)}(\Omega - \Omega_*)^{3/2} + F_{[2]}^{(5,5/2)}(\Omega - \Omega_*)^{5/2} + {\cal O}\left[(\Omega - \Omega_*)^{7/2}\right].
\end{align}
From this, we see that the 0PG plunge trajectory must asymptote to a circular orbit with $r_{\{0\}}\to6M$ and $\dot\Omega\to0$---i.e., a circular geodesic infinitesimally below the geodesic ISCO.

\subsubsection{Geodesic order}\label{sec:0PG} 

At leading order, we can use the $t$ component of the equation of motion~\eqref{eom} to solve for $\partial_\Omega^2 r_{\{0\}}$:
\begin{equation}\label{ddrG}
\begin{split}
    \partial_\Omega^2 r_{\{0\}} =&\, \frac{3M (\partial_\Omega r_{\{0\}})^2}{r_{\{0\}}(r_{\{0\}}-2M)} - \frac{\partial_\Omega F^\Omega_{\{0\}} \partial_\Omega r_{\{0\}}}{F^\Omega_{\{0\}}}\\ 
    &+ \frac{2M^2 - r_{\{0\}}(r_{\{0\}}^2\Omega^2(r_{\{0\}}-4M) + M) - r_{\{0\}}^4\Omega(r_{\{0\}}-2M)/\partial_\Omega r_{\{0\}}}{r_{\{0\}}^3 (F^\Omega_{\{0\}})^2}.
\end{split}
\end{equation}
Plugging this result into the radial component of the equation of motion allows us the obtain the following simple expression for $\partial_\Omega r_{\{0\}}$:
\begin{equation}\label{partialrG}
    \partial_\Omega r_{\{0\}} = -\frac{r_{\{0\}} (r_{\{0\}} - 2M)}{2\Omega (r_{\{0\}} - 3M)}.
\end{equation}
We can use this to simplify $U_{\{0\}}$, which we obtain from Eq.~\eqref{4velnorm} at order ${\cal O}(\varepsilon^0)$,
\begin{equation}\label{UG}
    U_{\{0\}} = \frac{2\Omega\,r_{\{0\}}^{1/2}(r_{\{0\}} - 3M)}{\sqrt{4\Omega^2(r_{\{0\}} - 3M)^2(r_{\{0\}} - 2M - r_{\{0\}}^3\Omega^2) - (r_{\{0\}} - 2M)r_{\{0\}}^4 (F^\Omega_{\{0\}})^2}}.
\end{equation}

The geodesic orbital energy and angular momentum per unit mass of the plunging particle, $E_{\{0\}}$ and $L_{\{0\}}$, are given by
\begin{subequations}\label{EGLG}
\begin{align}
    E_{\{0\}} &= -u_t^{\{0\}} = -g_{tt}(r_{\{0\}}) U_{\{0\}} = f(r_{\{0\}})U_{\{0\}},
    \\[1ex]
    L_{\{0\}} &= u_\phi^{\{0\}} = g_{\phi\phi}(r_{\{0\}}, \pi/2)\Omega\,U_{\{0\}}  = r_{\{0\}}^2\,\Omega\,U_{\{0\}}.
\end{align}
\end{subequations}
Since $u_t$ and $u_\phi$ are constant along any geodesic of Schwarzschild, $E_{\{0\}}$ and $L_{\{0\}}$ must be constant ``on shell'', when $\Omega$ satisfies $\dot\Omega = F^\Omega_{\{0\}}$. Moreover, making use of Eq.~\eqref{ddrG}, it is straightforward to verify that they are constants even if $\Omega$ satisfies $\dot\Omega=F^\Omega(\Omega,\varepsilon)$ because $dE_{\{0\}}/d\Omega=dL_{\{0\}}/d\Omega=0$. The values of these constants are fixed by the requirement for the plunge to asymptotically match the transition to plunge close to the ISCO. We can take the limit as $\Omega\to\Omega_*$ of Eqs.~\eqref{EGLG} keeping in mind that the geodesic radius reduces to the ISCO radius while $F^\Omega_{\{0\}}$ vanishes as prescribed by Eqs.~\eqref{rG_plunge_early} and \eqref{FG_plunge_early}. We get
\begin{equation}\label{E*L*}
    E_{\{0\}}=E_*\coloneqq \frac{2\sqrt{2}}{3}, \qquad L_{\{0\}}=L_*\coloneqq2\sqrt{3}M,
\end{equation}
which correspond to the (specific) energy and angular momentum of a test particle on the ISCO. The plunging geodesics with these values of $E$ and $L$ are the ``universal'' geodesics studied in the past~ \cite{Hadar:2009ip,Folacci:2018vtf,Folacci:2018cic,Rom:2022uvv,Strusberg:2025qfv}; at 0PG order, all quasicircular inspirals transition onto one of these geodesics, with the only difference between them being a constant shift in $\phi_p$.

\begin{figure}[tb]
\centering
\includegraphics[width=0.8\textwidth]{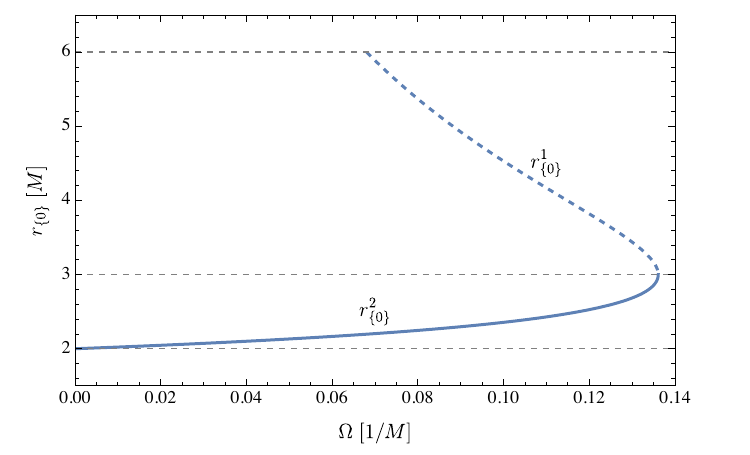}
\caption{Geodesic orbital radius $r_{\{0\}}(\Omega)$ on the two branches obtained by solving Eq.~\eqref{mastereq_plunge}: the early solution $r^1_{\{0\}}(\Omega)$ (dashed curve) from the ISCO to the light-ring frequency, and the late solution $r^2_{\{0\}}(\Omega)$ (solid curve) up to the horizon-crossing. The horizontal dashed lines at $r_{\{0\}}=2M,\, 3M,\, 6M$ indicate the horizon, the light ring and the ISCO, respectively.}
\label{fig:rG_Omega}
\end{figure}

By imposing $E_{\{0\}}/L_{\{0\}} = E_*/L_*$, where $E_{\{0\}}$ and $L_{\{0\}}$ are given in Eq.~\eqref{EGLG} and the constants $E_*$ and $L_*$ are defined in Eq.~\eqref{E*L*}, we obtain a cubic master equation for the orbital radius in terms of the orbital frequency,
\begin{equation}\label{mastereq_plunge}
    \sqrt{2}\Omega r_{\{0\}}^3 - 3 \sqrt{3} M r_{\{0\}} + 6\sqrt{3} M^2 = 0.
\end{equation}
The solution has two branches, as shown in Fig.~\ref{fig:rG_Omega}: on the first branch, $r^1_{\{0\}}(\Omega)$, the orbital frequency increases monotonically from $\Omega=\Omega_*\coloneqq 1/\left(6\sqrt{6}M\right)$ at the ISCO to its light ring value $\Omega=\Omega_{\rm LR}\coloneqq1/\left(3\sqrt{6}M\right)$. On the second branch, $r^2_{\{0\}}(\Omega)$, the orbital frequency monotonically decreases to $\Omega=0$ at the horizon-crossing.

The condition~$(E_{\{0\}})^2=(E_*)^2$, together with Eqs.~\eqref{UG} and \eqref{EGLG}, also yields the expression for the geodesic forcing term,
\begin{equation}\label{FOmegaG}
    F^\Omega_{\{0\}} = \frac{\Omega (r_{\{0\}} - 3M)\left(36 M^2 - 20 M r_{\{0\}} + r_{\{0\}}^2 + 8\Omega^2 r_{\{0\}}^4\right)^{1/2}}{\sqrt{2}r_{\{0\}}^{5/2}\left(2M - r_{\{0\}}\right)^{1/2}}.
\end{equation}
Above the light ring, the frequency increases with $t$ ($F^\Omega_{\{0\}} (r_{\{0\}})>0$ for $3M < r_{\{0\}} < 6M$), while it decreases between the light ring and the horizon ($F^\Omega_{\{0\}} (r_{\{0\}})<0$ for $2M < r_{\{0\}} < 3M$), consistently with Fig.~\ref{fig:rG_Omega}.

\subsubsection{First post-geodesic order} 

At first post-geodesic order, the $t$ and $r$ components of the equation of motion~\eqref{eom} lead to hierarchical differential equations for $r_{\{1\}}$ and $F^\Omega_{\{1\}}$,
\begin{subequations}\label{1PGeqsfixedOmega}
\begin{equation}
\begin{split}
    \partial_\Omega r_{\{1\}} + \frac{r_{\{0\}}^2-6Mr_{\{0\}}+6M^2}{2\Omega(r_{\{0\}}-3M)^2}&r_{\{1\}} =\, \frac{1}{32 r_{\{0\}}^4\Omega^3(r_{\{0\}}-3M)^2}\Biggl[9r_{\{0\}}^3(r_{\{0\}}-2M)^3f^r_{\{1\}}\\
    &- \frac{18\sqrt{2}r_{\{0\}}^{5/2}(2M-r_{\{0\}})^{7/2}\left(r_{\{0\}}^3\Omega^2 - r_{\{0\}} + 2M\right)}{(36 M^2 - 20 M r_{\{0\}} + r_{\{0\}}^2 + 8\Omega^2 r_{\{0\}}^4)^{1/2}}f^t_{\{1\}}\Biggr],
\end{split}
\end{equation}
\begin{equation}
\begin{split}
    &\partial_\Omega F^\Omega_{\{1\}} - \frac{4\Omega \left(3M^2 + \Omega^2 r_{\{0\}}^4 - M (3 \Omega^2 r_{\{0\}}^3 + r_{\{0\}})\right) + r_{\{0\}}^4 F^\Omega_{\{0\}} \partial_\Omega F^\Omega_{\{0\}}}{r_{\{0\}}^4  (F^\Omega_{\{0\}})^2} F^\Omega_{\{1\}} 
    \\[1ex]
    =&\; \frac{1}{4 \Omega^2 r_{\{0\}}^5 (r_{\{0\}}-3M)^2 (r_{\{0\}}-2M)^2 (F^\Omega_{\{0\}})^2}\Biggl[8 \Omega^3 r_{\{0\}}^4 (r_{\{0\}} - 2M) (r_{\{0\}} - 3M)^3 (F^\Omega_{\{0\}})^3 \partial^2_\Omega r_{\{1\}}
    \\[1ex]
    &- 8 \Omega^2 r_{\{0\}}^3 (r_{\{0\}} - 3M)^2 F^\Omega_{\{0\}} \left[4\Omega^4 (r_{\{0\}} - 3M)^3 - 3 M r_{\{0\}} (r_{\{0\}} - 2M) (F^\Omega_{\{0\}})^2\right.\\
    &\left. - \Omega\,r_{\{0\}} (r_{\{0\}}-3M) (r_{\{0\}}-2M) F^\Omega_{\{0\}} \partial_\Omega F^\Omega_{\{0\}}\right]\partial_\Omega r_{\{1\}}
    \\[1ex]
    &- 4\Omega (r_{\{0\}} - 3M) F^\Omega_{\{0\}} \left[2\Omega^2 (r_{\{0\}} - 3M)^2 \left(12M^3 + 3 \Omega^2 r_{\{0\}}^5 + 2 M^2 r_{\{0\}} (6\Omega^2 r_{\{0\}}^2 - 5)\right.\right.\\
    \end{split}
\end{equation}    
\begin{equation}
\begin{split}
    &\left.\left.-M (14 \Omega^2 r_{\{0\}}^4 - 2 r_{\{0\}}^2)\right) - 3 M r_{\{0\}}^4 (r_{\{0\}}^2 - 3M r_{\{0\}} + 2M^2) (F^\Omega_{\{0\}})^2\right] r_{\{1\}}
    \\[1ex]
    &+(r_{\{0\}} - 2M) \left[4\Omega^2 (r_{\{0\}} - 3M)^2 (\Omega^2 r_{\{0\}}^3 - r_{\{0\}} + 2M) + r_{\{0\}}^4 (r_{\{0\}} - 2M) (F^\Omega_{\{0\}})^2\right]^2 f^t_{\{1\}}\Biggr].
\end{split}
\end{equation}
\end{subequations}
Contrary to the 0PG equations, which describe a geodesic in the background spacetime, the motion at 1PG order is driven by both the $t$ and $r$ components of the first-order self-force, $f^t_{\{1\}}$ and $f^r_{\{1\}}$. 

We notice that the post-geodesic expansion derived in this section breaks down at the light ring, where the evolution equations~\eqref{1PGeqsfixedOmega} encounter a pole. This is a consequence of the fact that the frequency does not provide a global coordinate for the whole plunge, as mentioned previously. It motivates an alternative choice of coordinates, which we develop in Sec.~\ref{sec:expfixedrp}.

Like the 0PG dynamics, the 1PG dynamics is not fully specified by the orbital equations of motion~\eqref{1PGeqsfixedOmega}. Since those equations are first-order differential equations, they only determine $F^\Omega_{\{1\}}$ up to a constant. Fixing that constant requires additional input from the transition-to-plunge regime. Moreover, we note that Eq.~\eqref{1PGeqsfixedOmega} becomes singular at the ISCO as $F^\Omega_{\{0\}}$ vanishes there. In general, we expect the post-geodesic expansion to become increasingly singular at the ISCO at higher PG orders. Matching to the transition-to-plunge solution then becomes increasingly important.

\subsection{Asymptotic match with the transition to plunge}\label{sec:match}

In this section, following Ref.~\cite{Kuchler:2023jbu}, we derive the asymptotic match with the late-time transition-to-plunge motion of Ref.~\cite{Kuchler:2024esj}. Specifically, we generalize the brief description in the previous section to 1PG order and high PLT orders. This will allow us to determine how the transition to plunge selects the correct solution to the 1PG equations~\eqref{1PGeqsfixedOmega}, analogously to what we have done at geodesic order in Sec.~\ref{sec:0PG}.

To explain the structure of the asymptotic matching, we consider a function $g(\varepsilon,J^a)$ that admits both a transition-to-plunge expansion (a regular expansion in powers of $\lambda\coloneqq\varepsilon^{1/5}$ at fixed $\Delta J^a$) and a plunge expansion (a regular expansion in powers of $\varepsilon$ at fixed $J^a$). We write the PLT expansion as $g(\lambda, \Delta J^a) = g_* + \lambda^p\sum_{n\geq0}\lambda^n g_{[n]}(\Delta J^a)$; for examples, see Eqs.~\eqref{rp_transition} and~\eqref{FDO_transition} for $r_p$ and $d\Omega/dt$. At late times, where $\Delta\Omega\to+\infty$, we can expand in half-integer powers of $\Delta\Omega$. Since $\Delta\Omega=(\Omega-\Omega_*)/\varepsilon^{2/5}$, 
re-expanding $g$ for large $\Delta\Omega$ is equivalent to re-expanding for small $\varepsilon$ at fixed~$\Omega>\Omega_*$:
\begin{subequations}\label{g late expansion}
\begin{align}
    g(\lambda, \Delta J^a) &= g_* + \sum_{n\geq0}\sum_{m} \lambda^{p+n} \Delta\Omega^{m/2} g_{[n]}^{(p+n,m/2)}(\delta M^\pm) 
    \label{g late expansion - DOmega}\\[1ex]
    &= g_* + \sum_{n\geq0}\sum_{m} \varepsilon^{(p+n-m)/5} (\Omega-\Omega_*)^{m/2} g_{[n]}^{(p+n,m/2)}(\delta M^\pm).\label{g late expansion - Omega}
\end{align}
\end{subequations}
In principle, there is no restriction on the integers $m$. However, the range of $m$ is restricted by the requirement of  asymptotically matching the early-time behaviour of the plunge expansion, which contains only integer powers of the mass ratio, $\varepsilon^i$ with $i=0,1,2,\dots$; see Sec.~\ref{sec:expfixedOmega}. This restricts the powers of $\Delta\Omega$ that can appear in the late-time expansion of a given $n$PLT order:
\begin{equation}
    m=p+n-5i, \qquad n,i\geq0.
\end{equation}
As an example, if we consider the 0PLT forcing term $F^{\Delta\Omega}_{[0]}$ we have the following: as anticipated in Sec.~\ref{sec:leading-order match}, since $p=3$ and $n=0$, the late-time expansion is given by a series expansion with powers $\Delta\Omega^{(3-5i)/2}$. We obtain the coefficients $r_{[n]}^{(p+n,m/2)}$ for $r_p$ and $F_{[n]}^{(p+n,m/2)}$ for $\dot\Omega$ by substituting expansions of the form~\eqref{g late expansion - DOmega} into the transition-regime equations for $r_{[n]}$ and $F^{\Delta\Omega}_{[n]}$ (see Ref.~\cite{Kuchler:2024esj}). These coefficients are given explicitly in Appendix~\ref{app:earlylate_transition}.

We can follow the analogous procedure for the plunge. The plunge expansion of a function $g$ is given by $g(\varepsilon, J^a)=\sum_{i\geq0}\varepsilon^i g_{\{i\}}(J^a)$. Re-expanding close to the ISCO and substituting $(\Omega-\Omega_*)=\lambda^2\Delta\Omega$ we obtain
\begin{equation}\label{g near ISCO}
    g(\varepsilon, J^a) = g_* + \sum_{i\geq0}\sum_{j} \lambda^{5i + j} \Delta\Omega^{j/2} g_{\{i\}}^{(5i+j,j/2)}(\delta M^\pm).
\end{equation}
We can restrict the range of $j$ by requiring the powers of $\lambda$ in the near-ISCO plunge solution to match those in the late-time transition-to-plunge solution~\eqref{g late expansion - DOmega}. This yields the condition
\begin{equation}\label{j plunge}
    j=p+n-5i, \qquad n,i\geq0.
\end{equation}
At geodesic order ($i=0$), we plug the asymptotic expansions~\eqref{g near ISCO} into Eqs.~\eqref{mastereq_plunge} and \eqref{FOmegaG} and solve for the coefficients $g_{\{0\}}^{(j,j/2)}$ with $j\geq p+n$ and $n\geq0$. These coefficients are given explicitly in Appendix~\ref{app:nearISCO_plunge}.

At first post-geodesic order, we need to know the near-ISCO structure of the self-force. The first-order field equations in the plunge regime yield a metric perturbation that contains pieces which are smooth at the ISCO and pieces $\sim F^\Omega_{\{0\}}\sim(\Omega-\Omega_*)^{3/2}$ due to the presence of time derivatives (see Sec.~\ref{sec:RWZt}, where the first-order field equations are presented in a slightly different expansion scheme, but the general statement made here remains unchanged). The self-force inherits this structure through Eq.~\eqref{sfexpr}, where at most one time derivative acts on the metric perturbations and gives rise to half-integer powers $\geq3/2$ of $(\Omega-\Omega_*)$ in the near-ISCO limit. We can therefore write the first-order plunge self-force as
\begin{equation}\label{plungesf_early_Omega}
    f^\mu_{\{1\}}(\Omega\to\Omega_*, \delta M^\pm) = f^\mu_{\{1\},0}(\delta M^\pm) + \sum_{n=2}^\infty f^\mu_{\{1\},n}(\delta M^\pm)(\Omega - \Omega_*)^{n/2},
\end{equation}
where $f^t_{\{1\},n}$ and $f^\phi_{\{1\},n}$ are independent of $\delta M^\pm$ for $n=0,2$.\footnote{For $\mu=t,\phi$, linear-in-$\delta M^\pm$ perturbations towards Kerr appear at first order in Eq.~\eqref{sfexpr} always accompanied by a factor $u^r=\partial_\Omega r_{\{0\}} F^\Omega_{\{0\}} U_{\{0\}}$. Close to the ISCO, $u^r$ is ${\cal O}\left[(\Omega-\Omega_*)^{3/2}\right]$ or higher, as follows from Eqs.~\eqref{rG_plunge_early} and \eqref{FG_plunge_early}. The coefficients in Eq.~\eqref{plungesf_early_Omega} therefore start to depend on $\delta M^\pm$ from $n=3$. This is consistent with the results of Ref.~\cite{Kuchler:2024esj} (i.e., that $f^\mu_{[5]}$ and $f^\mu_{[7]A}$ in that paper do not depend on $\delta M^\pm$ for $\mu=t,\phi$), which becomes manifest from the match of the self-force in Eq.~\eqref{matchingsf} below.} 

Knowing the structure of the self-force, we can solve Eq.~\eqref{1PGeqsfixedOmega} in the near-ISCO limit. From Eqs.~\eqref{g near ISCO} and \eqref{j plunge} we get
\begin{align}\label{r1nearISCO}
    \varepsilon\, r_{\{1\}}(\Omega\to\Omega_*, \delta M^\pm) &= \lambda^5\sum_{j\geq-3} r_{\{1\}}^{(5+j,j/2)} \lambda^j \Delta\Omega^{j/2},
    \\[1ex]\label{F1nearISCO}
    \varepsilon\, F^\Omega_{\{1\}}(\Omega\to\Omega_*, \delta M^\pm) &= \lambda^5\sum_{j\geq-2} F^{(5+j,j/2)}_{\{1\}} \lambda^j \Delta\Omega^{j/2}.
\end{align}
From the transition to plunge we have that $r_{[0]}^{(2,-3/2)}=0$ and $r_{[1]}^{(3,-1)}=0$; see Appendix~\ref{app:earlylate_transition}. Through the asymptotic match this also implies that $r_{\{1\}}^{(2,-3/2)}=r_{\{1\}}^{(2,-1)}=0$, and the series in Eq.~\eqref{r1nearISCO} therefore starts at $\Delta\Omega^{-1/2}$.
Solving Eq.~\eqref{1PGeqsfixedOmega} iteratively in the limit $\Omega\to\Omega_*$ determines all these coefficients, with the exception of $r_{\{1\}}^{(5,0)}$ in Eq.~\eqref{r1nearISCO}. We fix this coefficient by enforcing the match with the transition-to-plunge solution: ${r_{\{1\}}^{(5,0)}=r_{[3]}^{(5,0)} = -54 M^2 f^r_{\{1\},0}}$. The coefficients appearing in the near-ISCO solutions~\eqref{r1nearISCO} and \eqref{F1nearISCO} are given explicitly in Appendix~\ref{app:nearISCO_plunge}. The asymptotic solutions~\eqref{r1nearISCO} and \eqref{F1nearISCO} can then be used to obtain initial conditions to solve Eq.~\eqref{1PGeqsfixedOmega} (once the plunge self-force has been computed) and obtain 1PG solutions that asymptotically match with the transition to plunge close to the ISCO.

In summary, the asymptotic matching has (i) restricted the form of the plunge expansion by restricting the powers of $(\Omega-\Omega_*)$ in the plunge's near-ISCO behaviour, and (ii) fixed the value of a single near-ISCO coefficient, $r_{\{1\}}^{(5,0)}$. This minimal information from the matching (together with the orbital equations of motion, field equations, and boundary conditions) serves to completely fix the plunge solution.

Although the plunge solution requires only a small amount of information from the transition regime, we can use the other term-by-term matching conditions as consistency checks. The conditions for Eqs.~\eqref{g late expansion} and \eqref{g near ISCO} to match term by term are obtained by equating the powers of $\lambda$ and $\Delta\Omega$:
\begin{equation}
    p + n = 5i + j, \qquad m=j.
\end{equation}
Therefore, 
\begin{itemize}
    \item the near-ISCO expansion of the 0PG forcing term $F^\Omega_{\{0\}}$ ($p=3$, $i=0$) is matched by the terms with $m=3+n$ in Eq.~\eqref{g late expansion}, that is, a term $\sim\Delta\Omega^{3/2}$ from 0PLT, a (vanishing) term $\sim\Delta\Omega^2$ from 1PLT, a term $\sim\Delta\Omega^{5/2}$ from 2PLT, \dots
    \item the near-ISCO expansion of the 0PG forcing term $F^\Omega_{\{1\}}$ ($p=3$, $i=1$) is matched by the terms with $m=-2+n$ in Eq.~\eqref{g late expansion}, that is, a term $\sim\Delta\Omega^{-1}$ from 0PLT, a (vanishing) term $\sim\Delta\Omega^{-1/2}$ from 1PLT, a constant term $\sim\Delta\Omega^0$ from 2PLT, \dots
\end{itemize}
and analogously for the orbital radius. This structure of the asymptotic match between the transition-to-plunge and plunge orbital motions is summarized in Table~\ref{tab:matchstructureTP}.
\begin{table}[tb]
    \centering
    \begin{NiceTabularX}{.725\textwidth}{c *2{>{\centering\arraybackslash}X}  c}
    \CodeBefore
    \rowcolors{1}{}{gray!7}
    \Body
    \toprule
        & \text{0PG} & \text{1PG} & $\cdots$  \\ \cmidrule{2-3}
        0PLT & \thead{$r^{(2,1)}_{\{0\}}=r^{(2,1)}_{[0]}$ \\[1ex] $F^{(3,3/2)}_{\{0\}}=F^{(3,3/2)}_{[0]}$}  & \thead{$-$ \\[1ex] $F^{(3,-1)}_{\{1\}}=F^{(3,-1)}_{[0]}$} & $\cdots$\\
        1PLT\vphantom{\thead{$r^{(2,1)}_{(0)}=r^{(2,1)}_{[0]}$ \\[1ex] $F^{(3,-1)}_{(0)}=F^{(3,-1)}_{[0]}$}} & $-$ & $-$ & $\cdots$\\
        2PLT & \thead{$r^{(4,2)}_{\{0\}}=r^{(4,2)}_{[2]}$ \\[1ex] $F^{(5,5/2)}_{\{0\}}=F^{(5,5/2)}_{[2]}$} & \thead{$r^{(4,-1/2)}_{\{1\}}=r^{(4,-1/2)}_{[2]}$ \\[1ex] $F^{(5,0)}_{\{1\}}=F^{(5,0)}_{[2]}$} & $\cdots$\\
        3PLT &  $-$ & \thead{$r^{(5,0)}_{\{1\}}=r^{(5,0)}_{[3]}$ \\[1ex] $F^{(6,1/2)}_{\{1\}}=F^{(6,1/2)}_{[3]}$} & $\cdots$\\
        4PLT & \thead{$r^{(6,3)}_{\{0\}}=r^{(6,3)}_{[4]}$ \\[1ex] $F^{(7,7/2)}_{\{0\}}=F^{(7,7/2)}_{[4]}$} & \thead{$r^{(6,1/2)}_{\{1\}}=r^{(6,1/2)}_{[4]}$ \\[1ex] $F^{(7,1)}_{\{1\}}=F^{(7,1)}_{[4]}$} & $\cdots$\\  
        5PLT & $-$ & \thead{$r^{(7,1)}_{\{1\}}=r^{(7,1)}_{[5]}$ \\[1ex] $F^{(8,3/2)}_{\{1\}}=F^{(8,3/2)}_{[5]}$} & $\cdots$\\
        6PLT & \thead{$r^{(8,4)}_{\{0\}}=r^{(8,4)}_{[6]}$ \\[1ex] $F^{(9,9/2)}_{\{0\}}=F^{(9,9/2)}_{[6]}$} & \thead{$r^{(8,3/2)}_{\{1\}}=r^{(8,3/2)}_{[6]}$ \\[1ex] $F^{(9,2)}_{\{1\}}=F^{(9,2)}_{[6]}$} & $\cdots$\\
        7PLT & $-$ & \thead{$r^{(9,2)}_{\{1\}}=r^{(9,2)}_{[7]}$ \\[1ex] $F^{(10,5/2)}_{\{1\}}=F^{(10,5/2)}_{[7]}$} & $\cdots$\\
        $\vdots$ & $\vdots$ & $\vdots$ & $\ddots$\\
        \bottomrule
    \end{NiceTabularX}
    \caption{Matching conditions between plunge and transition to plunge for the asymptotic coefficients ($r_{\{n\}}^{(p,q)}$, $r_{[n]}^{(p,q)}$, $F_{\{n\}}^{(p,q)}$, and $F_{[n]}^{(p,q)}$) of $\lambda^p(\Delta\Omega)^q$ in the solutions for the orbital radius $r_p$ and the rate of change $d\Omega/dt$. The coefficients are labeled with their PLT ($[n]$) or PG order ($\{n\}$) in addition to the powers of $\lambda=\varepsilon^{1/5}$ and of $\Delta\Omega=(\Omega-\Omega_*)/\lambda^2$ in the asymptotic expansions~\eqref{g late expansion} and \eqref{g near ISCO}. Here we only show nonzero coefficients.}
    \label{tab:matchstructureTP}
\end{table}

We have explicitly verified the matching conditions in Table~\ref{tab:matchstructureTP} for all terms involved in the match between the transition to plunge to 7PLT order and the plunge to 1PG order, using the coefficients listed in Appendices~\ref{app:earlylate_transition} and \ref{app:nearISCO_plunge}. In order to do so successfully, we required the match of the self-force. In practice, one proceeds order by order, obtaining matching conditions for both the orbital motion and the self-force. The matching conditions for the self-force are presented in Appendix~\ref{app:selfforce_matchingcond}.

\subsection{Expansion at fixed orbital radius}\label{sec:expfixedrp}

In this section we derive an alternative to the formulation of Sec.~\ref{sec:expfixedOmega}. Instead of parametrizing the orbit using the orbital frequency $\Omega$, we can recast the equations describing the plunge in terms of the orbital radius $r_p$, which is monotonically decreasing during the plunge---in contrast to $\Omega$, which encounters a maximum at the light ring (see Fig.~\ref{fig:rG_Omega}). We refer to the two approaches as the fixed-$\Omega$ and the fixed-$r_p$ formulation, respectively. In Appendix~\ref{app:equivalence} we show how to transform between the formulations. Numerous other quantities would serve equally well as global coordinates for the plunge; an example is the orbital frequency with respect to advanced time, $\Omega_v \coloneqq d\phi_p/dv$. However, we find the plunge dynamics is particularly simple when written in terms of $r_p$. 

If using $r_p$ as a phase-space coordinate, we expand all orbital quantities in powers of $\varepsilon$ holding the coordinates $(\phi_p,r_p,\delta M^\pm)$ fixed. The orbital frequency $\Omega$ and the redshift $U$ are therefore expanded as
\begin{align}
    \Omega(r_p,\delta M^\pm,\varepsilon) &= \Omega_{\{0\}}(r_p) + \sum_{n=1}^\infty \varepsilon^n \Omega_{\{ n\}}(r_p,\delta M^\pm),
    \\[1ex]
    U(r_p,\delta M^\pm,\varepsilon) &= U_{\{0\}}(r_p) + \sum_{n=1}^\infty \varepsilon^n U_{\{ n\}}(r_p,\delta M^\pm).
\end{align}
The expansion of the rate of change of the orbital radius is given by
\begin{equation}\label{dOmegafixedrp}
    \dot r_p(r_p,\delta M^\pm,\varepsilon) = F^{r_p}_{\{0\}}(r_p) + \sum_{n =1}^\infty \varepsilon^n F^{r_p}_{\{ n\} }(r_p,\delta M^\pm),
\end{equation}
while the corrections to the background mass and spin evolve as
\begin{equation}\label{ddeltaMpmfixedrp}
    \delta \dot{M}{}^\pm(r_p,\delta M^\pm,\varepsilon) = \sum_{n=1}^\infty \varepsilon^n F^{\pm}_{\{ n\} }(r_p,\delta M^\pm).
\end{equation}
The expansion of the self-force finally reads
\begin{equation}\label{fmufixedrp}
    f^\mu(r_p,\delta M^\pm,\varepsilon) = \sum_{n=1}^\infty \varepsilon^n f^\mu_{\{n\}}(r_p,\delta M^\pm).
\end{equation}
As we have done for the fixed-$\Omega$ formulation, we now perform the post-geodesic expansion of the equation of motion~\eqref{eom} and the normalization of the four-velocity~\eqref{4velnorm}, and obtain equations describing the plunging motion at each $n$PG order.

\subsubsection{Geodesic order}

We follow a procedure analogous to the one used for the fixed-$\Omega$ formulation. At leading order, we then find the geodesic quantities
\begin{align}\label{OmegarG}
    \Omega_{\{0\}}(r_p)=\sqrt{\frac{3}{2}}\frac{3M}{r_p^2}f(r_p), \qquad U_{\{0\}}(r_p) = \frac{2\sqrt{2}}{3f(r_p)}.
\end{align}    
The leading-order rate of change of the orbital radius, $\dot r_{\{0\}}$, is defined as  
\begin{equation}\label{rGdot}
    \dot r_{\{0\}}(r_p)\coloneqq F_{\{0\}}^{r_p}(r_p) = -\frac{1}{2^{3/2}}(6 M/r_p - 1)^{3/2}f(r_p),
\end{equation}
and the leading-order radial acceleration, $\ddot r_{\{0\}}$, is then given by 
\begin{align}\label{ddotrG}
    \ddot r_{\{0\}}(r_p) \coloneqq \dot r_{\{0\}}\partial_{r_p} \dot r_{\{0\}} = \frac{M(30M/r_p-11)(6M/r_p-1)^2}{8r_p^2}f(r_p).
\end{align}
We also obtain expressions for the coordinate time and the azimuthal angle at geodesic order:
\begin{equation}\label{trG}
\begin{split}
    t_G(r_p) = \int^{r_p} \frac{dr_p'}{\dot r_{\{0\}}(r_p')} =&\, \frac{2\sqrt{2}(r_p-24M)}{(6M/r_p-1)^{1/2}}-44\sqrt{2}M\arctan\left[\left(6M/r_p-1\right)^{1/2}\right]\\
    &+4M{\rm arctanh}\left[\frac{1}{\sqrt{2}}(6M/r_p-1)^{1/2}\right] + t_0,
\end{split}
\end{equation}
\begin{equation}\label{phirG}
    \phi_G(r_p) = \int^{r_p} \Omega_{\{0\}}(r_p') \frac{dr_p'}{\dot r_{\{0\}}(r_p')} = -\frac{2\sqrt{3}}{(6M/r_p - 1)^{1/2}} + \phi_0,
\end{equation}
where $t_0$ and $\phi_0$ are integration constants. The asymptotic behaviours close to the horizon and to the ISCO are given by
\begin{align}\label{tG2M}
    t_G(r_p\to2M) &=-2M \log\left(\frac{r_p}{2M}-1\right)+{\cal O}\left[(r_p-2M)^0\right],
    \\[1ex]
    t_G(r_p\to6M) &=-72\sqrt{3}M^{3/2}(6M-r_p)^{-1/2}+{\cal O}\left[(6M-r_p)^{1/2}\right].\label{tG6M}
\end{align}

\begin{figure}[tb]
\centering
\includegraphics[width=0.6\textwidth]{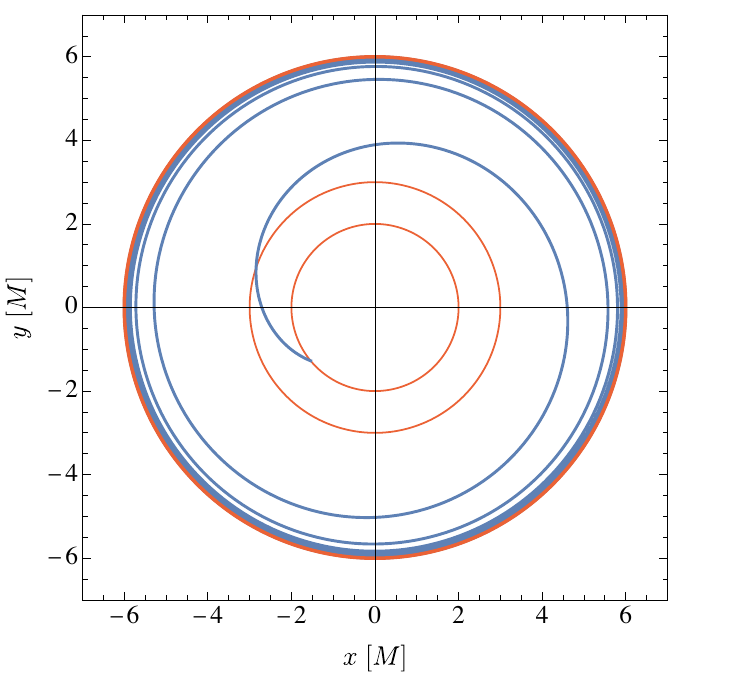}
\caption{The trajectory of the geodesic plunge obtained from Eq.~\eqref{plunge_trajectory} (blue curve). In this plot we set $\phi_0=0$. The red lines at $r=2M,\, 3M,\, 6M$ mark the event horizon, light ring and ISCO, respectively.}
\label{fig:plunge_trajectory}
\end{figure}

By inverting Eq.~\eqref{phirG}, we obtain a simple relationship for the plunging trajectory at geodesic order:
\begin{equation}\label{plunge_trajectory}
    r_p(\phi_G) = \frac{6M}{1+12/(\phi_G-\phi_0)^2}.
\end{equation}
This is a known result at least since Ref.~\cite{Chandrasekhar:579245}. The plunging trajectory is displayed in Fig.~\ref{fig:plunge_trajectory}.
Our expressions at geodesic order match those in Refs.~\cite{Hadar:2009ip, Folacci:2018cic}.

\subsubsection{First post-geodesic order} 

At first post-geodesic order we obtain the following nested differential equations for $\Omega_{\{1\}}$ and $F^{r_p}_{\{1\}}$ from the time and radial components of Eq.~\eqref{eom}:
\begin{subequations}\label{1PGeqsfixedrp}
\begin{equation}\label{eq:pG1}
\begin{split}
    \partial_{r_p}\Omega_{\{1\}}+\frac{2(r_p-3M)}{r_p(r_p-2M)}\Omega_{\{1\}} = -\frac{\sqrt{6}}{8M}f^r_{\{1\}}-\frac{\sqrt{3}(r_p-2M)(2r_p^3-27M^2 r_p+54M^3)}{4Mr_p^{5/2}(6M-r_p)^{3/2}} f^t_{\{1\}},
\end{split}
\end{equation}
\begin{equation}\label{eq:pG2}
\begin{split}
    \partial_{r_p}F^{r_p}_{\{1\}} - \frac{M(5r_p+6M)}{r_p(r_p-2M)(6M-r_p)}F^{r_p}_{\{1\}} =& -\frac{12\sqrt{3}M(r_p-2M)}{r_p^{1/2}(6M-r_p)^{3/2}}\Omega_{\{1\}} -\frac{9(r_p-2M)^2}{8r_p^2}f^t_{\{1\}}\\
    &-\frac{9\sqrt{2}r_p^{1/2}(r_p-2M)}{4(6M-r_p)^{3/2}}f^r_{\{1\}}.
\end{split}
\end{equation}    
\end{subequations}
We notice that these equations are simpler and more compact than the ones in Eq.~\eqref{1PGeqsfixedOmega} obtained from the fixed-$\Omega$ formulation. More importantly, Eq.~\eqref{1PGeqsfixedrp} admits a smooth evolution across the light ring. 

Solving Eq.~\eqref{1PGeqsfixedrp} requires initial conditions close to the ISCO. We start by considering the near-ISCO behaviour of the plunge self-force:
\begin{equation}\label{plungesf_early_rp}
    f^\mu_{\{1\}}(r_p\to6M, \delta M^\pm) = f^\mu_{\{1\},0}(\delta M^\pm) + \sum_{n=2}^\infty f^\mu_{\{1\},n}(\delta M^\pm)(6M - r_p)^{n/2}.
\end{equation}
This follows from Eq.~\eqref{sfexpr} and the field equations~\eqref{RWZ} below: the first-order field contains a piece that is a smooth function of $r_p$ at the ISCO and a piece that is not smooth $\sim \dot r_{\{0\}} \sim (6M-r_p)^{3/2}$. Equation~\eqref{sfexpr} does not alter this structure since it introduces at most one time derivative, which yields terms of the form $\sim \dot r_{\{0\}} \sim (6M-r_p)^{3/2}$ and with higher half-integer powers. The coefficients in Eq.~\eqref{plungesf_early_rp} can be obtained via asymptotic matching with the transition-to-plunge regime in an analogous manner to what is shown in Appendix~\ref{app:selfforce_matchingcond}. Using Eq.~\eqref{plungesf_early_rp}, we then get the following asymptotic solutions from the near-ISCO limit of Eq.~\eqref{1PGeqsfixedrp} (these solutions are obtained by enforcing the match with the late-time transition to plunge, similarly to the procedure described in Sec.~\eqref{sec:match}):
\begin{align}
    \Omega_{\{1\}}(r_p\to6M, \delta M^\pm) &= -\frac{9\sqrt{2}M^{1/2}f^t_{\{1\},0}}{(6M-r_p)^{1/2}} - \frac{3}{4}\sqrt{\frac{3}{2}}f^r_{\{1\},0} + {\cal O}(6M-r_p)^{1/2},
    \\[1ex]
    F^{r_p}_{\{1\}}(r_p\to6M, \delta M^\pm) &= -\frac{864M^2 f^t_{\{1\},0}}{(6M-r_p)} + 24M \left(f^t_{\{1\},0} + 12M f^t_{\{1\},2}\right) + {\cal O}(6M-r_p)^{1/2}.
\end{align}
These solutions can be used to obtain initial conditions to solve Eq.~\eqref{1PGeqsfixedrp} once the plunge self-force has been computed.

\section{Plunge: field equations at first and second order}\label{sec:field_equations}

We now turn to deriving the field equations for the metric perturbation~\eqref{eq:h plunge}. We first consider the simple case in which the global time $s$ reduces to $t$ along the particle's trajectory, as assumed in Sec.~\ref{sec:merger-ringdown_phase-space}. We then describe how to lift that restriction.

The essential idea, regardless of which regime we consider and which choice of $s$ we make, is to convert derivatives with respect to time into differential operators on phase space. When acting on a function of $(\phi_p(s,\varepsilon),J^a(s,\varepsilon),x^i)$, assuming $s=t$ on the trajectory, we apply the chain rule together with $d\phi_p/ds = \Omega$ and $dJ^a/ds=F^a(J^b,\varepsilon)$:
\begin{equation}\label{chain rule}
    \frac{\partial}{\partial s} = \Omega \frac{\partial }{\partial \phi_p} + F^a(J^b,\varepsilon)\frac{\partial}{\partial J^a}.
\end{equation}
We then expand this in the form appropriate to each regime. In the inspiral,
\begin{equation}
    F^a\frac{\partial}{\partial J^a} = \varepsilon\left( F^\Omega_{(0)}\frac{\partial}{\partial\Omega} + F^\pm_{(1)}\frac{\partial}{\partial{\delta M^\pm}}\right)+ {\cal O}(\varepsilon^2),    
\end{equation}
where $F^\pm_{(1)}\partial_{\delta M^\pm} \coloneqq F^+_{(1)}\partial_{\delta M^+} + F^-_{(1)}\partial_{\delta M^-}$. In the transition to plunge, with an obvious change of notation,
\begin{multline}
    F^a\frac{\partial}{\partial \Delta J^a} = \varepsilon^{1/5}\left( F^{\Delta\Omega}_{[0]} + \varepsilon^{2/5}F^{\Delta\Omega}_{[2]} + \varepsilon^{3/5}F^{\Delta\Omega}_{[3]} + \varepsilon^{4/5}F^{\Delta\Omega}_{[4]} \right)\frac{\partial}{\partial\Delta\Omega}\\ + \varepsilon\,F^\pm_{[3]}\frac{\partial}{\partial\delta M^\pm}+ {\cal O}(\varepsilon^{6/5}).
\end{multline}
Finally, in the plunge, 
\begin{equation}
    F^a\frac{\partial}{\partial J^a} = F^\Omega_{\{0\}}\frac{\partial}{\partial\Omega} + \varepsilon\left( F^\Omega_{\{1\}}\frac{\partial}{\partial\Omega} + F^\pm_{\{1\}}\frac{\partial}{\partial{\delta M^\pm}}\right)+ {\cal O}(\varepsilon^2).
\end{equation}

Focusing on the plunge, we can then write time derivatives as
\begin{equation}\label{dds chain rule}
    \frac{\partial}{\partial s} = \Omega \frac{\partial}{\partial \phi_p} + F^\Omega_{\{0\}}(\Omega)\frac{\partial}{\partial \Omega} + \varepsilon\,F^a_{\{1\}}(J^b)\frac{\partial}{\partial J^a} + {\cal O}(\varepsilon^2),
\end{equation}
assuming we work in phase-space coordinates $(\phi_p,\Omega,\delta M^\pm)$. If instead we work with $(\phi_p,r_p,\delta M^\pm)$, then $\Omega$ is expanded, and time derivatives become
\begin{equation}
    \frac{\partial}{\partial s} = \Omega_{\{0\}}(r_p) \frac{\partial}{\partial \phi_p} + F^{r_p}_{\{0\}}(r_p)\frac{\partial}{\partial r_p} + \varepsilon \left[\Omega_{\{1\}}(J^a)\frac{\partial}{\partial \phi_p} +F^a_{\{1\}}(J^b)\frac{\partial}{\partial J^a}\right] + {\cal O}(\varepsilon^2).
\end{equation}
When acting on a perturbation of the form~\eqref{eq:h plunge}, which has been expanded in discrete Fourier modes, we also use
\begin{equation}
    \frac{\partial}{\partial\phi_p} = - im.
\end{equation}

If $s$ does \emph{not} reduce to $t$ along the particle's trajectory, then the above treatment must be generalized. Two options present themselves. First, one can formulate the orbital equations of motion directly in terms of $s$, defining an orbital frequency $\Omega_s\coloneqq d\phi_p/ds$ and deriving equations for the forcing functions appearing in $dJ^a/ds$. Alternatively, one can continue to use $t$ as the parameter along the trajectory and account for how $t$ changes with $s$ when applying the chain rule for $\partial/\partial s$. In the next section, we take the latter approach. We define $t_p(s)$ as the value of $t$ on the particle at global time $s$ (i.e., at the point where the slice of constant $s$ intersects the trajectory). The definition $s=t-\kappa(x)$ then implies $t_p(s)=s+\kappa(x_p(t_p(s)))$. Differentiating with respect to $s$ and rearranging, we obtain
\begin{equation}\label{dtpds}
    \frac{dt_p}{ds} = \frac{1}{1-f^{-1}(r_p)H(r_p)\dot r_p},
\end{equation}
with $H\coloneqq d\kappa/dx$, $r_p=r_p(t_p(s))$, and $\dot r_p = \dot r_p(t_p(s))$. Equation~\eqref{chain rule} then becomes
\begin{equation}\label{dds chain rule v2}
    \frac{\partial}{\partial s} = \frac{dt_p}{ds}\left[\Omega \frac{\partial }{\partial \phi_p} + F^a(J^b,\varepsilon)\frac{\partial}{\partial J^a}\right].
\end{equation}
This formula holds in all regimes. Equation~\eqref{dtpds} is additionally expanded for small $\varepsilon$, using the expansion appropriate to the regime (inspiral, transition to plunge, or plunge) and to the choice of phase-space coordinates ($\Omega$ or $r_p$).

This treatment of time derivatives implies an expansion
\begin{align}
    \delta G_{\alpha\beta}[h^{\{n\}}] =      \delta G^{\{0\}}_{\alpha\beta}[h^{\{n\}}] + \varepsilon\,\delta G^{\{1\}}_{\alpha\beta}[h^{\{n\}}] + \varepsilon^2\delta G^{\{2\}}_{\alpha\beta}[h^{\{n\}}] + {\cal O}(\varepsilon^3).
\end{align}
Here $h^{\{n\}}_{\alpha\beta}=\sum_{m=-\infty}^\infty h^{\{n\},m}_{\alpha\beta}(J^a,x^i)e^{-im\phi_p}$, and the expansion in powers of $\varepsilon$ is at fixed $(\phi_p,J^a,x^i)$. The leading term, $\delta G^{\{0\}}_{\alpha\beta}[h^{\{n\}}]$ comes from omitting all order-$\varepsilon$ and higher terms in $\partial/\partial s$. Subleading terms arise from the subleading terms in $\partial/\partial s$. Analogously, 
\begin{align}
    \delta^2 G_{\alpha\beta}[h^{\{1\}},h^{\{1\}}] =      \delta^2 G^{\{0\}}_{\alpha\beta}[h^{\{1\}},h^{\{1\}}] + \varepsilon\,\delta^2 G^{\{1\}}_{\alpha\beta}[h^{\{1\}},h^{\{1\}}]+ {\cal O}(\varepsilon^2).
\end{align}

The stress-energy tensor~\eqref{eq:TDet} is likewise expanded at fixed phase-space coordinates:
\begin{equation}    
    T_{\alpha\beta} = m_p \frac{\tilde u_\alpha \tilde u_\beta}{\tilde u^t} \frac{\delta^3(x^i-x^i_p(\phi_p,J^a,\varepsilon))}{\sqrt{-\tilde g}} = \varepsilon\,T^{\{1\}}_{\alpha\beta} + \varepsilon^2 T^{\{2\}}_{\alpha\beta} + {\cal O}(\varepsilon^3).
\end{equation}
Here $T^{\{n\}}_{\alpha\beta}$ is a function of $(\phi_p,J^a,x^i)$, given at lowest order by
\begin{equation}\label{T1}
    T^{\{1\}}_{\alpha\beta} = M g_{\alpha\mu}\,g_{\beta\nu}\,\dot x^\mu_{\{0\}} \dot x^\nu_{\{0\}}\, U_{\{0\}}\frac{\delta(r-r_{\{0\}})}{r_{\{0\}}^2} \delta(\theta-\pi/2)\delta(\phi-\phi_p),
\end{equation}
with $\dot x^\mu_{\{0\}} = \left(1,\frac{\partial r_{\{0\}}}{\partial \Omega} F^{\Omega}_{\{0\}},0,\Omega\right)$ in a fixed-$\Omega$ expansion or by $\dot x^\mu_{\{0\}} = \left(1,F^{r_p}_{\{0\}},0,\Omega_{\{0\}}\right)$ in a fixed-$r_p$ expansion.

Once we have converted time derivatives to phase-space operators in this way, and suitably expanded the stress-energy tensor, we can then equate coefficients of powers of $\varepsilon$ in the expansion of the Einstein equation~\eqref{eq:EFE self-consistent}, treating $(\phi_p,J^a,x^i)$ as independent variables. This leads to a sequence of field equations for each $h^{\{n\}}_{\alpha\beta}$,
\begin{subequations}\label{eq:EFE plunge}
\begin{align}
    \delta G^{\{0\}}_{\alpha\beta}[h^{\{1\}}] &= 8\pi T^{\{1\}}_{\alpha\beta},\label{eq:EFE1 plunge}\\*
    \delta G^{\{0\}}_{\alpha\beta}[h^{\{2\}}] &= 8\pi T^{\{2\}}_{\alpha\beta} - \delta^2 G^{\{0\}}_{\alpha\beta}[h^{\{1\}},h^{\{1\}}] - \delta G^{\{1\}}_{\alpha\beta}[h^{\{1\}}].\label{eq:EFE2 plunge}
\end{align}
\end{subequations}
These are equations on a seven-dimensional product manifold made up of phase space (with coordinates $\phi_p$ and $J^a$) and space (with coordinates $x^i$). However, the dimensionality is quickly reduced. By virtue of the background's stationarity, the Fourier basis functions $e^{-im\phi_p}$ are eigenfunctions of $\delta G^{\{0\}}_{\alpha\beta}$, meaning the  discrete Fourier modes decouple in $\delta G^{\{0\}}_{\alpha\beta}[h^{\{n\}}]$, reducing the problem to six-dimensional partial differential equations for each of the mode coefficients $h^{\{n\},m}_{\alpha\beta}$. Moreover, by virtue of the background's spherical symmetry, the angular dependence can be separated by expanding the metric perturbations in a basis of tensor or spin-weighted spherical harmonics~\cite{Martel:2005ir,Spiers:2023mor}. This reduces the equations to four-dimensional partial differential equations in $J^a$ and $r$ for each $\ell m$ mode coefficient. Finally, we note that derivatives with respect to $\delta M^\pm$ do not appear on the left-hand side of the field equations because they are suppressed by a power of $\varepsilon$ in the chain rule~\eqref{dds chain rule}. This means they only enter Eq.~\eqref{eq:EFE plunge} in the source term $\delta G^{\{1\}}_{\alpha\beta}[h^{\{1\}}]$. Hence, the field equations are reduced to two-dimensional partial differential equations in $\Omega$ (or $r_p$) and $r$.

In the next section, we work through a concrete example of deriving mode-expanded phase-space field equations. We also describe how to recast them in the more familiar form of frequency-domain Regge-Wheeler-Zerilli (RWZ) equations.

\section{First-order Regge-Wheeler-Zerilli waveforms}\label{sec:first-order_wf}

We restrict our attention to the first-order field equations for $h^{\{1\}}_{\alpha\beta}$. Rather than working directly with the Einstein equation~\eqref{eq:EFE1 plunge}, we develop the associated RWZ equations for even- and odd-parity master functions. We first present the equations in a generic hyperboloidal slicing $s$ and then restrict the analysis to our frequency-domain implementation in $t$ slicing, obtained from transforming the phase-space equations to the frequency domain.

Our analysis in the remainder of the paper is restricted to 0PG order. For simplicity, we omit labels $\{n\}$ on the field variables, sources, and linear operators, with the understanding that all quantities and operations are restricted to leading order.

\subsection{RWZ equations and sources}\label{sec:RWZt}
 
We consider the RWZ equations describing the odd- and even-parity perturbations in Schwarzschild spacetime~\cite{PhysRev.108.1063,PhysRevD.2.2141},
\begin{equation}\label{RWZeqs}
    \left(\partial_x^2 - \partial_t^2 - V^\text{e/o}_\ell(r)\right) \Psi^\text{e/o}_{\ell m}(t, r) = \tilde S^\text{e/o}_{\ell m}(t,r).
\end{equation}
Following Ref.~\cite{Hopper:2010uv}, in the odd-parity sector, we use the Cunningham-Price-Moncrief master function~\cite{Cunningham:1978zfa}, while in the even-parity sector we use the Zerilli-Moncrief master function~\cite{Moncrief:1974am,Cunningham:1979px}. The even- and odd-parity potentials are given by~\cite{Hopper:2010uv}
\begin{subequations}
\begin{align}
    V^\text{e}_\ell(r) &\coloneqq \frac{f(r)}{r^2 \Lambda_\ell^2}\left[2\gamma_\ell^2 \left(\gamma_\ell+1+\frac{3M}{r}\right)+\frac{18M^2}{r^2}\left(\gamma_\ell+\frac{M}{r}\right)\right],
    \\[1ex]
    V^\text{o}_\ell(r) &\coloneqq \frac{f(r)}{r^2}\left[\ell(\ell+1)-\frac{6M}{r}\right],
\end{align}
\end{subequations}
where $\gamma_\ell\coloneqq(\ell+2)(\ell-1)/2$ and $\Lambda_\ell(r)\coloneqq \gamma_\ell+3M/r$. We write the master functions and the sources as
\begin{align}\label{sourceDecomp}
    \Psi^\text{e/o}_{\ell m}(t,r) = R^\text{e/o}_{\ell m}(r_p(t_p(s)),r)e^{-im\phi_p(t_p(s))}, \quad \tilde S^\text{e/o}_{\ell m}(t,r) = S^\text{e/o}_{\ell m}(r_p(t_p(s)), r)e^{-im\phi_p(t_p(s))},
\end{align}
where we recall that $t_p$ is the value of $t$ where the slice of constant $s$ intersects the worldline, $t_p = s + \kappa(x(r_p(s)))$. The master functions are related to the tensor spherical-harmonic modes of $h^{\{1\}}_{\alpha\beta}$ through a linear operation, in which we neglect subleading terms in the chain rules~\eqref{dds chain rule} and \eqref{dds chain rule v2}; likewise, the sources are constructed through a linear operation on $T^{\{1\}}_{\alpha\beta}$, again neglecting subleading terms in the chain rule.

Our primary goal in this paper is to calculate the GW strain, which is expressed in terms of the two GW polarizations as the limit $r \rightarrow \infty$ of the expression
\begin{equation}
    r( h_+ - i h_\times) = r h_{\mu\nu}\bar m^\mu \bar m^\nu = \sum_{\ell=2}^\infty\sum_{m=-\ell}^\ell r h_{\bar m \bar m}^{\ell m}  \;  \mbox{}_{-2}Y_{\ell m}(\theta,\phi).
\end{equation} 
Here $\bar m=\frac{1}{\sqrt{2}}(0,0,1,-\text{i}\csc\theta)$ and ${}_{-2}Y_{\ell m}$ is a spin-weighted spherical harmonic. The limit is taken at fixed $(\phi_p,J^a)$, assuming slices of global time $s$ smoothly connect to future null infinity. We define $h_{\ell m}\coloneqq \lim_{r\to\infty}(r\, h^{\ell m}_{\bar m \bar m})$ and write the asymptotic $\ell m$ mode of the waveform as~\cite{Pound:2021qin}
\begin{equation}\label{hlm conventions}
    h_{\ell m} = \varepsilon\,\frac{\sqrt{D_\ell}}{2} \left(\Psi^\text{e}_{\ell m} - i\Psi^\text{o}_{\ell m}\right) = \varepsilon\,\frac{\sqrt{D_\ell}}{2} \left(R^\text{e}_{\ell m} - iR^\text{o}_{\ell m}\right)e^{-im\phi_p}\coloneqq H_{\ell m}e^{-im\phi_p}\coloneqq |H_{\ell m}|e^{-i\Phi_{\ell m}},
\end{equation}
where $D_\ell\coloneqq(\ell-1)\ell(\ell+1)(\ell+2)$ and $\Phi_{\ell m} = m\phi_p - {\rm arg}(H_{\ell m})$. Only $m \geq 0$ modes need to be computed as the $m<0$ modes are deduced from the usual symmetry relation $h_{\ell,-m}=(-1)^\ell h_{\ell m}^*$.

We allow a generic choice of time $s$, which might not reduce to $t$ on the particle trajectory. By plugging Eq.~\eqref{sourceDecomp} into Eq.~\eqref{RWZeqs} and applying the chain rule~\eqref{dds chain rule v2}, we put the RWZ equations in the following form:
\begin{equation}\label{RWZs}
\begin{split}
    &\left(\partial_x^2 -  V_\ell^\text{e/o}\right) R^\text{e/o}_{\ell m} - 2H\frac{dt_p}{ds}\left(\dot r_{\{0\}} \partial_x\partial_{r_p} - im\Omega_{\{0\}}\partial_x\right)R^\text{e/o}_{\ell m} - \frac{dH}{dx}\frac{dt_p}{ds}\left(\dot r_{\{0\}} \partial_{r_p} - im\Omega_{\{0\}}\right)R^\text{e/o}_{\ell m}
    \\[1ex]
    &+ \left(1 - H^2\right)\left[im\Omega_{\{0\}}\frac{d^2t_p}{ds^2} + \left(m^2\Omega_{\{0\}}^2 + im \partial_{r_p}\Omega_{\{0\}} \dot r_{\{0\}}\right)\left(\frac{dt_p}{ds}\right)^2 \right.
    \\[1ex]
    &\left. - \dot r_{\{0\}}\frac{d^2t_p}{ds^2}\partial_{r_p} - \dot r_{\{0\}}^2\left(\frac{dt_p}{ds}\right)^2\partial_{r_p}^2 - \left(\ddot r_{\{0\}} - 2im\Omega_{\{0\}}\dot r_{\{0\}}\right)\left(\frac{dt_p}{ds}\right)^2\partial_{r_p}\right]R^\text{e/o}_{\ell m} = S^\text{e/o}_{\ell m}.
\end{split}
\end{equation}
The quantities $dt_p/ds$ and $d^2t_p/ds^2$ are functions of $r_p$, as given in Eq.~\eqref{dtpds} (and the $s$ derivative thereof). The geodesic quantities $\Omega_{\{0\}}(r_p)$, $\dot r_{\{0\}}(r_p)$ and $\ddot r_{\{0\}}(r_p)$ are given in Eqs.~\eqref{OmegarG}, \eqref{rGdot} and \eqref{ddotrG}, respectively. 

While the original, time-domain RWZ equations are hyperbolic, our phase-space equations~\eqref{RWZs} have an unusual character. They are hyperbolic for all $2M<r_p<6M$, as is easily confirmed by calculating the discriminant, which is proportional to $\left(\dot r_{\{0\}}\frac{dt_p}{ds}\right)^2>0$ for any $H$. However, this discriminant vanishes in the limit to $r_p=6M$, since $\dot r_{\{0\}}$ vanishes in that limit. One might conclude that the equations are consequently parabolic at $r_p=6M$, but $\dot r_{\{0\}}(6M)=0$ implies they actually reduce to radial ordinary differential equations there. Fundamentally, this singular behaviour is a consequence of the fact that the plunging geodesic asymptotes to $r_p=6M$ in the infinite past but never actually reaches $6M$; the physical domain for the plunge solution is the open region $2M<r_p<6M$. To obtain the physically correct solution in that domain, one must ensure that the solution appropriately matches to the transition-to-plunge solution; we discuss this further in the next section.

In the remainder of this section and the practical implementation of first-order waveforms in this paper, we will consider $t$ slicing, $\kappa(x)=H(x)=0$. Appendix~\ref{sec:RWZ s slicing} provides the parallel development of the field equations in generic slicing, which are important in establishing the validity of our approach. With $t$ slicing, the RWZ equations become
\begin{multline}\label{RWZ}
    \left[\partial_x^2 +\left(m^2\Omega_{\{0\}}^2 + i m\,\partial_{r_p}\Omega_{\{0\}} \, \dot r_{\{0\}}\right) - \dot r_{\{0\}}^2 \partial_{r_p}^2\right.\\ \left. - \left(\ddot r_{\{0\}} - 2im\Omega_{\{0\}}\dot r_{\{0\}}\right) \partial_{r_p}- V^\text{e/o}_\ell(r)\right] R^\text{e/o}_{\ell m}(r_p, r) = S^\text{e/o}_{\ell m}(r_p, r).
\end{multline}
The source appearing on the right-hand side of this equation is that of the plunging point-particle. We construct this source in the even and odd sectors using the formalism described in Appendix~C of Ref.~\cite{Hopper:2010uv}. We write
\begin{subequations}\label{ppSourcePlunge}
\begin{align}
    \begin{split}\tilde S^\text{e}_{\ell m}(t,r) \coloneqq&\, \frac{1}{(\gamma_\ell+1)\Lambda_\ell}\left[r^2 f\left(f^2\partial_r Q_{\ell m}^{tt} - \partial_r Q_{\ell m}^{rr}\right) + r\left(\Lambda_\ell - f\right)Q_{\ell m}^{rr} + r f^2 Q_{\ell m}^\mathfrak{b}\right.\\
    &\left.- \frac{f^2}{r\Lambda_\ell}\left(\gamma_\ell(\gamma_\ell-1)r^2 + (4\gamma_\ell-9)M r + 15M^2\right)Q_{\ell m}^{tt}\right] + \frac{2f}{\Lambda_\ell}Q_{\ell m}^r - \frac{f}{r}Q_{\ell m}^\#,\end{split}
    \\[1ex]\label{Sodd}
    \tilde S^\text{o}_{\ell m}(t,r) \coloneqq&\, \frac{r f}{\gamma_\ell}\left(\frac{1}{f}\partial_t P_{\ell m}^r + f\partial_r P_{\ell m}^t + \frac{2M}{r^2}P_{\ell m}^t\right),
\end{align}
\end{subequations}
with
\begin{subequations}\label{PandQ}
\begin{align}
    P_{\ell m}^a(t, r) &\coloneqq \frac{16\pi r^2}{\ell(\ell+1)}\int T_{\{1\}}^{aB}(t,r,\theta,\phi) X^{\ell m*}_{B}(\theta, \phi) \sin\theta \, d\theta d\phi,
    \\
    Q_{\ell m}^a(t, r) &\coloneqq \frac{16\pi r^2}{\ell(\ell+1)}\int T_{\{1\}}^{aB}(t,r,\theta,\phi) Y^{\ell m*}_{B}(\theta, \phi) \sin\theta \, d\theta d\phi,
    \\
    Q_{\ell m}^{ab}(t, r) &\coloneqq 8\pi\int T_{\{1\}}^{ab}(t,r,\theta,\phi) Y^{\ell m*}(\theta, \phi) \sin\theta \, d\theta d\phi,
    \\
    Q_{\ell m}^\mathfrak{b}(t, r) &\coloneqq 8\pi r^2 \int T_{\{1\}}^{AB}(t,r,\theta,\phi) \Omega_{AB} Y^{\ell m*}(\theta, \phi) \sin\theta \, d\theta d\phi,
    \\
    Q_{\ell m}^\#(t, r) &\coloneqq 32\pi r^4\frac{(\ell-2)!}{(\ell+2)!}\int T_{\{1\}}^{AB}(t,r,\theta,\phi) Y^{\ell m*}_{AB}(\theta, \phi) \sin\theta \, d\theta d\phi.
\end{align}
\end{subequations}
In these expressions, lowercase Latin letters stand for the coordinates $\{t,r\}$, while uppercase Latin letters indicate the coordinates $\{\theta,\phi\}$. The scalar ($Y^{\ell m}$), vector ($Y^{\ell m}_A$ and $X^{\ell m}_A$) and tensor ($Y^{\ell m}_{AB}$ and $X^{\ell m}_{AB}$) spherical harmonics are defined in Appendix~\ref{app:vecten_sphericalharm}, and $\Omega_{AB}=\text{diag}(1,\sin^2\theta)$ is the metric on the unit 2-sphere. 

We evaluate the integrals in Eq.~\eqref{PandQ} using the point-particle stress-energy tensor~\eqref{T1}. The differential operator $\partial_t$ in Eq.~\eqref{ppSourcePlunge} should therefore be understood as the operator $\dot r_{\{0\}} \partial_{r_p} + \Omega_{\{0\}} \partial_{\phi_p}$. Recalling Eq.~\eqref{sourceDecomp}, we can then write the sources~\eqref{ppSourcePlunge} as
\begin{equation}\label{sourceplunget}
    S^\text{e/o}_{\ell m}(r_p,r) = A^\text{e/o}_{\ell m}(r_p) \delta(r-r_p) +  B^\text{e/o}_{\ell m}(r_p) \partial_r\delta(r-r_p),
\end{equation}
where
\begin{subequations}
\begin{align}
    \begin{split}A^\text{e}_{\ell m}(r_p) \coloneqq& -\frac{2\sqrt{2}\pi M (r_p-2M)^2}{(\gamma_\ell+1)r_p^7 \Lambda_\ell(r_p)^2}\Bigl[324 M^4 + 36 M^3 (3 + 8 \gamma_\ell) r_p + 9 M^2 (1 + 4 \gamma_\ell^2) r_p^2\\
    &+2 M(1 + 9 \gamma_\ell) r_p^3 + 3 \gamma_\ell (1 + \gamma_\ell) r_p^4 \Bigr] Y^{\ell m}(\pi/2,0)\\
    &- \frac{16\sqrt{6}\pi M^2 (r_p - 2M)^2(6M-r_p)^{3/2}}{\ell(\ell+1)r_p^{11/2}\Lambda_\ell(r_p)}Y^{\ell m*}_\phi(\pi/2,0)\\
    &- \frac{288\sqrt{2}\pi M^3(\ell-2)!(r_p-2M)^2}{(\ell+2)!\,r_p^5}Y^{\ell m*}_{\phi\phi}(\pi/2,0),\end{split}
    \\[1ex]
    \begin{split}A^\text{o}_{\ell m}(r_p) \coloneqq& -\frac{4 \sqrt{3}\pi M^2 (r_p-2M)^2}{\ell(\ell+1)\gamma_\ell \, r_p^7}\Bigl[756M^3 - 6\sqrt{3} i M m\,r_p^{1/2}(6M-r_p)^{3/2}\\
    &- 216M^2r_p + 45M r_p^2 + 7r_p^3\Bigr]X_\phi^{\ell m*}(\pi/2,0),\end{split}
    \\[1ex]
    B^\text{e}_{\ell m}(r_p) \coloneqq& \, \frac{6\sqrt{2} \pi M \left(r_p - 2M\right)^3 \left(r_p^2 + 12M^2\right)}{(\gamma_\ell+1) r_p^5 \Lambda_\ell(r_p)}Y^{\ell m}(\pi/2,0),
    \\[1ex]
    B^\text{o}_{\ell m}(r_p) \coloneqq& \, \frac{36\sqrt{3}\pi M^2 \left(r_p - 2M\right)^3 \left(r_p^2 + 12M^2\right)}{\ell(\ell+1)\gamma_\ell\,r_p^6}X_\phi^{\ell m*}(\pi/2,0).
\end{align}
\end{subequations}
Here $Y^{\ell m *}_\phi(\pi/2,0)=-im Y^{\ell m}(\pi/2,0)$ and $Y^{\ell m}_{\phi\phi}(\pi/2,0)=\left(-m^2+\frac{\ell(\ell+1)}{2}\right)Y^{\ell m}(\pi/2,0)$. The even- and odd-parity sectors are sourced by respectively even and odd $\ell+m$ modes since
\begin{subequations}
\begin{align}
    S^\text{e}_{\ell m}(r_p,r) &=0 \quad \text{ for } \ell+m \; \text{odd}, 
    \\[1ex]
    S^\text{o}_{\ell m}(r_p,r) &=0 \quad \text{ for } \ell+m \; \text{even}. 
\end{align}
\end{subequations}

\subsection{Punctured RWZ equations}

As highlighted above, the plunge field equations become singular at $r_p=6M$. Our physical boundary condition is that at early times, in a neighbourhood of the singular surface $r_p=6M$, our plunge solution must asymptotically match the transition-to-plunge solution. 

This asymptotic matching was discussed at the level of the orbital dynamics in Sec.~\ref{sec:match}. For the metric perturbations~\eqref{eq:h transition} and~\eqref{eq:h plunge}, the matching condition applies at the level of Fourier coefficients $h^{[n],m}_{\alpha\beta}$ and $h^{\{n\},m}_{\alpha\beta}$. At 0PG order, the condition is that if (i) $h^{\{1\},m}_{\alpha\beta}(\Omega,x^i)$ is expanded for small $(\Omega-\Omega_*)$, and (ii) $h^{[n],m}_{\alpha\beta}(\Delta \Omega,x^i)=h^{[n],m}_{\alpha\beta}((\Omega-\Omega_*)/\varepsilon^{2/5},x^i)$ is expanded for small $\varepsilon$ at fixed $\Omega$, then (iii) the small-$(\Omega-\Omega_*)$ expansion of $\varepsilon\,h^{\{1\},m}_{\alpha\beta}(\Omega,x^i)$ must agree, term by term, with the linear-in-$\varepsilon$ terms in the re-expansion of $\sum_{n=5}^\infty \varepsilon^{n/5}h^{[n],m}_{\alpha\beta}$. Here and in many expressions below, we suppress dependence on $\delta M^\pm$ for simplicity.

From Eq.~\eqref{eq:h transition}, we see that the re-expansion of the transition-to-plunge metric takes the form
\begin{equation}\label{htransition_reexp}
    h_{\alpha\beta} = \sum_{m=-\infty}^\infty\sum_{n=5}^\infty\sum_{k=5-n}^\infty\varepsilon^{n/5}\frac{\varepsilon^{k/5}}{(\Omega-\Omega_*)^{k/2}}h^{[n,k],m}_{\alpha\beta}(\delta M^\pm,x^i)e^{-im\phi_p}.
\end{equation} 
For $\Omega>\Omega_*$, this re-expansion for small $\varepsilon$ at fixed $\Omega$ is equivalent to an expansion in the limit $\Delta\Omega\to+\infty$ (i.e., the ``late transition to plunge''). The expansion contains a power series in $\Delta\Omega^{1/2}$ (or, equivalently, $(\Omega-\Omega_*)^{1/2}/\varepsilon^{1/5}$) rather than $\Delta\Omega$ due to the structure of the orbital forcing functions in the transition to plunge.\footnote{The reason for this can be understood as follows: in the transition-to-plunge expansion, the Fourier coefficients $h^{[n],m}_{\alpha\beta}$ can always be written as a sum of terms that are factored into a $\Delta\Omega$-dependent and a $\Delta\Omega$-independent piece~\cite{Kuchler:2024esj}. The $\Delta\Omega$-dependent factors consist in integer powers of $\Delta\Omega$ itself, the forcing terms $F^{\Delta\Omega}_{[n]}$ and their $\Delta\Omega$ derivatives. For example, in the notation of Ref.~\cite{Kuchler:2024esj}, the 4PLT metric perturbation is given by $h^{[9],m}_{\alpha\beta}(\Delta\Omega,x^i)=\Delta\Omega^2 h^{[9]A,m}_{\alpha\beta}(x^i) + F^{\Delta\Omega}_{[0]}\partial_{\Delta\Omega}F^{\Delta\Omega}_{[0]}h^{[9]B,m}_{\alpha\beta}(x^i)$. In the large-$\Delta\Omega$ limit, the forcing terms $F^{\Delta\Omega}_{[n]}$ admit asymptotic solutions in half-integer powers of $\Delta\Omega$, as can be seen from Eq.~\eqref{g late expansion} or Table~\ref{tab:matchstructureTP}.}
Here we have restricted the range of $k$ using the fact that this must match the plunge expansion, which begins at linear order in $\varepsilon$; hence, $(k+n)/5\geq 1$. Moreover, $(k+n)/5$ must be an integer, equal to the order $\{n\}$ in the plunge metric, implying $h^{[n,k],m}_{\alpha\beta}$ vanishes except when $k=(5\{n\}-n)$ for integers $\{n\}\geq 1$. Hence, the terms that must match the near-ISCO expansion of $h^{\{1\}}_{\alpha\beta}$ are $1/(\Omega-\Omega_*)^{(5-n)/2}$:  a constant term $\propto(\Omega-\Omega_*)^0$ from $h^{[5],m}_{\alpha\beta}$, a linear term $\propto(\Omega-\Omega_*)$ from $h^{[7],m}_{\alpha\beta}$, a term $\propto(\Omega-\Omega_*)^{3/2}$ from $h^{[8],m}_{\alpha\beta}$, and so on. 

From the above analysis, we see that the leading term in the near-ISCO expansion of $h^{\{1\},m}_{\alpha\beta}$ must be a constant, equal to $h^{[5,0],m}_{\alpha\beta}$. In fact, the analysis in Ref.~\cite{Kuchler:2024esj} shows that the first two terms in the transition-to-plunge regime are exactly equal to the first two terms in the near-ISCO expansion of the first-order inspiral solution:
\begin{equation}
    \varepsilon\,h^{[5],m}_{\alpha\beta} + \varepsilon^{7/5}h^{[7],m}_{\alpha\beta} = \varepsilon\left[h^{(1),m}_{\alpha\beta}(\Omega_*,x^i)+(\Omega-\Omega_*)\partial_\Omega h^{(1),m}_{\alpha\beta}(\Omega_*,x^i)\right].
\end{equation}
The first two terms in the near-ISCO expansion of $h^{\{1\},m}_{\alpha\beta}$ must agree with these two terms from the inspiral solution. This immediately extends to the RWZ master variables as functions of $r_p$:
\begin{equation}\label{R near ISCO}
    R^{\rm e/o}_{\ell m}(r_p,r) = R^{{\rm e/o}(1)}_{\ell m}(6M,r)+(r_p-6M)\partial_{r_p} R^{{\rm e/o}(1)}_{\ell m}(6M,r) + {\cal O}[(r_p-6M)^2],
\end{equation}
where $R^{{\rm e/o}(1)}_{\ell m}(r_p,r)$ are the first-order master functions in the inspiral. 

To enforce this near-ISCO matching to the transition-to-plunge solution, we adopt a puncture scheme~\cite{Miller:2023ers}. In a region near the ISCO, we write the physical field as the sum of two terms,
\begin{equation}\label{puncture_t}
    R^\text{e/o}_{\ell m}(r_p,r) = R^{\text{e/o}\,\mathcal{R}}_{\ell m}(r_p,r) + R^{\text{e/o}\,\mathcal{P}}_{\ell m}(r_p,r).
\end{equation}
The residual field $R^{\text{e/o}\,\mathcal{R}}_{\ell m}$ has vanishing boundary conditions at the ISCO, $R^{\text{e/o}\,\mathcal{R}}_{\ell m}(6M,r)=0$. The puncture field $R^{\text{e/o}\,\mathcal{P}}_{\ell m}$ consists of the near-ISCO solution~\eqref{R near ISCO} truncated at some order in $(r_p-6M)$, and it ``lives'' until an orbital radius $r_\mathcal{P}$ such that $2M<r_\mathcal{P}<6M$. As an example, here we truncate at leading order, meaning 
\begin{equation}\label{puncture_field}
    R^{\text{e/o}\,\mathcal{P}}_{\ell m}(r_p,r) \coloneqq R^{\text{e/o}\,(1)}_{\ell m}(6M,r)\theta(r_p - r_\mathcal{P}).
\end{equation}
The inspiral field $R^{\text{e/o}\,(1)}_{\ell m}(6M, r)$ satisfies Eq.~\eqref{RWZ} evaluated at $r_p=6M$. To understand this, recall that $\dot r_{\{0\}}(6M)=0=\partial_{r_p} \dot r_{\{0\}}(6M)$ (from Eq.~\eqref{rGdot}), implying that the limit of Eq.~\eqref{RWZ} to $r_p=6M$ is simply the RWZ equation for a particle on a circular orbit at the ISCO:
\begin{equation}\label{teomR1}
    \left[\partial_x^2 + m^2\left.\Omega_{\{0\}}^2\right\vert_* - V^{\text{e/o}}_\ell(r)\right] R^{\text{e/o}\,(1)}_{\ell m}(6M,r) = S^\text{e/o}_{\ell m}(6M, r),
\end{equation}
where $\vert_*$ indicates evaluation at $r_p=6M$. We now plug Eq.~\eqref{puncture_t} into Eq.~\eqref{RWZ}, move the puncture fields to the right-hand side and use Eq.~\eqref{teomR1}. We then obtain the following equation for the residual field:
\begin{multline}\label{RWZpunctured}
    \left[\partial_x^2 +\left(m^2\Omega_{\{0\}}^2 + i m\,\partial_{r_p}\Omega_{\{0\}} \, \dot r_{\{0\}}\right) - \dot r_{\{0\}}^2 \partial_{r_p}^2\right.\\* \left. - \left(\ddot r_{\{0\}} - 2im\Omega_{\{0\}}\dot r_{\{0\}}\right) \partial_{r_p}- V^{\text{e/o}}_\ell(r)\right] R^{\text{e/o}\,\mathcal{R}}_{\ell m}(r_p, r) = S^\text{e/o eff}_{\ell m}(r_p, r).
\end{multline}

The effective source, $S^\text{e/o eff}_{\ell m}$, can be split into two pieces: the ordinary point particle source in the region $r_p< r_{\cal P}^-$ and a more complicated, extended source in the punctured region $6M\geq r_p\geq r_{\cal P}^-$, where we use the notation $r_{\cal P}^-= r_{\cal P} - 0^+$ to indicate that the interval $[r_{\cal P}^-,6M]$ encloses the point $r_p=r_{\cal P}$. Explicitly, we write 
\begin{equation}
    S^\text{e/o eff}_{\ell m}(r_p, r) = S^\text{e/o pp}_{\ell m}(r_p, r) + S^\text{e/o ext}_{\ell m}(r_p, r),
\end{equation}
where $S^\text{e/o pp}_{\ell m}$ is the source outside the punctured region,
\begin{equation}\label{pppSource}
    S^{\text{e/o pp}}_{\ell m}(r_p, r) \coloneqq S^{\text{e/o}}_{\ell m}(r_p, r)\theta(r_\mathcal{P}-r_p),
\end{equation}
and $S^\text{e/o ext}_{\ell m}$ is the source inside the punctured region,
\begin{equation}\label{Seffext}
\begin{split}
    S^\text{e/o ext}_{\ell m}(r_p, r) \coloneqq&\; \theta(r_p - r_\mathcal{P})\bigl[S^{\text{e/o}}_{\ell m}(r_p, r) - S^{\text{e/o}}_{\ell m}(6M, r)\bigr]
    \\[1ex]
    &\,-\theta(r_p - r_\mathcal{P})\left(m^2\Omega_{\{0\}}^2 - m^2\left.\Omega_{\{0\}}^2\right\vert_* + i m \partial_{r_p}\Omega_{\{0\}} \, \dot r_{\{0\}}\right)R^{\text{e/o}\,(1)}_{\ell m}(6M,r)
    \\[1ex]
    &\,-\delta(r_p - r_\mathcal{P})\left.\dot r_{\{0\}}\right\vert_\mathcal{P}\left(2im\left.\Omega_{\{0\}}\right\vert_\mathcal{P} +\partial_{r_p}\left. \dot r_{\{0\}}\right\vert_\mathcal{P}\right) R^{\text{e/o}\,(1)}_{\ell m}(6M,r)
    \\[1ex]
    &\;+\partial_{r_p}\delta(r_p - r_\mathcal{P})\left.\dot r_{\{0\}}^2\right\vert_\mathcal{P} R^{\text{e/o}\,(1)}_{\ell m}(6M,r).
\end{split}
\end{equation}
We use $\vert_\mathcal{P}$ to indicate quantities evaluated at $r_p=r_\mathcal{P}$. In arriving at this result we have used $\dot r_{\{0\}}^2(r_p)\partial_{r_p}\delta(r_p-r_{\cal P}) = \dot r_{\{0\}}^2(r_\mathcal{P})\partial_{r_p}\delta(r_p-r_\mathcal{P})-\partial_{r_p}\dot r_{\{0\}}^2(r_{\cal P})\delta(r_p-r_\mathcal{P})$.

The essential feature of the extended source is that it vanishes as $r_p\to6M$ (i.e., in the infinite past), consistent with our initial condition that the residual field vanishes there; the total solution (residual field plus puncture) then reduces to the transition-to-plunge solution, as desired. We also explain in the next section how this behaviour of the effective source eliminates ill-defined integrals that would otherwise appear. 

Since the puncture vanishes for $r_p<r_{\cal P}$, the residual field reduces to the physical field in that region:
\begin{equation}\label{residual field reduces to physical}
   R^{{\rm e/o}\,\cal R}_{\ell m}(r_p,r) = R^{{\rm e/o}}_{\ell m}(r_p,r) \quad\text{for }r_p<r_{\cal P}.
\end{equation}
We also note that the Dirac $\delta$ and $\delta'$ terms in Eq.~\eqref{Seffext} can be discarded and replaced with the junction condition that 
\begin{equation}
    \lim_{r_p\to r_{\cal P}^-}R^{{\rm e/o}}_{\ell m}(r_p,r) = \lim_{r_p\to r_{\cal P}^+}\left[R^{{\rm e/o}\,\cal R}_{\ell m}(r_p,r)+R^{{\rm e/o}\,\cal P}_{\ell m}(r_p,r)\right].
\end{equation}
Moreover, although we said that the punctured region $[r_{\cal P}^-,6M]$ is ``near the ISCO'', the physical solution~\eqref{puncture_t} must be independent of $r_{\cal P}$, so long as $2M<r_{\cal P}<6M$. This independence follows from the same type of arguments as in Ref.~\cite{Miller:2023ers}. Finally, we comment that we have only worked with a leading-order puncture in this section, but a higher-order (in $(r_p-6M)$) puncture could be required. It is possible for the puncture to be sufficiently high order to obtain a convergent solution for the residual field, without being high enough order to obtain the \emph{correct} solution; see again Ref.~\cite{Miller:2023ers}. We leave investigation of this, and derivation of the required order of the puncture, to future work.

\subsection{Frequency-domain equations}\label{sec:f domain eqs}

In order to solve Eq.~\eqref{RWZpunctured} for the residual field we work in the frequency domain. This allows us to use well-known Green-function techniques to solve the problem. 

We start by defining the following forward and inverse transforms:
\begin{subequations}
\begin{align}\label{transformforw}
    \hat g(\omega,r) =&\, \int_{2M}^{6M} \frac{dr_p}{\dot r_{\{0\}}(r_p)}g(r_p,r)e^{i\omega t_G(r_p) - im\phi_G(r_p)},
    \\[1ex]\label{transforminv}
    g(r_p,r) =&\, -\frac{1}{2\pi}\int_{-\infty}^{+\infty} d\omega \, \hat g(\omega,r)e^{-i\omega t_G(r_p) + im\phi_G(r_p)},
\end{align}
\end{subequations}
with $t_G$ and $\phi_G$ defined in Eqs.~\eqref{trG} and \eqref{phirG}, respectively. For simplicity we have set $t_0=\phi_0=0$, noting that $t_0$ and $\phi_0$ cancel between the forward and inverse transforms. We can verify that the above transforms are self-consistent in the range of interest $2M \leq r_p \leq 6M$:
\begin{equation}
\begin{split}
    g(r_p, r) =&\, -\frac{1}{2\pi}\int^{+\infty}_{-\infty} d\omega \, e^{-i\omega t_G(r_p) + i m \phi_G(r_p)}  \int_{2M}^{6M} \frac{dr_p'}{\dot r_{\{0\}}(r_p')}g(r_p', r)e^{i\omega t_G(r_p') - i m \phi_G(r_p')}
    \\[1ex]
    =&\, \frac{1}{2\pi}\int_{2M}^{6M}  \frac{dr_p'}{\vert \dot r_{\{0\}}(r_p')\vert }g(r_p', r) e^{im\left(\phi_G(r_p) - \phi_G(r_p')\right)} \int^{+\infty}_{-\infty} d\omega \, e^{i\omega\left(t_G(r_p') - t_G(r_p)\right)}
    \\[1ex]
    =&\, \int_{2M}^{6M} \frac{dr_p'}{\vert \dot r_{\{0\}}(r_p')\vert }g(r_p', r) e^{im\left(\phi_G(r_p) - \phi_G(r_p')\right)} \vert\dot r_{\{0\}}(r_p') \vert \delta(r_p' - r_p) = g(r_p, r),
\end{split}
\end{equation}
where we have used $\dot r_{\{0\}}=-|\dot r_{\{0\}}|$ since $\dot r_{\{0\}}\leq0$ between the ISCO and the event horizon.

We now take the transform of Eq.~\eqref{RWZpunctured}. This transform is delicate because we arrive at ill-defined boundary terms if we apply the transform naively and integrate by parts to move derivatives onto the exponential. If we naively discard those boundary terms, we obtain
\begin{equation}\label{tRWZtransformed}
    \left(\partial_x^2 + \omega^2 - V_\ell^{\text{e/o}}(r)\right)\hat R^{\text{e/o}\,\mathcal{R}}_{\ell m}(\omega,r) = \hat S^\text{e/o eff}_{\ell m}(\omega, r) 
\end{equation}
where $\hat S^\text{e/o eff}_{\ell m}$ is the transform of $S^\text{e/o eff}_{\ell m}$. To properly justify this equation, we treat each side of Eq.~\eqref{RWZpunctured} as a tempered distribution and adopt the standard definition $\langle \varphi,{\cal F}[f]\rangle\coloneq \langle {\cal F}[\varphi],f\rangle$, where $f$ is any tempered distribution, ${\cal F}$ is our forward transform, $\varphi$ is any element of the Schwartz space, and $\langle \varphi, f\rangle$ is the integral over the real line (for integrable $f$) or the action of $f$ on $\varphi$ (for distributional $f$). Writing Eq.~\eqref{RWZpunctured} schematically as $(\partial_x^2+{\cal D})R_{\ell m}=S^{\rm eff}_{\ell m}$, we apply the forward transform to obtain $\bigl\langle {\cal F}[\varphi],(\partial_x^2+{\cal D})R_{\ell m}\bigr\rangle=\bigl\langle \varphi,{\cal F}[S^{\rm eff}_{\ell m}]\bigr\rangle$. On the left, we apply the standard definition $\langle \varphi,{\cal D}f\rangle\coloneqq \langle {\cal D}^\dagger\varphi,f\rangle$ for any linear operator ${\cal D}$, where ${\cal D}^\dagger$ is the adjoint of ${\cal D}$. Defining the frequency-domain operator ${\cal D}^\dagger_\omega$ via ${\cal D}^\dagger{\cal F}[\varphi]={\cal F}[{\cal D}^\dagger_\omega\varphi]$, we obtain $\bigl\langle {\cal F}[\varphi],{\cal D}R_{\ell m}\bigr\rangle = \bigl\langle {\cal F}[{\cal D}^\dagger_\omega\varphi],R_{\ell m}\bigr\rangle = \bigl\langle{\cal D}^\dagger_\omega\varphi,{\cal F}[R_{\ell m}]\bigr\rangle = \bigl\langle\varphi,{\cal D}_\omega{\cal F}[R_{\ell m}]\bigr\rangle$. Therefore, $(\partial_x^2+{\cal D}_\omega){\cal F}[R_{\ell m}]={\cal F}[S^{\rm eff}_{\ell m}]$ as a distribution, which is simply Eq.~\eqref{tRWZtransformed}.

Equation~\eqref{tRWZtransformed} has the standard form of the frequency-domain RWZ equation. Our method of reaching this form applies equally well at higher PG orders. At all orders, we use $t_G$ and $\phi_G$ in the transforms, rather than $t_p$ and $\phi_p$, since we require known functions of $r_p$; $t_p$ and $\phi_p$ are never determined until the online waveform-generation stage.

\subsection{Inhomogeneous solutions}\label{sec:inhomogsol}

We solve Eq.~\eqref{tRWZtransformed} using the standard Green-function method (or equivalently, variation of parameters), building a retarded Green function from a basis of homogeneous solutions.

In the limits $r\to+\infty$ and $r\to2M$ (or, equivalently, $x\to\pm\infty$), the homogeneous solutions to Eq.~\eqref{tRWZtransformed} are complex exponentials of the form $e^{\pm i\omega x}$. We consider two independent homogeneous solutions to Eq.~\eqref{tRWZtransformed}, the so-called ``in'' and ``up'' solutions, with the following asymptotic behaviours:
\begin{subequations}\label{inup}
\begin{equation}
    \hat R_{\ell}^\text{e/o in}(\omega,x) \sim \left\{ \begin{array}{ll}
    A_{\ell}^\text{e/o in}(\omega) e^{-i\omega x} + A_{\ell}^\text{e/o out}(\omega) e^{+i\omega x} & \text{as } r\to+\infty \, (x\to+\infty),\\
    e^{-i\omega x} & \text{as } r\to2M \, (x\to-\infty),
    \end{array} \right.
\end{equation}
\begin{equation}
    \hat R_{\ell}^\text{e/o up}(\omega,x) \sim \left\{ \begin{array}{ll}
    e^{i\omega x} & \text{as } r\to+\infty \, (x\to+\infty),\\
    B_{\ell}^\text{e/o in}(\omega) e^{-i\omega x} + B_{\ell}^\text{e/o out}(\omega) e^{+i\omega x} & \text{as } r\to2M \, (x\to-\infty).
    \end{array} \right.
\end{equation}
\end{subequations}
Here and below we freely write functions of $r$ as functions of $x(r)$. According to the Chandrasekhar-Detweiler transformation between the Regge-Wheeler and Zerilli-Moncrief solutions~\cite{1975RSPSA.344..441C}, the homogeneous even and odd ``in'' solution are related by
\begin{align}\label{defRe}
    \left(D_\ell- 12 i M \omega\right)\hat R_{\ell}^\text{e in}(\omega,x) = \left[D_\ell + \frac{72M^2 f(r)}{(\ell-1)(\ell+2)r^2+6Mr}+12 M  \frac{\partial}{\partial x}\right] \hat R_{\ell}^\text{o in}(\omega,x).
\end{align}
As a consequence, the coefficients $A_{\ell}^\text{e/o in}(\omega)$ and $A_{\ell}^\text{e/o out}(\omega)$ obey the relations 
\begin{equation}\label{AinAout_eo}
    A_{\ell}^\text{e in}(\omega) = A_{\ell}^\text{o in}(\omega), \qquad A_{\ell}^\text{e out}(\omega) = \frac{D_\ell +12 i M \omega }{D_\ell -12 i M \omega} A_{\ell}^\text{o out}(\omega).  
\end{equation}
From here on we will therefore omit the label $\text{e/o}$ on $A_{\ell}^\text{e/o in}(\omega)$. As a consequence of the RWZ equation we have $\hat R_\ell^\text{e/o in}(-\omega^*,x)=\hat R_\ell^\text{e/o in*}(\omega,x)$. This implies
\begin{equation}\label{sAin}
    A_\ell^{\rm in}(-\omega^*) = A_\ell^{\rm in *}(\omega), \qquad A_\ell^\text{e/o out}(-\omega^*) = A_\ell^\text{e/o out*}(\omega). 
\end{equation}

Using the two independent homogeneous solutions~\eqref{inup} we can construct the following Green function:
\begin{equation}
\begin{split}
    \hat G^\text{e/o}_{\ell}(\omega, x, x') = \frac{1}{W_\ell (\omega)}&\left[\theta(x-x')\hat R_{\ell}^\text{e/o in}(\omega, x')\hat R_{\ell}^\text{e/o up}(\omega, x)\right.\\
    &\;\;\left.+ \theta(x'-x)\hat R_{\ell}^\text{e/o in}(\omega, x)\hat R_{\ell}^\text{e/o up}(\omega, x')\right],
\end{split}
\end{equation}
where $W_\ell$ is the Wronskian
\begin{equation}
    W_\ell(\omega) = \hat R_{\ell}^\text{e/o in}(\omega, x)\partial_x \hat R_{\ell}^\text{e/o up}(\omega, x) - \hat R_{\ell}^\text{e/o up}(\omega, x)\partial_x \hat R_{\ell}^\text{e/o in}(\omega, x),
\end{equation}
which we anticipate is identical for the even and odd sector. This Green function satisfies
\begin{equation}
    \left(\partial_x^2 + \omega^2 -  V_{\ell}^{\text{e/o}}\right) \hat G^{\text{e/o}}_{\ell}(\omega, x, x') = \delta(x-x').
\end{equation}
It is straightforward to verify that $W_\ell$ is independent of the field point where it is computed (since $\partial_x W_\ell=0$). We can therefore compute it in the limit $r\to+\infty$ using the asymptotic solutions~\eqref{inup}, yielding
\begin{equation}\label{Wronskian}
    W_\ell(\omega) = 2i\omega A_\ell^{\rm in}\left(\omega\right).
\end{equation}
This justifies $W_\ell$ having no $\text{e/o}$ label. Evaluating the Wronskian in the limit $r\to2M$, we deduce $B_\ell^\text{e/o out}\left(\omega\right)=A_\ell^{\rm in}\left(\omega\right)$. The inhomogeneous solution to Eq.~\eqref{tRWZtransformed} can then be obtained as 
\begin{equation}
\begin{split}
    \hat R^{\text{e/o}\,\mathcal{R}}_{\ell m}(\omega, r) =&\;
    \int_{-\infty}^{+\infty}dx' \hat G^\text{e/o}_\ell (\omega, x, x') \hat S^\text{e/o eff}_{\ell m}(\omega, x')
    \\[1ex]
    =&\;\frac{\hat R^\text{e/o up}_{\ell}(\omega, r)}{W_\ell(\omega)}\int_{2M}^r \frac{dr'}{f(r')} \hat R^\text{e/o in}_{\ell}(\omega, r')\hat S^\text{e/o eff}_{\ell m}(\omega, r')
    \\[1ex]
    &\;+ \frac{\hat R^\text{e/o in}_{\ell}(\omega, r)}{W_\ell(\omega)}\int_r^{+\infty}\frac{dr'}{f(r')} \hat R^\text{e/o up}_{\ell}(\omega, r')\hat S^\text{e/o eff}_{\ell m}(\omega, r').
\end{split}
\end{equation}
As written, the integrals over the (transform of the) extended effective source~\eqref{pppSource} fail to converge due to the source's behaviour at $r=2M$ and $r\to\infty$; this is an artifact of $t$ slicing, which we address in Appendix~\ref{sec:RWZ s slicing}.

The goal of our implementation in this paper is to construct the 0PG asymptotic waveform. Moreover, since we are interested in the merger and ringdown and not in very early times when the particle is near the ISCO, we restrict our attention to the region  $r_p<r_{\cal P}$, outside the puncture window. In this region the physical and residual fields are identical, and we can omit the superscript ${\cal R}$ on the field; see Eq.~\eqref{residual field reduces to physical}. To obtain the waveform for $r_p<r_{\cal P}$, we now focus on the only source in that region: the point-particle portion of the effective source, $S^{\text{e/o pp}}_{\ell m}$. At large distances the field generated by this source, $\left.\hat R^{\text{e/o}}_{\ell m}\right\vert_{{\rm pp}\,\infty}(\omega) \coloneqq \left.\hat R^{\text{e/o}}_{\ell m}\right\vert_\text{pp}(\omega, r\to\infty)$, is given by
\begin{equation}
\begin{split}
    \left.\hat R^{\text{e/o}}_{\ell m}\right\vert_{{\rm pp}\,\infty}(\omega) =&\; \frac{e^{i\omega x}}{W_\ell(\omega)}\int_{2M}^{\infty}\frac{dr'}{f(r')} \hat R^\text{e/o in}_{\ell}(\omega, r')\hat S^\text{e/o pp}_{\ell m}(\omega, r')
    \\[1ex]
    =&\; \frac{e^{i\omega x}}{W_\ell(\omega)}\int_{2M}^{\infty}\frac{dr'}{f(r')} \hat R^\text{e/o in}_{\ell}(\omega, r')\int_{2M}^{r_{\cal P}}\frac{dr_p}{\dot r_{\{0\}}(r_p)} \, e^{i\omega t_G(r_p) - im\phi_G(r_p)} S^{\text{e/o pp}}_{\ell m}(r_p, r').
\end{split}
\end{equation}
Here we have used the asymptotic behaviour of the ``up'' solution~\eqref{inup}. We evaluate the $r_p$ integral using the properties of the Dirac delta function, recalling that $\delta'(r'-r_p)=-\partial_{r_p}\delta(r'-r_p)$. We are then left with
\begin{equation}\label{ppintegral_tslicing}
   \left.\hat R^{\text{e/o}}_{\ell m}\right\vert_{{\rm pp}\,\infty}(\omega) = \frac{\,e^{i\omega x}}{2i\omega A_\ell^{\rm in}\left(\omega\right)}\int_{2M}^{r_{\cal P}}\frac{dr'}{\dot r_{\{0\}}(r')} e^{i\omega t_G(r') - im\phi_G(r')} K^\text{e/o}_{\ell m}(\omega, r'),
\end{equation}
where
\begin{equation}\label{Klm}
\begin{split}
    K_{\ell m}^\text{e/o}(\omega, r) = \frac{A_{\ell m}^\text{e/o}(r)\hat R_{\ell}^\text{e/o in}(\omega, r)}{f(r)} - B_{\ell m}^\text{e/o}(r)\frac{d}{dr}\left(\frac{\hat R_{\ell}^\text{e/o in}(\omega, r)}{f(r)}\right).
\end{split}
\end{equation}

The solution on phase space is obtained after taking the inverse transform~\eqref{transforminv}:
\begin{equation}\label{waveformF}
\begin{split}
    \left.\Psi^{\text{e/o}}_{\ell m}\right\vert_{{\rm pp}\,\infty}(\phi_p,r_p) &= \left. R^{\text{e/o}}_{\ell m}\right\vert_{{\rm pp}\,\infty}(r_p)e^{-i m \phi_p}
    \\[1ex]
    &= -\frac{1}{2\pi}\left[\int_{-\infty}^{+\infty} d\omega\, e^{-i\omega[t_G(r_p) - x]} \frac{C^\text{e/o}_{\ell m}(\omega)}{2i\omega A_\ell^{\rm in}(\omega)}\right]e^{im[\phi_G(r_p) - \phi_p]},
\end{split}
\end{equation}
where
\begin{equation}\label{Clm}
    C^\text{e/o}_{\ell m}(\omega) \coloneqq \int_{2M}^{r_{\cal P}}\frac{dr'}{\dot r_{\{0\}}(r')} e^{i\omega t_G(r') - im\phi_G(r')} K^\text{e/o}_{\ell m}(\omega, r').
\end{equation}
This becomes a time-domain waveform when we substitute the solutions $\phi_p(t)$ and $r_p(t)$ to the orbital equations of motion, $\left.\Psi^{\text{e/o}}_{\ell m}\right\vert_{{\rm pp}\,\infty}(t) = \left.\Psi^{\text{e/o}}_{\ell m}\right\vert_{{\rm pp}\,\infty}(\phi_p(t),r_p(t))$. At geodesic order, the difference $\phi_G(r_p(t)) - \phi_p(t)$ is a constant. At 1PG and higher orders, Eq.~\eqref{waveformF} still represents the leading-order waveform, but $\phi_p(t)$ and $r_p(t)$ become $\varepsilon$ dependent, and $\phi_p(t)$ no longer cancels $\phi_G(r_p(t))$; this incomplete cancellation will be true even at leading, 0PG order if $\phi_p(t)$ is obtained from an equation of motion that is hybridized with the transition-to-plunge dynamics.

We can now point to the singular behaviour cured by the puncture. If there were no puncture, the integral~\eqref{Clm} would run from $2M$ to $6M$. That integral would not converge, due to the fact that $\dot r_{\{0\}}(r')\sim (6M-r')^{3/2}$ near the ISCO. Obtaining the final physical solution requires adding the piece of the field sourced by the effective source~\eqref{Seffext} in the puncture region, call it $\left.R^{\text{e/o}\,\mathcal{R}}_{\ell m}\right\vert_\text{ext}$. However, as shown in Appendix~\ref{app:Sext}, a large part of $\left.R^{\text{e/o}\,\mathcal{R}}_{\ell m}\right\vert_\text{ext}$ vanishes in the limit $r_{\cal P}\to 6M$, and the remainder makes a small contribution to the final waveform for values of $r_p$ sufficiently far below  $r_{\cal P}$; this is to be expected because any small interval of $r_p$ near $6M$ corresponds to an asymptotically large interval of coordinate time. As a first approximation, we therefore place $r_{\cal P}$ very near the ISCO, at $r_{\cal P}=5.999M$, and we neglect $\left.R^{\text{e/o}\,\mathcal{R}}_{\ell m}\right\vert_\text{ext}$. We expect that a complete implementation of the puncture scheme will become more important at higher PG orders, as the early-time behaviour becomes more singular at each successive order.

As a final remark in this section, we highlight that Eq.~\eqref{waveformF} does not precisely yield the waveform at future null infinity. This is because we work with $t$ as our time coordinate. Our waveform therefore represents the coefficient of $1/r$ in a large-$r$ expansion of $h^{\{1\}}_{\bar m\bar m}$ as a function of $t$ along a timelike surface at large $r$. A consequence is the explicit appearance of the tortoise coordinate $x$ in Eq.~\eqref{waveformF}, which makes our waveform phase depend on the value of $r$ at which the waveform is extracted. Such a dependence is degenerate with a choice of initial time, meaning it has no physical consequence in the present paper. A more complete treatment will use hyperboloidal slicing or the sharp null slicing displayed in Fig.~\ref{fig:Penrose} to unambiguously obtain the waveform at future null infinity.

\subsection{Stationary-phase approximation}\label{sec:spa}

We can immediately extract qualitative information from the overall structure of the waveform~\eqref{waveformF}. The waveform comes with a total phase factor $e^{-i\omega t_G(r_p)+im\phi_G(r_p)-im\phi_p}$. At early times, the particle is near the ISCO on a very nearly circular orbit. We therefore expect the waveform to be strongly peaked around the orbital frequency $\omega=m\Omega$ at these early times, such that the first two terms in the exponential cancel, leaving an overall phase factor $e^{-im\phi_p}$; in words, the waveform phase closely mimics the orbital phase. On the other hand, at very late times, when the particle has plunged deep behind the effective potential and is near the black hole horizon, we expect the waveform to be dominated by QNM ringdown frequencies. In that case, $\omega t_G(r_p)$ is unrelated to $\phi_G(r_p)$; instead, the final two terms in the exponential cancel, leaving $e^{-i\omega t_G(r_p)}$.

In this section we apply a stationary-phase approximation that makes the above reasoning about the early-time behaviour more precise. We develop the complementary late-time approximation in the next section.

We can rewrite the $r_p$-domain solution (the coefficient of $e^{-im\phi_p}$ in Eq.~\eqref{waveformF}) as 
\begin{equation}\label{Rlmspa}
    \left.R^{\text{e/o}}_{\ell m}\right\vert_{{\rm pp}\,\infty}(r_p) = \int_{-\infty}^{+\infty} d\omega \int_{2M}^{r_{\cal P}} dr' g(\omega, r') e^{i \varphi(\omega,r',r_p)},
\end{equation}
where we have defined
\begin{align}
    g(\omega, r') &\coloneqq -\frac{1}{2\pi\,\dot r_{\{0\}}(r')}\frac{K^\text{e/o}_{\ell m}(\omega,r')}{2 i \omega A^{\rm in}_{\ell}(\omega)},
    \\[1ex]
    \varphi(\omega, r', r_p) &\coloneqq \omega \left[t_G(r')-t_G(r_p)\right] - m \left[\phi_G(r')-\phi_G(r_p)\right].
\end{align}
For simplicity, here we have set $x=0$. We notice that Eq.~\eqref{Rlmspa} has the form of a two-dimensional integral against a rapidly oscillating complex exponential, especially as the particle is close to the ISCO; recall that $t_G(r_p\to6M)\to-\infty$. We can therefore evaluate the integral using a leading-order SPA,
\begin{equation}
    \left.R^{\text{e/o}}_{\ell m}\right\vert_{{\rm pp}\,\infty}(r_p) \approx \left.\frac{2\pi g(\omega, r')}{|\text{det}\left(\text{Hess}(\varphi)\right)|^{1/2}} e^{i\varphi(\omega,r',r_p) + i \pi \sigma/4}\right\vert_{(\omega, r')=\text{stationary point}},
\end{equation}
where the stationary point $(\omega, r')=(m \Omega_{\{0\}}(r_p), r_p)$ is obtained from the condition $\nabla \varphi = (\partial_\omega \varphi,\partial_{r'}\varphi)=0$. $\text{Hess}(\varphi)$ is the Hessian of $\varphi$. Its signature, $\sigma$, vanishes, and its determinant evaluates to $|\text{det}\left(\text{Hess}(\varphi)\right)|^{1/2}=1/|\dot r_{\{0\}}(r_p)|=-1/\dot r_{\{0\}}(r_p)$ since $\dot r_{\{0\}}$ is negative; see Eq.~\eqref{rGdot}. The leading-order SPA to Eq.~\eqref{Rlmspa} then gives the final solution
\begin{align}\label{spa}
    \left.\Psi^{\text{e/o}}_{\ell m}\right\vert_{{\rm pp}\,\infty}(\phi_p,r_p) \approx \left[\frac{K^\text{e/o}_{\ell m}(\omega,r')}{2 i \omega A^{\rm in}_{\ell}(\omega)}\right]_{\omega=m \Omega_{\{0\}}(r_p), r'=r_p} e^{-im\phi_p}.
\end{align}

Equation~\eqref{spa} is an ``instantaneous'' representation of the waveform, in which the state of the binary system is instantaneously transmitted to infinity. It relies on the waveform frequency and amplitude changing slowly relative to the waveform phase, just as in the inspiral and transition-to-plunge regimes. In Sec.~\ref{sec:spaqnm} we show that the SPA is a good approximation almost up to merger.

\subsection{Quasinormal mode sum}\label{sec:QNMs}

At late times, we expect our 0PG solution to be well described by a sum of QNM modes. In this section we describe how to uniquely extract the amplitudes of these modes.

We can evaluate the real-line $\omega$ integral in Eq.~\eqref{waveformF} by expressing it in terms of residues at poles in the complex $\omega$ plane. 
Figure~\ref{fig:integrationcontour} shows the analyticity properties of the Green function, which has a branch cut along the negative imaginary axis and poles in the lower half plane; the poles, at zeros of the Wronskian~\eqref{Wronskian}, correspond to the QNM frequencies. Following the standard procedure~\cite{Berti:2009kk}, we define a contour that runs along the real axis and closes in the lower half plane, with a detour around the branch cut. By virtue of the residue theorem, the integral along the real line in Eq.~\eqref{waveformF} is equal to the sum of residues of the enclosed poles, minus the integral along the high-frequency arc and the integral around the branch cut:
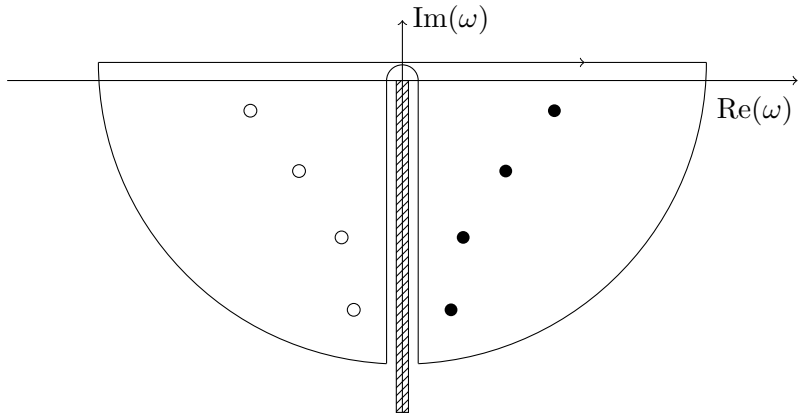
\begin{figure}[tb]
\centering
    \begin{tikzpicture}[scale=0.8][>=stealth]
        \draw[->] (-6.5,0) -- (6.5,0) node[below] {};
        \node at (5.8,-.5) {$\text{Re}(\omega)$};
        \draw[->] (0,-5.5) -- (0,1) node[right] {$\text{Im}(\omega)$};
        \draw[->] (-5,0.3) -- (3,0.3) node[right] {};
        \draw[-] (3,0.3) -- (5,0.3) node[right] {};
        \draw (-5,0.3) arc[start angle=180, end angle=267, radius=5];
        \draw (5,0.3) arc[start angle=0, end angle=-87, radius=5];
        \draw[-] (-0.26,-4.69) -- (-0.26,0) node[right] {};
        \draw[-] (0.26,-4.69) -- (0.26,0) node[right] {};
        \draw (0.26,0) arc[start angle=0, end angle=180, radius=0.26];
        \draw[pattern=north east lines] (-.1,0) rectangle (.1,-5.5);
        \fill (2.5,-.5) circle (3pt);
        \fill (1.7,-1.5) circle (3pt);
        \fill (1,-2.6) circle (3pt);
        \fill (.8,-3.8) circle (3pt);
        \draw (-2.5,-.5) circle (3pt);
        \draw (-1.7,-1.5) circle (3pt);
        \draw (-1,-2.6) circle (3pt);
        \draw (-.8,-3.8) circle (3pt);
    \end{tikzpicture}
\caption{Integration contour for Eq.~\eqref{waveformF} leading to Eq.~\eqref{QNM_BC_arcs}. The branch cut is indicated by the hatched area, while the zeros of the Wronskian~\eqref{Wronskian} are depicted as full (regular QNM frequencies) and empty (mirror QNM frequencies) circles.}
\label{fig:integrationcontour}
\end{figure}
\begin{equation}\label{QNM_BC_arcs}
    \left.\Psi^{\text{e/o}}_{\ell m}\right\vert_{{\rm pp}\,\infty} = \left.\Psi^{\text{e/o}}_{\ell m}\right\vert_\text{QNM} - \left.\Psi^{\text{e/o}}_{\ell m}\right\vert_\text{bc} - \left.\Psi^{\text{e/o}}_{\ell m}\right\vert_\text{arc}.
\end{equation}
Here we focus on the QNM contribution and leave assessing the impact of the branch-cut and arc pieces on the waveform to future work. Using the residue theorem, we then obtain
\begin{equation}\label{Psi_residues}
    \left.\Psi^{\text{e/o}}_{\ell m}\right\vert_\text{QNM} = i\sum_{n=0}^\infty\,{\rm Res} \left[\frac{C^{\text{e/o}}_{\ell m}(\omega)}{2i\omega A_\ell^{\rm in}(\omega)}e^{-i\omega[t_G(r_p)-x]}\right]_{\omega=\omega_{\ell n},-\omega_{\ell n}^*} e^{im[\phi_G(r_p) - \phi_p]}.
\end{equation}
A minus sign due to the clockwise orientation of the contour cancels the overall minus sign in Eq.~\eqref{waveformF}. $\omega_{\ell n}$ with $n=0,1,2,\dots$ are the QNM frequencies, for which $A^{\rm in}_\ell(\omega_{\ell n})=0$. The fundamental mode is indicated by $n=0$, while $n\geq 1$ label the $n$th overtone. Note that for all ``regular'' frequencies $\omega_{\ell n}$ (living in the quadrant where $\text{Re}(\omega_{\ell n})>0$), we also need to include ``mirror'' frequencies $-\omega_{\ell n}^*$, since the QNM frequency spectrum is symmetric with respect to the imaginary axis. 

Computing the residues explicitly and multiplying and dividing by $A^\text{e/o out}_\ell(\omega_{\ell n})$, we can write
\begin{equation}\label{QNMsum}
\begin{split}
    \left.\Psi^{\text{e/o}}_{\ell m}\right\vert_\text{QNM} = \sum_{n=0}^\infty \Bigg[E^\text{e/o}_{\ell m}(\omega)e^{-i\omega(t_G(r_p) - x)}\Bigg]_{\omega=\omega_{\ell n},-\omega_{\ell n}^*}e^{im[\phi_G(r_p) - \phi_p]},
\end{split}
\end{equation}
where $E^\text{e/o}_{\ell m}(\omega_{\ell n})\coloneqq B^\text{e/o}_{\ell}(\omega_{\ell n})D^\text{e/o}_{\ell m}(\omega_{\ell n})$ are the so-called quasinormal excitation coefficients, given by the product of the quasinormal excitation factors $B^\text{e/o}_{\ell}(\omega_{\ell n})$~\cite{Berti:2006wq,Zhang:2013ksa,Oshita:2021iyn} and the coefficients $D^\text{e/o}_{\ell m}(\omega_{\ell n})$ defined as 
\begin{equation}\label{Bl_Dlm}
    B^\text{e/o}_{\ell}(\omega_{\ell n}) \coloneqq \frac{1}{2\omega_{\ell n}}\frac{A^\text{e/o out}_\ell(\omega_{\ell n})}{\frac{dA^{\rm in}_\ell}{d\omega}(\omega_{\ell n})}, \qquad D^\text{e/o}_{\ell m}(\omega_{\ell n})\coloneqq \frac{C^\text{e/o}_{\ell m}(\omega_{\ell n})}{A^\text{e/o out}_\ell(\omega_{\ell n})}.
\end{equation}
While the excitation factors only depend on the Schwarzschild geometry, the coefficients $D^\text{e/o}_{\ell m}(\omega_{\ell n})$ depend on the nature of the perturbation, that is, in our case, on the plunging point-particle source.
As a consequence of Eq.~\eqref{AinAout_eo}, the even and odd excitation factors are related by
\begin{equation}
    B_{\ell}^\text{e}(\omega_{\ell n}) = \frac{D_\ell +12 i M \omega_{\ell n} }{D_\ell -12 i M \omega_{\ell n}} B_{\ell}^\text{o}(\omega_{\ell n}).      
\end{equation}
Furthermore, the properties~\eqref{sAin} imply that
\begin{equation}
    B^\text{e/o}_{\ell}(-\omega_{\ell n}^*) = \left[B^\text{e/o}_{\ell}(\omega_{\ell n})\right]^*.
\end{equation}

The integrand in $C^\text{e/o}_{\ell m}(\omega_{\ell n})$~\eqref{Clm} is exponentially convergent at the upper limit when $r_{\cal P}\approx 6M$, while at the horizon it behaves as $(r'-2M)^{-4iM\omega_{\ell n}}$. The integral therefore converges when $\text{Im}(\omega_{\ell n})>-1/(4M)$, which only holds for the fundamental QNM frequencies. In order to compute the excitation coefficients for the overtones we consider the following regularization procedure~\cite{Leaver:1986gd,Hadar:2009ip}:
\begin{equation}\label{ClmReg}
    \left.C^\text{e/o}_{\ell m}\right\vert_\text{reg}(\omega) \coloneqq \int_{2M}^{r_{\cal P}}dr'\left[\frac{e^{i\omega t_G(r') - im\phi_G(r')}}{\dot r_{\{0\}}(r')} K^\text{e/o}_{\ell m}(\omega, r') - q^\text{e/o}_{\ell m}(\omega, r')\right] + Q^\text{e/o}_{\ell m}(\omega, r_{\cal P}).
\end{equation}
Here we have introduced an auxiliary function $q^\text{e/o}_{\ell m}(\omega, r')$,
\begin{equation}\label{qlm}
    q^\text{e/o}_{\ell m}(\omega, r') = (r'-2M)^{-4iM\omega} \left[q^\text{e/o}_{0\ell m} + q^\text{e/o}_{1\ell m}(\omega) (r'-2M) + q^\text{e/o}_{2\ell m}(\omega) (r'-2M)^2 + \dots \right],
\end{equation}
where the coefficients $q^\text{e/o}_{i\ell m}$ with $i=0,1,2,\dots$ are obtained from the near-horizon expansion of the integrand in Eq.~\eqref{Clm}. Subtracting $q^\text{e/o}_{\ell m}$ from the integrand ensures that the $r'$ integral is finite. This regularized integral is then added to the antiderivative of $q^\text{e/o}_{\ell m}$, $Q^\text{e/o}_{\ell m}(\omega, r)=\int^r dr' q^\text{e/o}_{\ell m}(\omega, r')$ (neglecting any constant arising from the integration), evaluated at the integral's upper boundary, $r=r_{\cal P}$. We justify this regularization procedure in Appendix~\ref{app:regularization}.

Like for the SPA, we explore the accuracy of the QNM sum in Sec.~\ref{sec:spaqnm}.

\section{Numerical implementation and comparisons}\label{sec:imlpementation_and_comparison}

In this section we present our numerical implementation to obtain first-order plunge waveforms from the formalism presented in Sec.~\ref{sec:inhomogsol}, discussing and validating our implementation choices. We also outline the procedure to obtain SPA and QNM waveforms following Secs.~\ref{sec:spa} and \ref{sec:QNMs}, and assess how well they approximate the full waveform in their respective regimes of validity. Finally, we compare the 0PG waveforms to NR simulations. Our numerical implementation is freely accessible in the ancillary material~\cite{AncillaryFiles}.

\subsection{Implementation and validation}\label{sec:implementation}

We start by describing our numerical implementation to obtain first-order (0PG) waveforms from Eq.~\eqref{waveformF}, with $r_{\cal P}=5.999M$. Computing the waveforms consists in two major steps: we first tabulate the quantity $C^\text{e/o}_{\ell m}(\omega)$~\eqref{Clm} on a grid of $\omega$ values. The final waveform is then obtained as a function of $t_G(r_p)-x$ by evaluating Eq.~\eqref{waveformF} using a discrete Fourier transform (DFT). We almost exclusively present the time-domain waveform obtained using the geodesic orbital dynamics, $r_p(t)=r_{\{0\}}(t)$ and $\phi_p(t)=\phi_G(r_p(t))$. In this case, using the freedom to absorb $x$ into a choice of initial time, we are able to rename ``$t_G(r_p)-x$'' simply as ``geodesic time'' $t_G$. Our complete time-domain waveform~\eqref{waveformF} is then
\begin{equation}\label{full TD waveform}
    \left.\Psi^{\text{e/o}}_{\ell m}\right\vert_{{\rm pp}\,\infty}(t_G) =-\frac{1}{2\pi}\int_{-\infty}^{+\infty} d\omega\, e^{-i\omega t_G} \frac{C^\text{e/o}_{\ell m}(\omega)}{2i\omega A_\ell^{\rm in}(\omega)};
\end{equation}
our early-time SPA waveform~\eqref{spa} is
\begin{align}\label{SPA TD waveform}
    \left.\Psi^{\text{e/o}}_{\ell m}\right\vert_{{\rm pp}\,\infty}(t_G) \approx \left[\frac{K^\text{e/o}_{\ell m}(\omega,r_{\{0\}}(t_G))}{2 i \omega A^{\rm in}_{\ell}(\omega)}\right]_{\omega=m \Omega_{\{0\}}(t_G)} e^{-im\phi_G(t_G)};
\end{align}
and our late-time QNM waveform~\eqref{QNMsum} is
\begin{equation}\label{QNM TD waveform}
    \left.\Psi^{\text{e/o}}_{\ell m}\right\vert_\text{QNM} (t_G) = \sum_{n=0}^\infty \Bigg[E^\text{e/o}_{\ell m}(\omega)e^{-i\omega t_G}\Bigg]_{\omega=\omega_{\ell n},-\omega_{\ell n}^*}.
\end{equation}
All numerical results are obtained setting $M=1$.

We first construct the odd-parity ``in'' solution, which goes into $C^{\rm e/o}_{\ell m}(\omega)$ (as defined in Eq.~\eqref{Clm}) through the function $K^\text{o}_{\ell m}$. We use Mathematica's built-in $\texttt{HeunC}$ function (available in Mathematica version 12.1 or higher) as
\begin{equation}\label{RinOddHeun}
\begin{split}
    \hat R_{\ell}^\text{o in}(\omega,r)&=\frac{2M}{r}e^{-i(4M-r)\omega}\left(\frac{r}{2M}-1\right)^{-2 i M \omega} \!\texttt{HeunC}\biggl[\ell+\ell^2-2-8 i M\omega-16M^2\omega^2,
    \\[1ex] 
    &\quad -4M\omega(4M\omega-i),1-4iM\omega,-3,-4 i M \omega,1-\frac{r}{2M}\biggr].
\end{split}
\end{equation}
The even-parity ``in'' solution is then obtained from the odd-parity one using the relation~\eqref{defRe}. We have numerically checked that the definition~\eqref{RinOddHeun} matches with the current implementations of the ``in'' solution in the \texttt{ReggeWheeler} package within the Black Hole Perturbation Toolkit~\cite{BHPToolkit} (\texttt{ReggeWheelerRadial[2,$\ell$,$\omega$,Method -> "HeunC"]["In"][r]}). 
Our ``in'' solution differs from the one defined in Ref.~\cite{Leaver:1986gd} by an overall factor, $\hat R_{\ell}^\text{o in}(\omega,r) = e^{-2 i M \omega} \left.\hat R_{\ell}^\text{o in}(\omega,r)\right\vert_\text{Leaver}$. We get high-precision values for $A_\ell^{\rm in}(\omega)$ in the denominator of Eq.~\eqref{full TD waveform} using the \texttt{ReggeWheeler} package. 

Most of the computation time is spent in the radial numerical integration at a given value of $\omega$, which defines $C^\text{e/o}_{\ell m}(\omega)$. Through trial and error, we settled on the following integration strategy: 
\begin{equation}
  \int_{2}^{5.999} = \int_{2}^{2.1}\vert_{\text{GK}}+  \int_{2.1}^{3}\vert_{\text{GK}}+\int_{3}^{4}\vert_{\text{Levin}}+\int_{4}^{5}\vert_{\text{Levin}}+\int_{5}^{5.999}\vert_{\text{Levin}}.
\end{equation}
The individual integrals are performed either using the Gauss-Kronrod rule (GK) with 50 points or the Levin rule (Levin) with 50 points as options for Mathematica's \texttt{NIntegrate}. We set a timeout of 5000 seconds for each such integral. We found that for all $2\leq\ell\leq12$, $-\ell\leq m\leq\ell$ ($m\neq0$), and $\omega$ in the range $[-4,4]$ these integrals were performed successfully.

We tabulate $C^\text{e/o}_{\ell m}(\omega)$ on an evenly spaced grid of $\omega$ values with spacing $\Delta\omega=10^{-3}M^{-1}$ in the range $\omega \in [-\omega_\text{cutoff},\omega_\text{cutoff}]$ with $\omega_\text{cutoff}=4M^{-1}$. In order to attenuate the effects of the frequency cutoff on the Fourier transform, we used the low-pass filter described in Eq.~(16) of Ref.~\cite{Vega:2009qb} with parameters $(x_0,w,q,s)=(-4,0.2,-3.9,1)$ in the negative $\omega$ range, and a symmetrically defined low-pass filter in the positive $\omega$ range. With this filter, there are only marginally small high-frequency residual oscillations that remain visible in the deep QNM regime; see Fig.~\ref{fig:WFminusQNMs}. As a benchmark of the implementation, the maximal value for $\text{Re}\ \Psi_{\ell m}^{\text{e/o}}(t_G)$ is reached for the $(\ell,m)=(2,2)$ mode at $t_\text{peak} \approx -104.1M$ where $\Psi_{22}^{\text{e}}(t_\text{peak}) \approx (0.5877+0.0137 i)M$. 

\begin{figure}[tbp]
\centering
\includegraphics[width=.72\textwidth]{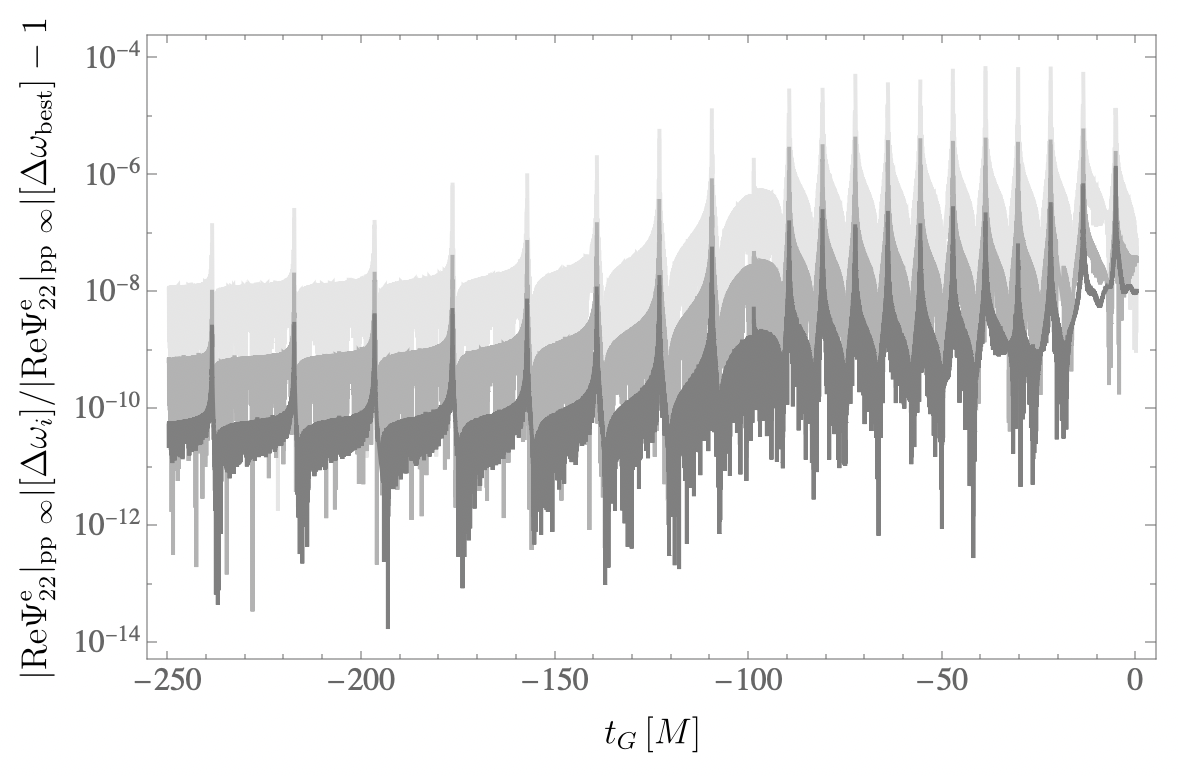}
\caption{Relative difference between the absolute value of the real part of the waveforms $ \vert \text{Re} \Psi^{\text{e}}_{22}\vert_{\text{pp}\,\infty}\vert[\Delta\omega_\text{best}]$ for the $(\ell,m)=(2,2)$ mode with $\omega_\text{cutoff}=4M^{-1}$ and sampling interval $\Delta\omega = \Delta\omega_\text{best} \coloneqq 1/(8000M)$ as compared with the corresponding waveform $ \vert \text{Re} \Psi^{\text{e}}_{22}\vert_{\text{pp}\,\infty}\vert[\Delta\omega_i]$ with sampling $\Delta \omega_1=1/(1000M)$ (light gray), $\Delta \omega_2=1/(2000M)$ (gray) and $\Delta \omega_3=1/(4000M)$ (dark gray).}
\label{fig:WF22samplingw}
\end{figure}

We have investigated how the choice of $\Delta\omega$ impacts the waveforms. We found that using a spacing $\Delta\omega=1/(500M)$ leads to qualitatively incorrect waveforms because the peak of the $(\ell,m)=(2,2)$ mode is not captured. In order to test the accuracy of the waveforms with our choice of $\Delta\omega=1/(1000M)$, we have computed the relative numerical difference between waveforms with different sampling intervals and a benchmark waveform with sampling interval $\Delta\omega=1/(8000M)$. The result is displayed in Fig.~\ref{fig:WF22samplingw} for the $(\ell,m)=(2,2)$ mode: the waveform with $\Delta\omega=1/(1000M)$ has a relative precision of $10^{-7}$ with respect to the waveform with $\Delta\omega=1/(8000M)$. The precision scales globally according $\Delta\omega$: shortening the interval by a factor 2 increases the accuracy by a factor 10.

\begin{figure}[tbp]
\centering
\includegraphics[width=.72\textwidth]{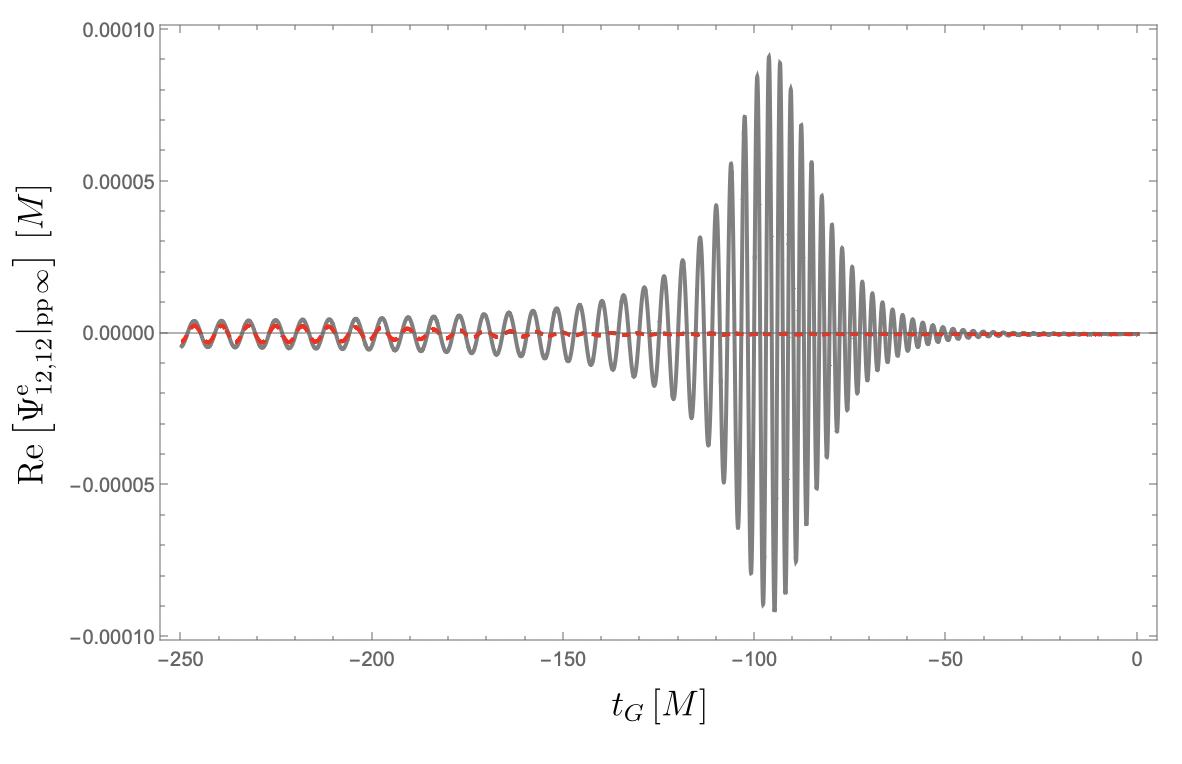}
\includegraphics[width=.72\textwidth]{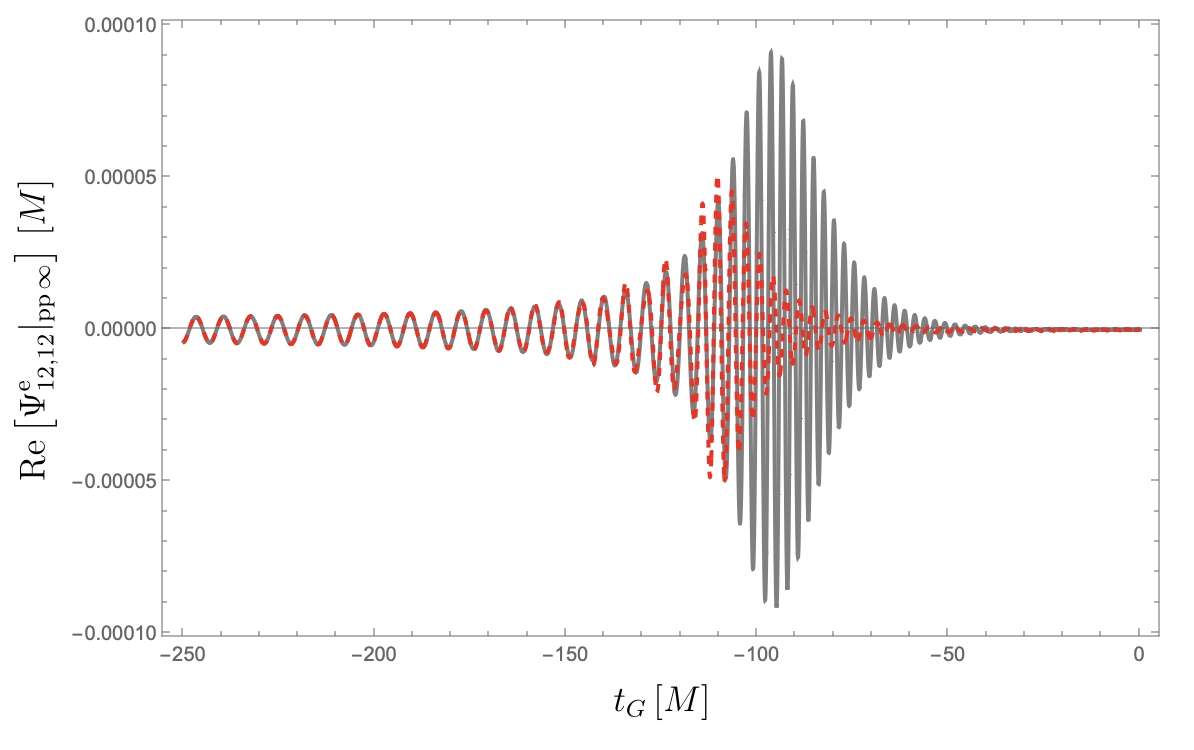}
\includegraphics[width=.72\textwidth]{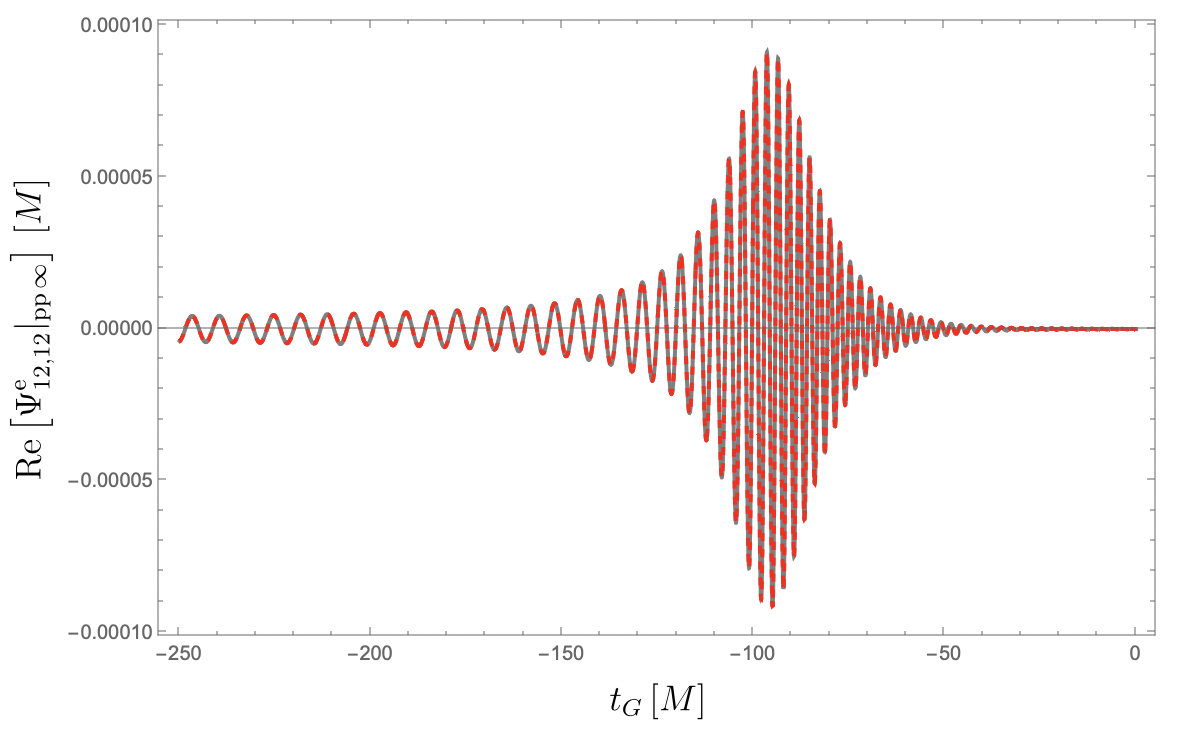}
\caption{Plunge waveforms for the $(\ell,m)=(12,12)$ mode with $\omega_\text{cutoff}=4M^{-1}$ in gray as compared with the corresponding waveforms with $\omega_\text{cutoff}=1M^{-1}$ (top figure), $\omega_\text{cutoff}=2M^{-1}$ (middle figure) and $\omega_\text{cutoff}=3M^{-1}$ (lower figure) in dashed red.}
\label{fig:WF1212w1}
\end{figure}

In order to study the frequency-cutoff dependence, we computed the relative error of the real amplitude as a function of $\omega_\text{cutoff}$ as 
\begin{equation}
    {\cal E}_{\ell m}(\omega_\text{cutoff}') = \frac{1}{\text{Re}\Psi_{22}^{\text{e}}(t_G)\vert_{\omega_\text{cutoff}=4/M}}\text{max}_{t_G} \left(\text{Re}\Psi_{\ell m}^{\text{e/o}}(t_G)\vert_{\omega_\text{cutoff}=\omega_\text{cutoff}'}-\text{Re}\Psi_{\ell m}^{\text{e/o}}(t_G)\vert_{\omega_\text{cutoff}=4/M} \right) .  
\end{equation}
We numerically obtained that ${\cal E}_{\ell m}(\omega_\text{cutoff}=2M^{-1}) < 10^{-4}$ for all $2 \leq \ell \leq 12$ and $-\ell \leq m \leq \ell $, $m \neq 0$ modes except $(\ell,m)=(11,11)$ where ${\cal E}_{11\, 11}(2)\approx2.1 \times 10^{-4}$. In that sense, it is barely sufficient to set the frequency cutoff to $\omega_\text{cutoff}=2M^{-1}$ in order to obtain accurate waveforms with a $10^{-4}$ precision. The cutoff $\omega_\text{cutoff}=1M^{-1}$ only allows us to have waveform with a precision of $10^{-2}$: the maximal value of ${\cal E}_{\ell m}(1)$ is reached for the $(\ell,m)=(5,5)$ mode with ${\cal E}_{55}(1)\approx6.3 \times 10^{-3}$. The analysis of the error at the cutoff $\omega_\text{cutoff}=3M^{-1}$ allows us to provide an estimate of the accuracy of our waveforms which are defined for $\omega_\text{cutoff}=4M^{-1}$. We find ${\cal E}_{\ell m}(\omega_\text{cutoff}=3M^{-1}) < 10^{-6}$ for all $2 \leq \ell \leq 12$ and $-\ell \leq m \leq \ell $, $m \neq 0$ modes except $(\ell,m)=(4,4)$ where ${\cal E}_{4\, 4}(3)\approx1.1 \times 10^{-6}$. In that sense, setting the cutoff to $4M^{-1}$ leads to a loss of precision less than $10^{-6}$. The higher the value of $\ell$, the higher the cutoff should  be in order to qualitatively capture the peak of the waveform. This is illustrated in Fig.~\ref{fig:WF1212w1}. Setting the cutoffs at $1M^{-1}$ or $2M^{-1}$ does not allow us to correctly capture the $(\ell,m)=(12,12)$ waveform, while it is correctly captured using a cutoff at $3M^{-1}$. We learn that even if the precision is $10^{-4}$ at cutoff value $2M^{-1}$, we need a larger cutoff frequency in order to qualitatively capture the peak of the waveforms at high $\ell$. After this analysis we conclude that our waveforms are qualitatively and quantitatively accurate upon setting $\omega_\text{cutoff}=4M^{-1}$. 

\begin{figure}[tbp]
\centering
\includegraphics[width=.72\textwidth]{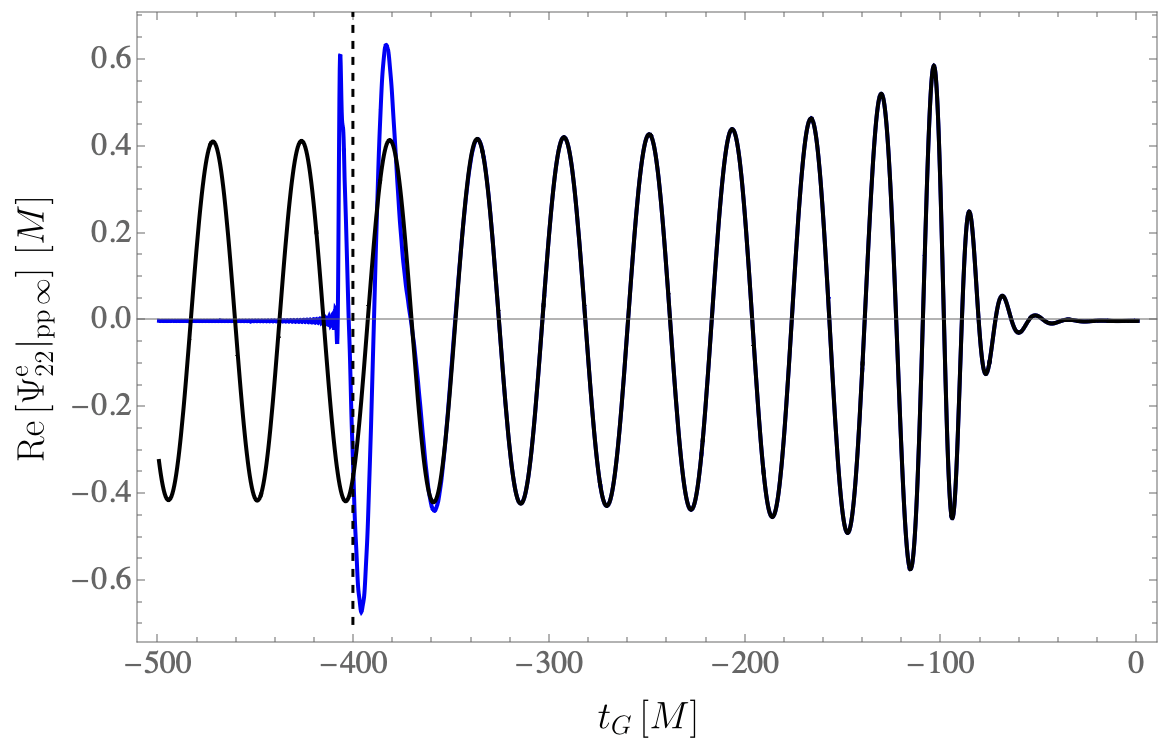}
\caption{The blue curve depicts the real part of the $(\ell, m)=(2, 2)$ mode waveform computed using the cutoff $r_{\cal P}=5.9M$. Around $t_G(r_{\cal P})\approx-401M$ the waveform is not valid due to effects of the cutoff. In black, the waveform is computed using the cutoff $r_{\cal P}=5.99M$ which extends the validity of the waveform to approximately $t_G(r_{\cal P})\approx-1249M$ (which lies much further to the left of the diagram).}
\label{fig:cutoffrc}
\end{figure}

We also investigated the effects of changing the radial cutoff $r_{\cal P}$. In Fig.~\ref{fig:cutoffrc} we plot the waveform corresponding to $r_{\cal P}=5.9M$ and $r_{\cal P}=5.99M$. In the case $r_{\cal P}=5.9M$, we observe that the waveform fails to capture the correct behaviour around the time corresponding to the cutoff $t_G(r_{\cal P})\approx-401M$, due to our omission of the field generated by the effective source in the punctured region. Putting a higher cutoff allows us to better capture the early-time behaviour since the geodesic gets closer and closer to the ISCO located at $r_p=6M$ as $t_G \to -\infty$. For $r_{\cal P}=5.99M$, the waveforms fail around $t_G=-1249M$, while for $r_{\cal P}=5.999$ they fail around $t_G=-3944M$. In the range $-750M < t_G < 100M$, the change in the waveform upon changing the cutoff from $r_{\cal P}=5.999M$ to $r_{\cal P}=5.99M$ is of the order of $10^{-8}M$ and therefore totally negligible. However, the smaller the mass ratio, the greater the time the particle spends near the ISCO, and for some sufficiently small mass ratio we might need to place the cutoff even closer to the ISCO. On the other hand, this is likely a moot point because at 1PLT order we require the first-order field for all $2M<r_p<6M$ as input to the second-order source, calling for a more complete implementation of the puncture scheme.

Next, SPA waveforms are obtained from Eq.~\eqref{SPA TD waveform}. Here no frequency integral is required. The waveform amplitudes (the coefficients of $e^{-im\phi_G(t_G)}$) are tabulated on a grid of $r_p$ values. In practice, $t_G$ is more simply expressed as a function of $r_p$ than vice versa, and we plot the waveforms parametrically using $t_G=t_G(r_p)$ and $\phi_G=\phi_G(r_p)$.

Finally, constructing the QNM sum~\eqref{QNM TD waveform} requires high-precision QNM frequencies $\omega_{\ell n}$, the excitation factors $B^\text{e/o}_{\ell}$, the coefficients $C^\text{e/o}_{\ell m}$, and $A^\text{e/o out}_\ell(\omega_{\ell n})$. We start with the QNM frequencies from Refs.~\cite{Bertiwebsite,Berti:2005ys,Berti:2009kk} and further increase the digits of accuracy using Leaver's continued fraction method~\cite{Leaver:1985ax}. The results for $\ell=2,3,\dots,7$ and $n=0,1,2,3$ are presented in Table~\ref{tab:QNMs}.
\begin{table}[tb]
    \centering
    \begin{NiceTabularX}{\textwidth}{c*4{>{\centering\arraybackslash}X}}
    \CodeBefore
    \rowcolors{1}{}{gray!7}
    \Body
    \toprule
        & $n=0$ & $n=1$ & $n=2$ & $n=3$ \\ \cmidrule{2-5}
        $\ell=2$ & $0.37367 - 0.08896i$  & $0.34671 - 0.27391 i$ & $0.30105 - 0.47828 i$ & $0.25150 - 0.70515 i$\\
        $\ell=3$ & $0.59944 - 0.09270 i$ & $0.58264 - 0.28130 i$ & $0.55168 - 0.47909 i$ & $0.51196 - 0.69034 i$\\
        $\ell=4$ & $0.80918 - 0.09416 i$ & $0.79663 - 0.28433 i$ & $0.77271 - 0.47991 i$ & $0.73984 - 0.68392 i$\\
        $\ell=5$ & $1.01230 - 0.09487 i$ & $1.00222 - 0.28582 i$ & $0.98270 - 0.48033 i$ & $0.95500 - 0.68056 i$\\
        $\ell=6$ & $1.21201 - 0.09527 i$ & $1.20357 - 0.28665 i$ & $1.18707 - 0.48056 i$ & $1.16327 - 0.67859 i$\\
        $\ell=7$ & $1.40974 - 0.09551 i$ & $1.40247 - 0.28716 i$ & $1.38818 - 0.48071 i$ & $1.36736 - 0.67735 i$\\
        \bottomrule
    \end{NiceTabularX}
    \caption{Schwarzschild QNM frequencies $M \omega_{\ell n}$ up to the third overtone for each $2\leq\ell\leq7$.}
    \label{tab:QNMs}
\end{table}
We take the excitation factors from the tabulated values in Refs.~\cite{Bertiwebsite,Zhang:2013ksa}. The coefficients $C^\text{e/o}_{\ell m}$ are computed from Eq.~\eqref{Clm} for the fundamental mode and Eq.~\eqref{ClmReg} for the overtones. We have kept the first five terms in Eq.~\eqref{qlm} (we have verified that increasing the number of terms to ten leads to a relative difference of $\sim10^{-6}$ or smaller in $C^\text{e/o}_{\ell m}$). We have tabulated all excitation coefficients $E^\text{e/o}_{\ell m}(\omega_{\ell n}) = B^\text{e/o}_{\ell}(\omega_{\ell n})D^\text{e/o}_{\ell m}(\omega_{\ell n})$ for $\ell=2$ up to the third overtone in Tables~\ref{tab:excitation_coefs_l2_omegaln} and \ref{tab:excitation_coefs_l2_-omegaln*}. The excitation coefficients for higher $\ell$ are provided in the ancillary material~\cite{AncillaryFiles}. Finally, we compute the asymptotic amplitudes $A^\text{e/o out}_\ell(\omega_{\ell n})$ using the procedure described in Ref.~\cite{Leaver:1986gd}.
\begin{table}[tb]
    \centering
    \begin{NiceTabularX}{1\textwidth}{c*4{>{\centering\arraybackslash}X}}
    \CodeBefore
    \rowcolors{1}{}{gray!7}
    \Body
    \toprule
        & $n=0$ & $n=1$ & $n=2$ & $n=3$ \\ \cmidrule{2-5}
        $m=2$ & \thead{$-1.14668\times10^{-4}$ \\[1ex] $+5.72179\times10^{-5}i$} & \thead{$-1.12799\times10^{-12}$ \\[1ex] $+4.13842\times10^{-13}i$} & \thead{$-8.03493\times10^{-22}$ \\[1ex] $-1.42463\times10^{-21}i$} & \thead{$2.52610\times10^{-31}$ \\[1ex] $-1.32057\times10^{-31}i$}\\
        $m=1$ & \thead{$6.53075\times10^{-5}$ \\[1ex] $-1.87158\times10^{-5}i$} & \thead{$1.33791\times10^{-12}$ \\[1ex] $+1.37608\times10^{-13}i$} & \thead{$-9.43025\times10^{-22}$ \\[1ex] $+2.48916\times10^{-21}i$} & \thead{$-3.96283\times10^{-31}$ \\[1ex] $-3.66314\times10^{-31}i$}\\
        $m=0$ & \thead{$2.06438\times10^{-5}$ \\[1ex] $+8.84288\times10^{-6}i$} & \thead{$4.91061\times10^{-13}$ \\[1ex] $+5.75437\times10^{-13}i$} & \thead{$-2.12623\times10^{-21}$ \\[1ex] $+4.28862\times10^{-22}i$} & \thead{$2.36628\times10^{-31}$ \\[1ex] $-4.82273\times10^{-31}i$}\\
        $m=-1$ & \thead{$-8.83249\times10^{-8}$ \\[1ex] $-4.89523\times10^{-6}i$} & \thead{$8.87604\times10^{-14}$ \\[1ex] $-2.20023\times10^{-13}i$} & \thead{$5.31923\times10^{-22}$ \\[1ex] $+7.96482\times10^{-22}i$} & \thead{$-2.92157\times10^{-31}$ \\[1ex] $-7.312004\times10^{-32}i$}\\
        $m=-2$ & \thead{$6.18182\times10^{-7}$ \\[1ex] $-8.08701\times10^{-8}i$} & \thead{$3.95622\times10^{415}$ \\[1ex] $+5.77111\times10^{-15}i$} & \thead{$-1.49044\times10^{-22}$ \\[1ex] $+1.63288\times10^{-22}i$} & \thead{$1.01126\times10^{-32}$ \\[1ex] $-8.95405\times10^{-32}i$}\\
    \bottomrule
    \end{NiceTabularX}
    \caption{Excitation coefficients $E^\text{e/o}_{\ell m}(\omega_{\ell n})$ for $\ell=2$ up to $n=3$.}
    \label{tab:excitation_coefs_l2_omegaln}
\end{table}
\begin{table}[tb]
    \centering
    \begin{NiceTabularX}{1\textwidth}{c*4{>{\centering\arraybackslash}X}}
    \CodeBefore
    \rowcolors{1}{}{gray!7}
    \Body
    \toprule
        & $n=0$ & $n=1$ & $n=2$ & $n=3$ \\ \cmidrule{2-5}
        $m=2$ & \thead{$6.18182\times10^{-7}$ \\[1ex] $+8.08701\times10^{-8}i$} & \thead{$3.95622\times10^{-14}$ \\[1ex] $-5.77111\times10^{-15}i$} & \thead{$-1.49044\times10^{-22}$ \\[1ex] $-1.63288\times10^{-22}i$} & \thead{$1.01126\times10^{-32}$ \\[1ex] $+8.95405\times10^{-32}i$}\\
        $m=1$ & \thead{$8.83249\times10^{-8}$ \\[1ex] $+4.89523\times10^{-6}i$} & \thead{$-8.87604\times10^{-14}$ \\[1ex] $-2.20023\times10^{-13}i$} & \thead{$-5.31923\times10^{-22}$ \\[1ex] $+7.96482\times10^{-22}i$} & \thead{$2.92157\times10^{-31}$ \\[1ex] $-7.31200\times10^{-32}i$}\\
        $m=0$ & \thead{$2.06438\times10^{-5}$ \\[1ex] $-8.84288\times10^{-6}i$} & \thead{$4.91061\times10^{-13}$ \\[1ex] $-5.75437\times10^{-13}i$} & \thead{$-2.12623\times10^{-21}$ \\[1ex] $-4.28862\times10^{-22}i$} & \thead{$2.36628\times10^{-31}$ \\[1ex] $+4.82273\times10^{-31}i$}\\
        $m=-1$ & \thead{$-6.53075\times10^{-5}$ \\[1ex] $-1.87158\times10^{-5}i$} & \thead{-1.33791$\times10^{-12}$ \\[1ex] $+1.37608\times10^{-13}i$} & \thead{$9.43025\times10^{-22}$ \\[1ex] $+2.48916\times10^{-21}i$} & \thead{$3.96283\times10^{-31}$ \\[1ex] $-3.66314\times10^{-31}i$}\\
        $m=-2$ & \thead{$-1.14668\times10^{-4}$ \\[1ex] $-5.72179\times10^{-5}i$} & \thead{$-1.12799\times10^{-12}$ \\[1ex] $-4.13842\times10^{-13}i$} & \thead{$-8.03493\times10^{-22}$ \\[1ex] $+1.42463\times10^{-21}i$} & \thead{$2.52610\times10^{-31}$ \\[1ex] $+1.32057\times10^{-31}i$}\\
    \bottomrule
    \end{NiceTabularX}
    \caption{Excitation coefficients $E^\text{e/o}_{\ell m}(-\omega_{\ell n}^*)$ for $\ell=2$ up to $n=3$.}
    \label{tab:excitation_coefs_l2_-omegaln*}
\end{table}

At this point, we can compare our excitations coefficients with the ones obtained by Hadar and Kol (HK) in Tables~1 and 2 of Ref.~\cite{Hadar:2009ip}, and by Folacci and Ould El Hadj (FO) in Table~II of Ref.~\cite{Folacci:2018cic}. In order to do so we first need to find dictionaries between the different conventions that are being used. Hadar and Kol fix the constant $t_0$ in Eq.~\eqref{trG} such that $t_G(2.2M)=0$, which leads to the relation $E^\text{e/o}_{\ell m}(\omega_{\ell n}) = \left.R_{n\ell m}\right\vert_\text{HK} e^{i\omega_{\ell n} t_G(2.2M)}$ between the excitation coefficients. We then find agreement with the results of Ref.~\cite{Hadar:2009ip} up to an overall minus sign with relative errors of $\sim10^{-4}$ or smaller (and a few isolated cases have relative errors of $10^{-3}$). We suspect the sign disagreement to stem from different conventions used to construct the Green function in Ref.~\cite{Hadar:2009ip}. It is however of secondary importance for the aim of building full IMR waveforms: the orbital phase during the plunge is known up to a constant, which will be determined by the orbital phase at the end of the transition to plunge. Using the conversion $\delta'(t-t_G) = -\left(\dot r_{\{0\}}(r_p)\right)^2\delta'(r-r_p) + (\cdots)\delta(r-r_p)$, we compare the terms proportional to $\delta'(r-r_p)$ in our sources~\eqref{sourceplunget} and those in Eqs.~(14) and (15) of Ref.~\cite{Folacci:2018cic}. We find that the sources differ by an overall factor, $S^\text{e/o}_{\ell m}=c_\text{FO}\left.S^\text{e/o}_{\ell m}\right\vert_\text{FO}$ with $c_\text{FO}=2\sqrt{2\pi}$. This leads to $E^\text{e/o}_{\ell m}(\omega_{\ell n}) = c_\text{FO}\left(-\sqrt{2\pi}\left.C^{(\text{e/o})}_{\ell mn}\right\vert_\text{FO}\right)$ and $E^\text{e/o}_{\ell m}(-\omega^*_{\ell n}) = c_\text{FO}\left(-\sqrt{2\pi}\left.D^{(\text{e/o})}_{\ell mn}\right\vert_\text{FO}\right)$. Up to an overall factor of $1/\sqrt{2\pi}$ and an overall minus sign, we find the first equality to be satisfied with relative errors of $\sim10^{-4}$ or smaller (except for the $\ell=m=2$ excitation coefficient, where we suspect a typo in the sign of the imaginary part in Ref.~\cite{Folacci:2018cic}). We also find the equality for the mirror modes to have larger relative errors especially for $\ell\geq4$, which originate from disagreements of the same magnitude already at the level of the QNM frequencies. We have investigated the source of the disagreement by an overall factor of $1/\sqrt{2\pi}$: in Ref.~\cite{Folacci:2018cic} the inverse Fourier transform is defined in Eqs.~(21) and (22) with the symmetric normalization $1/\sqrt{2\pi}$. Equations~(15) and (25) in Ref.~\cite{Folacci:2018cic} are however consistent with a forward transform of the form $S^{(e/o)}_{\omega\ell m}(r)=\int_{-\infty}^{+\infty}dt\,S^{(e/o)}_{\ell m}(t,r) e^{i\omega t}$, missing a factor $1/\sqrt{2\pi}$. Again, we have not found the source of the minus sign disagreement.

\subsection{Comparison with the stationary-phase approximation and the QNM sum}\label{sec:spaqnm}

In this subsection we perform internal consistency checks and compare the 0PG waveforms computed following the procedure described in Secs.~\ref{sec:inhomogsol} and \ref{sec:implementation} with the SPA and the QNM sum discussed in Secs.~\ref{sec:spa} and \ref{sec:QNMs}, respectively.

\begin{figure}[tb]
\centering
\includegraphics[width=\textwidth]{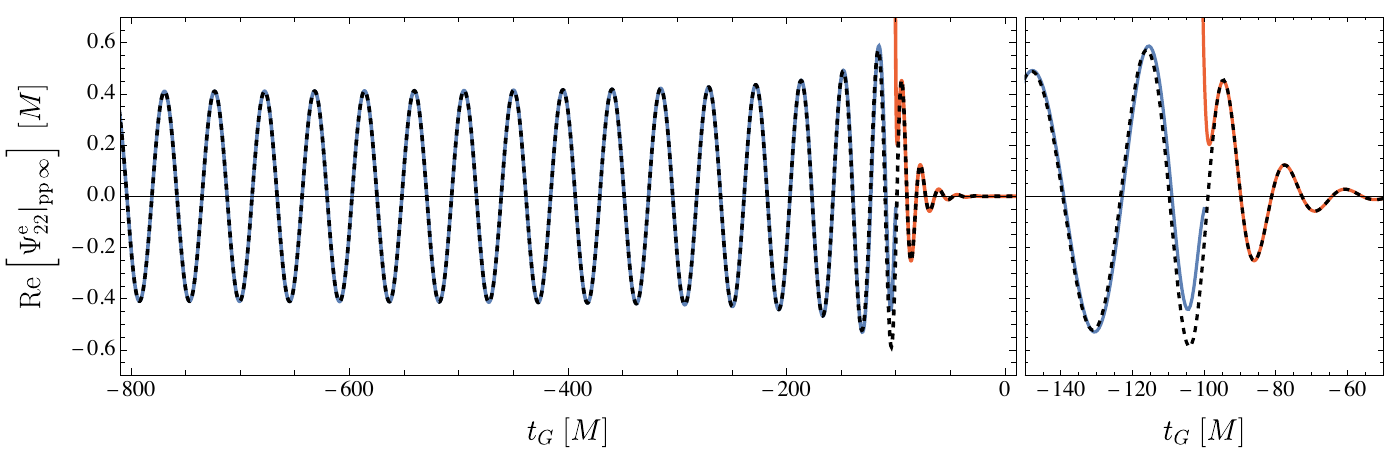}
\caption{The $(\ell,m)=(2,2)$ mode of the 0PG waveform (black dashed curve) compared to the corresponding SPA~\eqref{spa} (blue curve) and QNM~\eqref{QNMsum} (red curve) waveforms. The QNM sum includes overtones through $n=3$. We have truncated the SPA and QNM waveforms once they start to significantly differ from the 0PG waveform. The plot on the right zooms in on the region around the peak amplitude.}
\label{fig:WFvsSPA}
\end{figure}

In Fig.~\ref{fig:WFvsSPA} we compare the SPA~\eqref{spa} (blue curve) with the 0PG waveform (dashed black curve) for the $(\ell,m)=(2,2)$ mode. The waveforms agree very well at early times, with relative errors below $1\%$ up to four cycles before the peak amplitude. Around merger the two waveforms start to differ significantly, with the SPA being unable to capture the late-time QNM behaviour. Indeed, by construction, the SPA is expected to work as long as the integrand in Eq.~\eqref{Rlmspa} contains a rapidly oscillating factor, which occurs as the particle is close to the ISCO and corresponds to the early portion of the waveform. This is also consistent with the geodesic plunge trajectory displayed in Fig.~\ref{fig:plunge_trajectory}, where the particle spends a large number of orbits close to the ISCO before falling from $r\approx5.7M$ to the event horizon within $\sim2$ orbital cycles.

\begin{figure}[tbp]
\centering
\includegraphics[width=.48\textwidth]{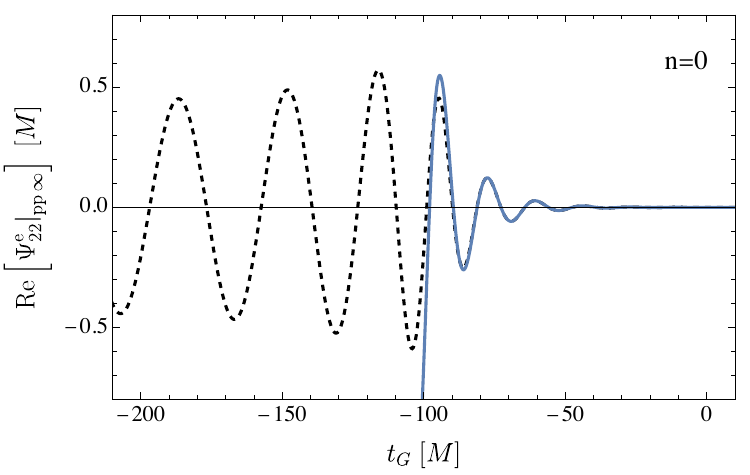}
\hspace{0.3cm}
\includegraphics[width=.48\textwidth]{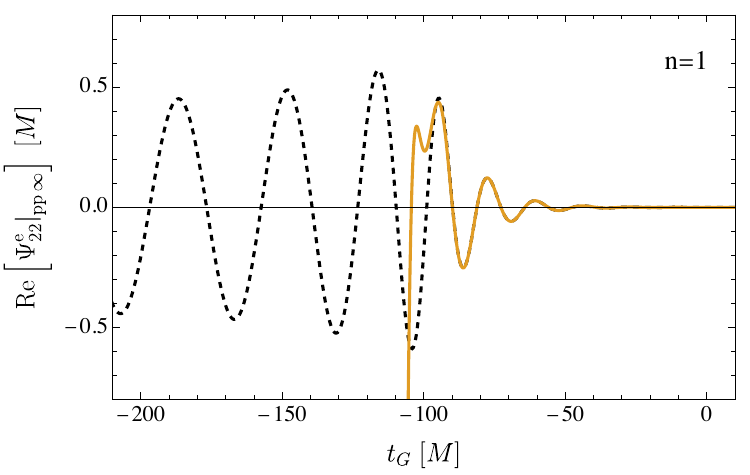}
\\[3ex]
\includegraphics[width=.48\textwidth]{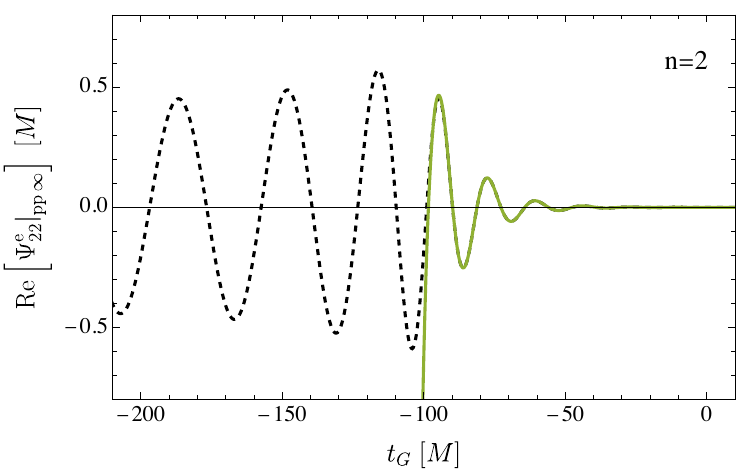}
\hspace{0.3cm}
\includegraphics[width=.48\textwidth]{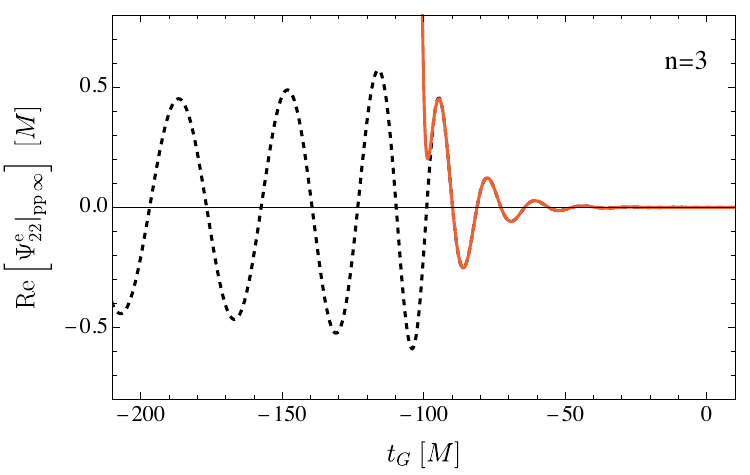}
\caption{0PG waveforms for the $(\ell,m)=(2,2)$ mode (dashed black curves) compared with the corresponding QNM sums that include the fundamental mode only ($n=0$) and an increasing number of overtones up to $n=3$. Note that we are only plotting the QNM waveforms up to a time where they start to significantly differ from the 0PG waveforms.}
\label{fig:WFvsQNMs}
\end{figure}

\begin{figure}[tbp]
\centering
\includegraphics[width=.8\textwidth]{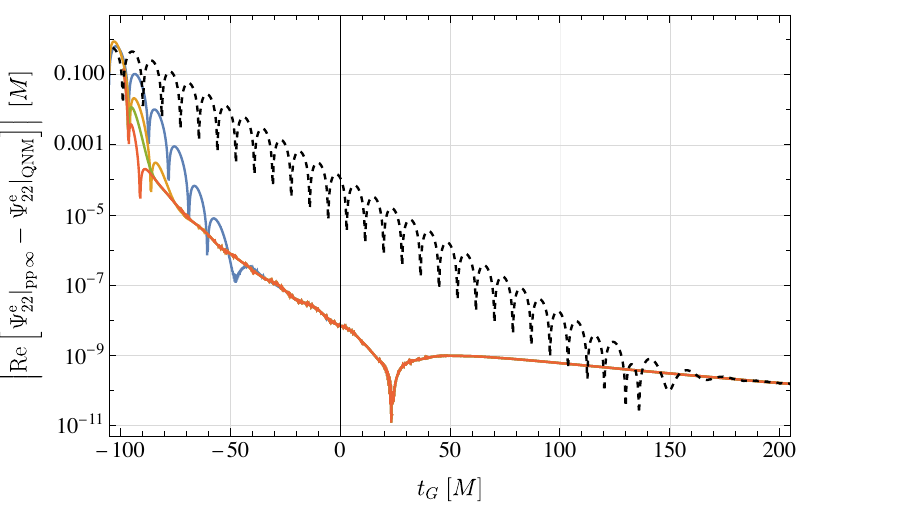}
\caption{Difference between the 0PG waveform for the $(\ell,m)=(2,2)$ mode and the corresponding QNM sums with an increasing number of overtones. Following the color coding of Fig.~\ref{fig:WFvsQNMs}, the blue curve corresponds to the QNM sum that includes only the fundamental mode, while the orange, green and red curves correspond to QNM sums truncated at the first, second and third overtone, respectively. For reference, we have also plotted the 0PG waveform as a dashed black curve. The $y$ axis is on a logarithmic scale.}
\label{fig:WFminusQNMs}
\end{figure}

As shown by the comparison with the red curve in Fig.~\ref{fig:WFvsSPA}, at late times the 0PG waveform is well described by the QNM sum~\eqref{QNMsum}. Figure~\ref{fig:WFvsQNMs} shows the comparison between the 0PG waveform for the $(\ell,m)=(2,2)$ mode and QNM sums that include a different number of overtones (up to the first three), while in Fig.~\ref{fig:WFminusQNMs} we plot the difference between the waveform and the QNM sums. These plots confirm that the QNM approximation correctly reproduces the late-time behaviour of the waveform, with increasing accuracy as higher overtones are included. Although including a larger number of overtones improves the accuracy of the QNM sum at late times, doing so does not extend its accuracy to earlier times. Crucially, the regime of validity breaks down well before reaching the waveform's peak amplitude. We can see from the right panel of Fig.~\ref{fig:WFvsSPA} that this breakdown occurs around $\approx 10M$ after the peak, which is consistent with studies of numerical merger-ringdown waveforms~\cite{Cheung:2023vki,Lim:2022veo,Carullo:2024smg,Mitman:2025hgy}. Those studies found that numerically fitting QNM sums to NR data becomes stable and robust around $10M$--$20M$ after the waveform peak.

We hence conclude that our first-order waveforms are internally consistent and agree with expectations in the sense that the SPA correctly reproduces the periodic behaviour at early times, while the QNM sum describes the ringdown at late times. The combination of the two approximations accurately covers almost the full 0PG waveform, only missing a few cycles around the peak amplitude.

\begin{figure}[tb]
\centering
\includegraphics[width=.8\textwidth]{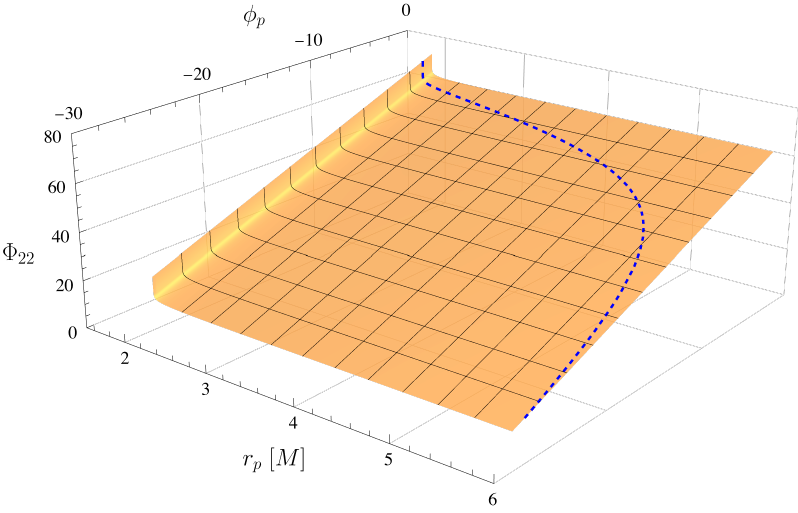}
\caption{Waveform phase of the $(\ell,m)=(2,2)$ mode as a function of the phase-space variables $(\phi_p,r_p)$, as defined from Eq.~\eqref{hlm conventions}. The blue dashed curve shows the path through phase space of the waveform phase at 0PG order, where $\phi_p=\phi_G(r_p)$ is given by Eq.~\eqref{phirG}}.
\label{fig:wfphase_phasespace}
\end{figure}

With these results in mind, we return to the phase-space description of the waveform. The SPA makes explicit that the waveform is a simple function of phase-space coordinates until near the waveform's peak amplitude, while the QNM sum, which involves frequencies unrelated to the orbit, suggests that the waveform becomes divorced from the phase-space trajectory after the peak amplitude. However, this intuition is somewhat misleading. Before generating the waveform as a function of time, we can directly examine its original form~\eqref{waveformF} as a function of the phase-space coordinates $(\phi_p,r_p)$. In Fig.~\ref{fig:wfphase_phasespace} we plot the waveform phase as a function of $\phi_p$ and $r_p$. Here we see there is no drama at the waveform peak, corresponding to $r_p\approx 3M$. The emitted waveform's phase is a simple, smooth function of the particle's orbital parameters until the particle gets extremely close to the event horizon; there, at $r_p\approx2.01M$, the waveform phase rapidly dissociates from the state of the particle's orbit. 

In the figure we also plot the particle's phase-space trajectory. From that trajectory, we see that at 0PG order, the waveform phase gradually deviates from $m \phi_p$ over the course of the plunge. When the trajectory approaches the horizon, it effectively freezes: approximately 0.7~radians after the peak, large changes in the waveform phase become associated with virtually no change in $\phi_p$ or $r_p$. It is this marked change in behaviour, when the particle is far below the light ring, that coincides with the time ($\approx10$--15$M$ after the peak) when the QNM sum begins to approximate the complete waveform.

At higher PG orders, the surface in Fig.~\ref{fig:wfphase_phasespace} will be deformed due to terms $h^{\{2\}}_{22}$ and higher in the waveform. The trajectory on the surface will also change due to 1PG and higher corrections to the orbital motion.

\subsection{Comparison with numerical relativity}

This subsection contains qualitative comparisons between our 0PG waveforms and NR simulations from the SXS catalog~\cite{sxscatalog,Boyle:2019kee}. All SXS data we work with here are drawn from the public python package \texttt{sxs}~\cite{sxspackage}. 

We start by re-expanding the relevant quantities for our waveform generation in powers of the symmetric mass ratio $\nu\coloneqq M m_p/(M+m_p)^2=\varepsilon/(1+\varepsilon)^2$ at fixed total mass $M_{\rm tot}\coloneqq M+m_p=M(1+\varepsilon)$. This yields the most accurate comparisons in the regime of comparable-mass binaries~\cite{LeTiec:2011bk,LeTiec:2011dp,LeTiec:2013uey,Nagar:2013sga,LeTiec:2017ebm,Rifat:2019ltp,vandeMeent:2020xgc,Wardell:2021fyy,Kuchler:2024esj}. Inverting the relation between $\nu$ and $\varepsilon$ gives $\varepsilon=(1-2\nu-\sqrt{1-4\nu})/(2\nu)$, which, in the small-mass-ratio expansion, leads to $\varepsilon=\nu+2\nu^2+{\cal O}(\nu^3)$. For our first-order waveforms this re-expansion simply amounts to substituting $\varepsilon\to\nu$ and $M\to M_{\rm tot}$. In order to compare the 0PG and NR waveforms, we align them in time and phase at a given waveform frequency. Referring to Eq.~\eqref{hlm conventions}, we define the $\ell m$-mode waveform phase and frequency from $h_{\ell m}=|H_{\ell m}|e^{-i\Phi_{\ell m}}$ as $\Phi_{\ell m}\coloneqq m\phi_p - \text{arg}(H_{\ell m})$ and $\omega_{\ell m}\coloneqq d\Phi_{\ell m}/dt$.

\begin{figure}[tbp]
\centering
\includegraphics[width=.48\textwidth]{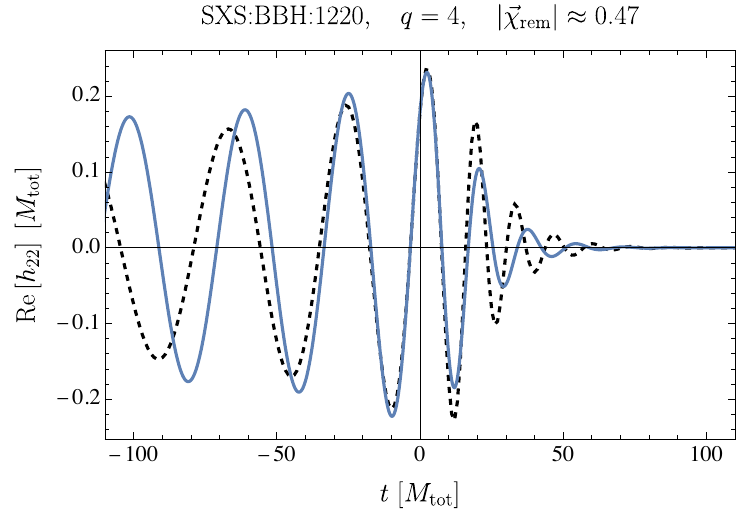}
\hspace{0.3cm}
\includegraphics[width=.48\textwidth]{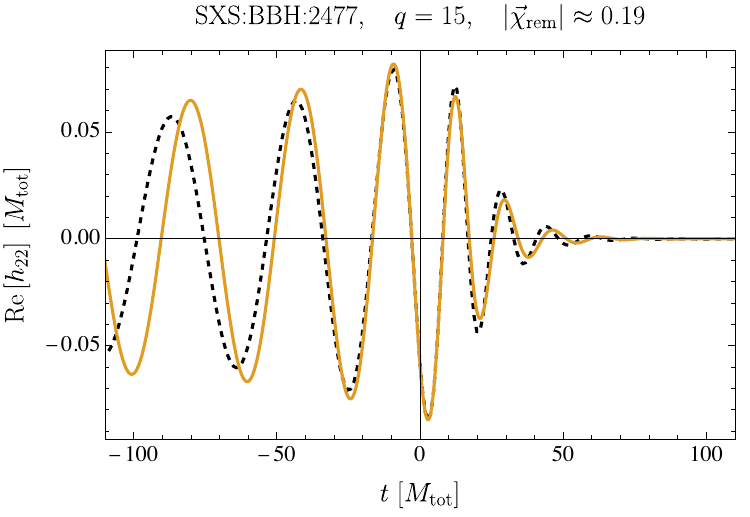}
\\[3ex]
\includegraphics[width=.48\textwidth]{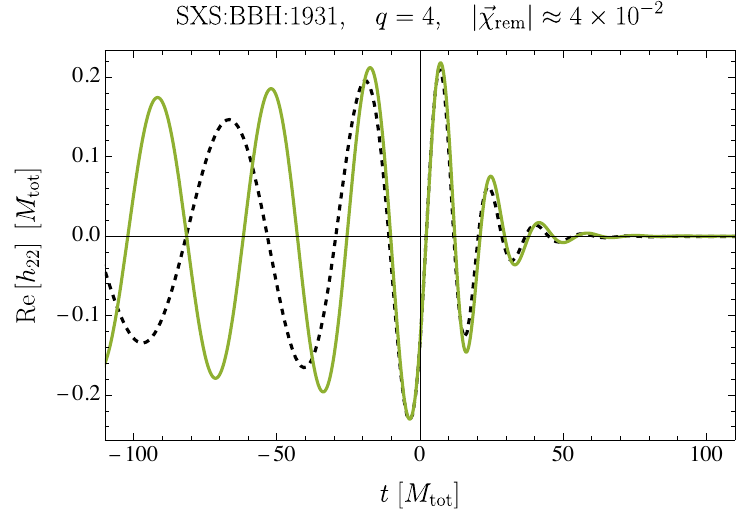}
\hspace{0.3cm}
\includegraphics[width=.48\textwidth]{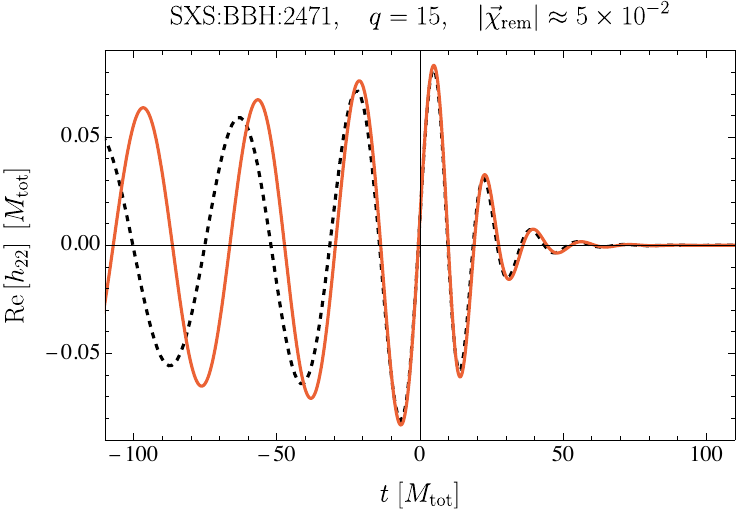}
\caption{Comparison between NR simulations (dashed black curves) and our 0PG waveforms. The waveforms are aligned in phase and frequency at $t=0$. Top panel: comparison with SXS:BBH:1220~\cite{jonathan_blackman_2024_13159569} and SXS:BBH:2477~\cite{sxs_collaboration_2024_13147606}, two quasi-circular, non-spinning black hole binaries with remnant dimensionless spin of $|\vec{\chi}_\text{rem}|\approx0.47$ and $|\vec{\chi}_\text{rem}|\approx0.19$, respectively. Bottom panel: comparison with SXS:BBH:1931~\cite{sxs_collaboration_2019_3273103} and SXS:BBH:2471~\cite{sxs_collaboration_2024_13147573}, two quasi-circular, spinning black hole binaries which coalesce to form a non-spinning remnant (remnant dimensionless spin $|\vec{\chi}_\text{rem}|\sim10^{-2}$).}
\label{fig:SXS}
\end{figure}

After the above preliminaries, we can begin to assess the accuracy of our 0PG waveforms, using SXS waveforms as our benchmark. The top panel of Fig.~\ref{fig:SXS} compares the $(\ell,m)=(2,2)$ mode of our 0PG waveforms with the NR simulations SXS:BBH:1220~\cite{jonathan_blackman_2024_13159569} and SXS:BBH:2477~\cite{sxs_collaboration_2024_13147606}. These are NR waveforms for quasi-circular, non-spinning black hole binaries with mass ratios $q=4$ and $q=15$, respectively. For both simulations, the reference dimensionless spin on each of the two black holes is $\sim10^{-5}$ or smaller, and the eccentricity is $\lesssim10^{-4}$. We align the waveforms in time and phase at $\omega_{22}=2\Omega_{\{0\}}(3M_{\rm tot})$, which corresponds to the peak amplitude $|H_{22}|$ of the plunge waveform. We also shift the relevant part of the waveforms in time such that $t=0$ corresponds to the alignment time. In the $q=4$ case the 0PG waveform is not able to accurately capture the late-time behaviour of the NR simulation, while the comparison improves drastically for $q=15$. The improvement with increasing $q$ is expected: our 0PG model omits 1PG and higher terms in the phase and amplitude, leading to relative errors that scale roughly linearly with $\varepsilon$. 

Numerous 1PG and higher effects contribute to our model's error, but the error after the waveform's peak appears to be dominated by the fact that our 0PG model does not account for the final, remnant black hole's spin. In the the top panel of Fig.~\ref{fig:SXS}, the NR simulations we compare against have small spins on the individual black holes during the inspiral, but the remnant black holes have substantial dimensionless spins $|\vec{\chi}_\text{rem}|\approx0.47$ for SXS:BBH:1220 and $|\vec{\chi}_\text{rem}|\approx0.19$ for SXS:BBH:2477. The spin of the remnant is largely determined by the dimensionless angular momentum of the plunging particle, which at 0PG order is given by its value at the ISCO, $2\sqrt{3}\nu$, equal to $\approx 0.55$ for $q=4$ and $\approx 0.20$ for $q=15$, in rough agreement with the SXS values. This significant final spin leads to QNM ringdown frequencies that differ significantly from the Schwarzschild ones captured by our 0PG model. In a 1PG model, these shifts in the $(\ell,m)=(2,2)$ ringdown frequencies should be accounted for primarily by the nonlinear source $\delta^2 G^{\{0\}}_{\alpha\beta}[h^{\{1\}},h^{\{1\}}]$, through couplings between the $m=2$ modes of $h^{\{1\}}_{\alpha\beta}$ and the $(\ell,m)=(1,0)$ mode of $h^{\{1\}}_{\alpha\beta}$; this is because the $(1,0)$ mode encodes the angular momentum content in $h^{\{1\}}_{\alpha\beta}$, and it necessarily asymptotes, at late times, to a perturbation toward the final, stationary Kerr metric of the remnant.

To investigate whether the remnant spin is really our main source of error in the post-peak regime, we now consider two alternative NR waveforms that have the same mass ratios as the ones used above but negligible remnant spin: the quasi-circular binary black hole simulations SXS:BBH:1931~\cite{sxs_collaboration_2019_3273103} with $q=4$ and SXS:BBH:2471 with $q=15$~\cite{sxs_collaboration_2024_13147573}. In these simulations the individual black holes are spinning prior to merger, but the remnant dimensionless spins are very small ($|\vec{\chi}_\text{rem}|\sim10^{-2}$). We compare these NR simulations to our 0PG waveforms in the bottom panel of Fig.~\ref{fig:SXS}. At late times the agreement has improved since the Schwarzschild QNM sum is now a better approximation to the remnants' ringdown. Before the peak amplitude the comparison instead becomes worse since before merger the NR waveform describes the inspiral of two spinning black holes, which is not the scenario described by our waveforms. In any case, we do not expect the plunge waveform to be accurate far before the peak amplitude and expect the transition to plunge to take over before reaching the ISCO.

\begin{figure}[tbp]
\centering
\includegraphics[width=.48\textwidth]{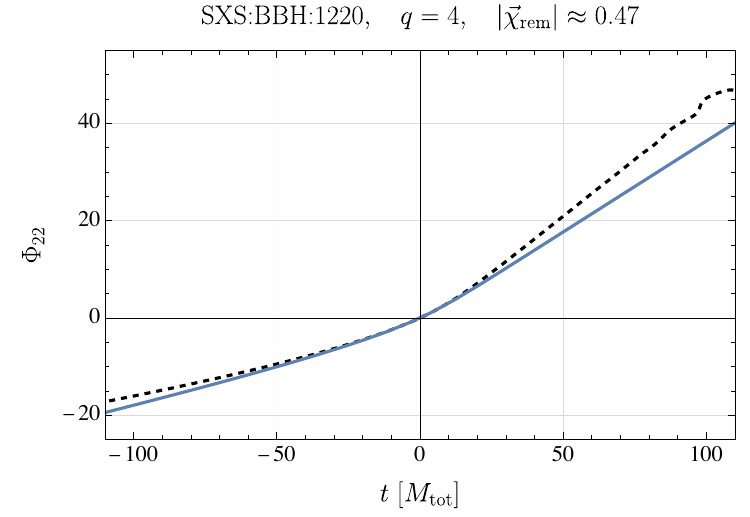}
\hspace{0.3cm}
\includegraphics[width=.48\textwidth]{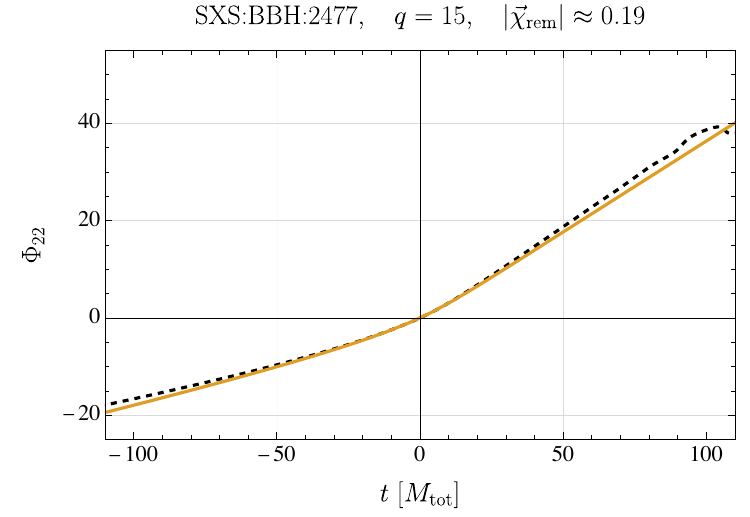}
\\[3ex]
\includegraphics[width=.48\textwidth]{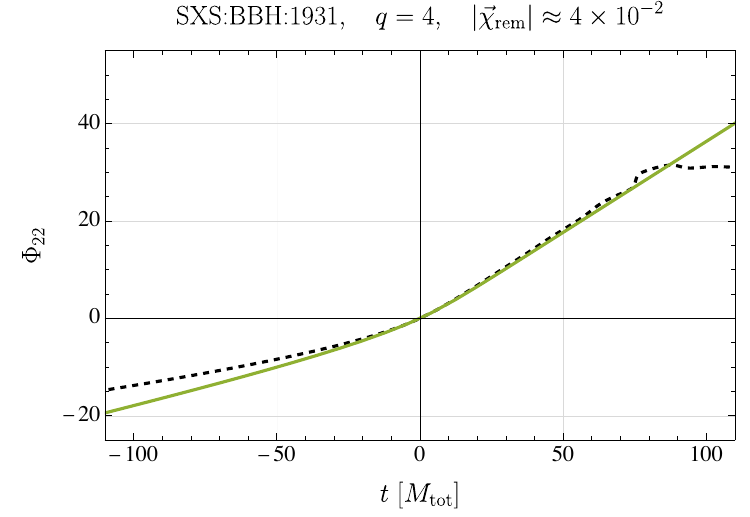}
\hspace{0.3cm}
\includegraphics[width=.48\textwidth]{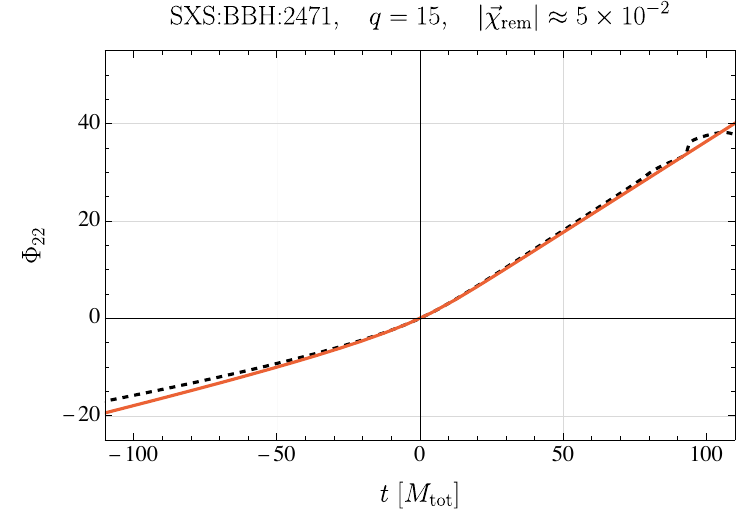}
\caption{Comparison of the $(\ell,m)=(2,2)$ waveform phase $\Phi_{\ell m}$ as a function of time between the 0PG waveforms (colored curves) and the NR simulations (dashed black curves) considered in Fig.~\ref{fig:SXS}. The color coding follows the one of Fig.~\ref{fig:SXS}. The phase difference vanishes at the alignment time $t=0$ by construction.}
\label{fig:SXSphase}
\end{figure}

\begin{figure}[tbp]
\centering
\includegraphics[width=.48\textwidth]{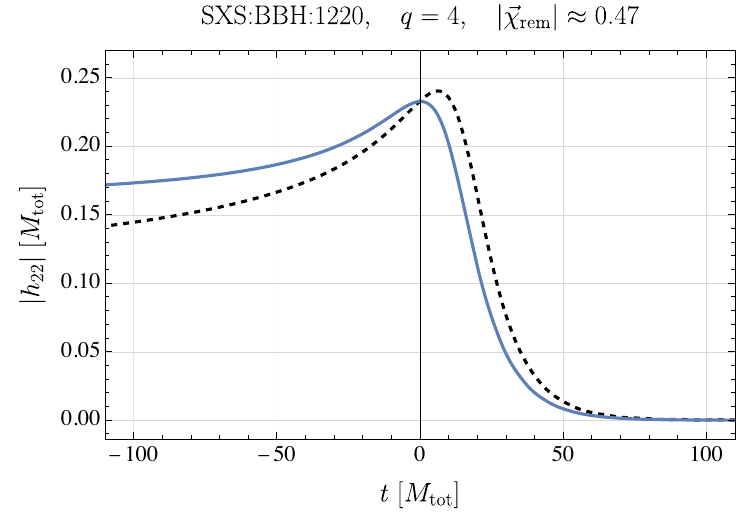}
\hspace{0.3cm}
\includegraphics[width=.48\textwidth]{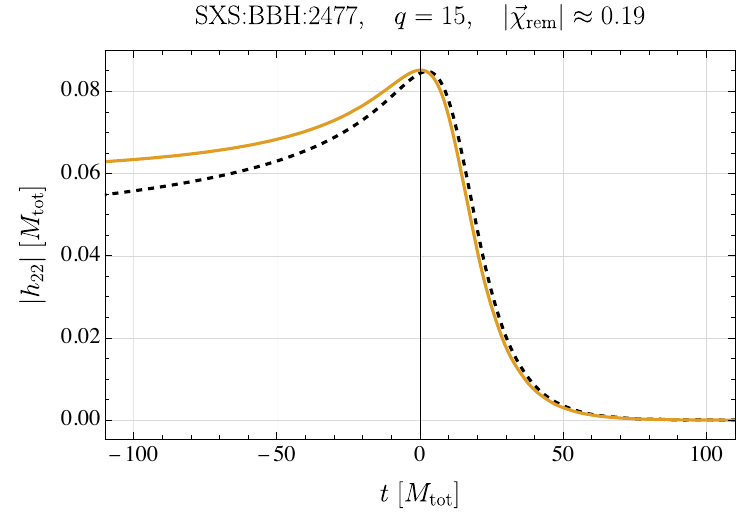}
\\[3ex]
\includegraphics[width=.48\textwidth]{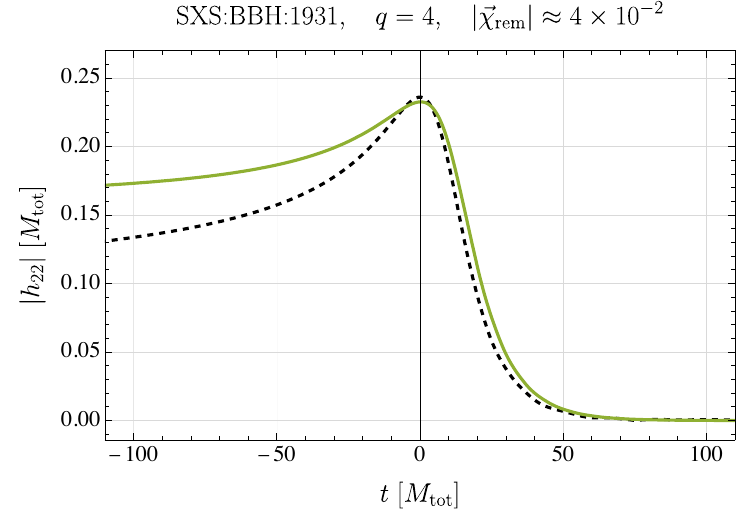}
\hspace{0.3cm}
\includegraphics[width=.48\textwidth]{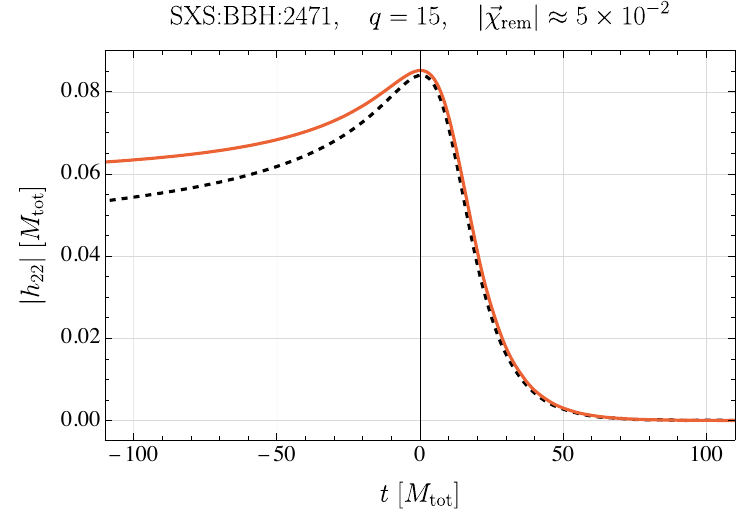}
\caption{Comparison of the $(\ell,m)=(2,2)$ waveform amplitude $|h_{\ell m}|$ as a function of time between the 0PG waveforms (colored curves) and the NR simulations (dashed black curves) considered in Fig.~\ref{fig:SXS}. The color coding follows the one of Fig.~\ref{fig:SXS}.}
\label{fig:SXSamplitude}
\end{figure}

\begin{figure}[tbp]
\centering
\includegraphics[width=.48\textwidth]{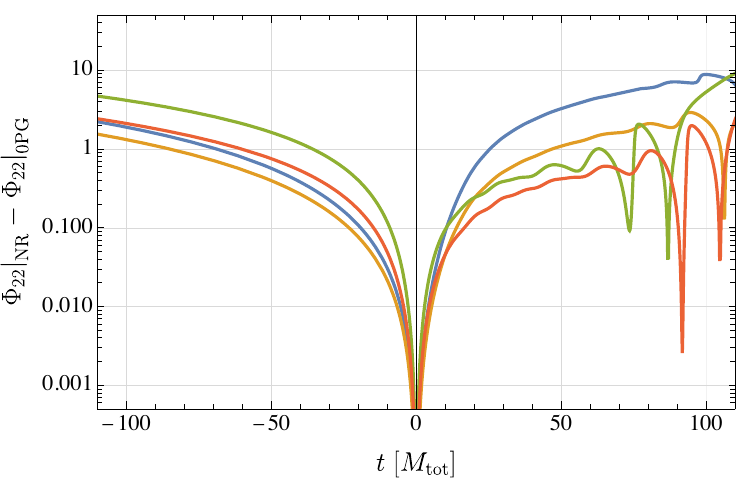}
\hspace{0.3cm}
\includegraphics[width=.48\textwidth]{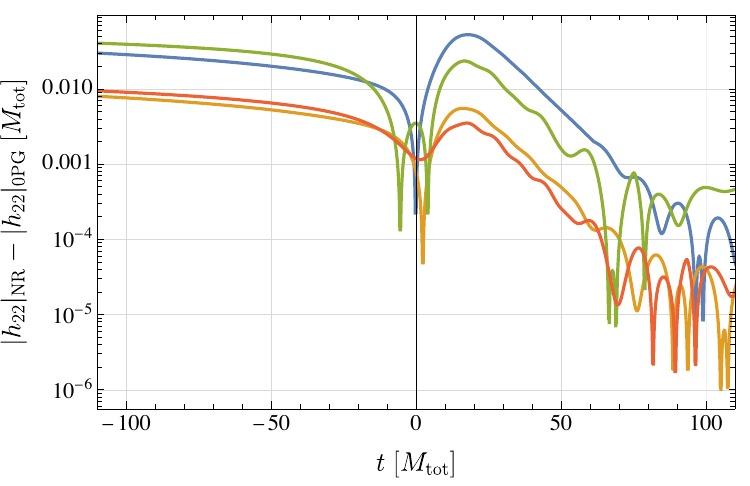}
\caption{Difference in waveform phase (left plot) and amplitude (right plot) between the $(\ell,m)=(2,2)$ modes of the 0PG waveform and the NR simulations SXS:BBH:1220 ($q=4$, $|\vec{\chi}_\text{rem}|\approx0.47$, blue), SXS:BBH:2477 ($q=15$, $|\vec{\chi}_\text{rem}|\approx0.19$, orange), SXS:BBH:1931 ($q=4$, $|\vec{\chi}_\text{rem}|\approx4\times10^{-2}$, green) and SXS:BBH:2471 ($q=15$, $|\vec{\chi}_\text{rem}|\approx5\times10^{-2}$, red). The $y$ axis is on a logarithmic scale.}
\label{fig:SXSphaseampldiff}
\end{figure}

We can illustrate our accuracy in more detail by separately comparing the 0PG waveform phases $\Phi_{\ell m}$ (Figs.~\ref{fig:SXSphase} and \ref{fig:SXSphaseampldiff}) and waveform amplitudes $|h_{\ell m}|$ (Figs.~\ref{fig:SXSamplitude} and \ref{fig:SXSphaseampldiff}) with NR. For a given mass ratio, we see that for $t\gtrsim0$ the ``zero-remnant-spin'' waveforms (SXS:BBH:1931 and SXS:BBH:2471) compare better than ``zero-initial-spin'' waveforms (SXS:BBH:1220 and SXS:BBH:2477), while the roles are reversed for $t\lesssim0$; graphically, in Fig.~\ref{fig:SXSphaseampldiff}, the green and red curves respectively lie below the blue and orange ones for $t\gtrsim0$ and vice versa for $t\lesssim0$. It is also the case that the comparison improves with increasing $q$: in Fig.~\ref{fig:SXSphaseampldiff}, the orange and red curves respectively lie below the blue and green ones for all $t$, modulo the parts of the plots dominated be the numerical noise. This also holds when considering the relative, rather than absolute, difference. In all cases except the one with large remnant spin, we observe that our 0PG model matches the peak amplitude of the SXS waveform to a high degree of accuracy.

\section{Discussion and conclusions}\label{sec:conclusions}

In this paper we have presented a framework for generating merger-ringdown waveforms within GSF theory. We have treated the final plunge, which gives rise to the merger-ringdown portion of the GW signal, using a post-geodesic expansion in a phase-space approach, mirroring our previous treatments of the inspiral and the transition to plunge~\cite{Miller:2020bft,Kuchler:2024esj}. The resulting waveform is summarized in Eqs.~\eqref{eq:ODEs plunge}--\eqref{eq:h plunge}, with $x^i=(r,\theta,\phi)$ and $r\to\infty$ in Eq.~\eqref{eq:h plunge}. All the ingredients needed for waveform generation---the waveform amplitudes $h^{\{n\},m}_{\alpha\beta}(J^a,r\to\infty)$ and the forcing functions $F^a_{\{n\}}(J^b)$---are pre-computed as functions of the binary's mechanical variables $J^a=(\Omega,\delta M^\pm)$. Evolving through the binary's mechanical phase space then yields the time-domain waveforms. This split between slow offline computations (solving field equations) and fast online ones (solving a very small number of ordinary differential equations) allows us to maintain the rapid waveform generation of inspiral and transition-to-plunge self-force models. We have outlined this framework through second order in $\varepsilon$ for the waveform, or equivalently, 1PG order in the orbital dynamics. Similarly, while we have worked with $t$ slicing in our calculations here, we have also emphasized that hyperboloidal slicing (or null slicing as in Fig.~\ref{fig:Penrose}) would offer advantages that become increasingly important at post-geodesic orders.

At early times, close to the ISCO, our plunge expansion asymptotically matches the late-time transition-to-plunge expansion of Ref.~\cite{Kuchler:2024esj}. We have detailed this asymptotic matching up to 1PG and 7PLT order and shown how it provides critical input for the plunge solution, both in the orbital dynamics and in the field equations. Without the asymptotic matching, the plunge expansion is under-determined and involves singularities at the ISCO. In the field equations, the matching conditions can be rigorously enforced through a puncture scheme, with an effective source appearing in a punctured region at early times, near the ISCO. For our implementation in this paper, we have safely neglected the effective source in the punctured region by focusing on times sufficiently far to the future of the puncture. However, calculations at 1PG order will likely require a more thorough treatment of the effective source, as the 0PG field is needed all the way back to the ISCO as input for the 1PG field equations.

Our framework involves field equations defined directly on the binary phase space, as opposed to the ordinary time domain, but we have shown how to transform these equations into a familiar frequency-domain form. Using frequency-domain methods, we have implemented our formalism at leading (0PG) order, solving  Regge-Wheeler-Zerilli equations to obtain the leading-order plunge waveform. Although this waveform has been obtained previously from the same frequency-domain equations~\cite{Folacci:2018cic}, we have shown that there is still new information to be mined from it. Specifically, we have utilized it to illuminate fundamental features of merger and ringdown, and we have begun to assess its accuracy through comparisons with NR waveforms.

First, to probe the structure of the merger and ringdown, we have focused on two questions: how well is the waveform described by an ``extended inspiral'' prior to the waveform's peak, as in EOB~\cite{Buonanno:2000ef}, and how well is it described by a QNM ringdown after the peak? Our key findings are illustrated in Figs.~\ref{fig:WFvsSPA} and~\ref{fig:wfphase_phasespace}.

In Fig.~\ref{fig:WFvsSPA} we see that the early part of the waveform, until very close to the peak, is well approximated by a stationary-phase approximation that expresses the waveform phase and amplitude as instantaneous functions of the orbital phase and radius, akin to EOB's extended-inspiral construction. In the future we will explore how much this can be improved by including higher-order terms in the SPA and whether our approach can be used to inform EOB's ``non-quasicircular corrections"~\cite{Damour:2007xr}. 

Figure~\ref{fig:WFvsSPA} also shows that at late times, $\gtrsim10M$ after the peak amplitude, the QNM ringdown becomes a good approximation to the full waveform. However, at earlier times the QNM sum diverges dramatically. This has bearing on a long line of studies. 
Many papers have investigated whether a QNM sum, including a variable number of overtones, is able to reproduce the GW strain even at the peak~\cite{Giesler:2019uxc,Bhagwat:2019dtm,Okounkova:2020vwu,Cook:2020otn,Dhani:2020nik,Mourier:2020mwa,Finch:2021iip,Forteza:2021wfq,Baibhav:2023clw,Nee:2023osy,Zhu:2023mzv,Cheung:2023vki,Qiu:2023lwo,Giesler:2024hcr}, following earlier evidence that a QNM model with up to $n=3$ overtones can fit the ringdown even at times \emph{preceding} the peak~\cite{Buonanno:2006ui}. Such investigations are mainly based on fitting NR data using QNM models built from a sum of damped sinusoids (with QNM frequencies). We have instead computed the QNM sum directly from first principles (analogous to Ref.~\cite{Oshita:2024wgt}'s study for a toy source term). Our results show that, at least in the case of a Schwarzschild primary and including the first three overtones, the QNM sum only becomes accurate significantly after the waveform's peak amplitude. Our finding of \emph{when} it becomes accurate is broadly consistent with the time at which QNM fits to NR ringdown waveforms become robust~\cite{Cheung:2023vki,Lim:2022veo,Carullo:2024smg,Mitman:2025hgy}. Figure~\ref{fig:WFvsQNMs} shows that this ringdown-onset time changes only marginally as we vary the number of overtones up to $n=3$, but we note that Ref.~\cite{Oshita:2024wgt}'s results suggest that the effective onset might appear at substantially earlier times if many more overtones were included.

A natural followup, in addition to including higher overtones, would be to further dissect the ringdown. We compute our full waveform through an inverse Fourier transform, involving an integral over all real frequencies. In Sec.~\ref{sec:QNMs} we have reviewed how this real-line integral can be equated to the sum of three contributions with complex frequencies: 
the QNM sum, which we have computed including the first 3 overtones; an integral around a branch cut, associated with power-law tails (which are clearly visible after $t_G\approx 150M$ in our full waveform in Fig.~\ref{fig:WFminusQNMs}); and an integral along a high-frequency arc, associated with a prompt response. 
We will explore the branch-cut and high-frequency arc contributions in future work, complementing recent studies such as Ref.~\cite{DeAmicis:2024not}.

At least at 0PG order, in Fig.~\ref{fig:WFvsSPA} we can identify the genuine ``merger'' regime as the brief interval when neither the SPA nor the QNM sum is accurate. However, Fig.~\ref{fig:wfphase_phasespace} shows that the waveform remains closely tied to the orbital motion in this interval, even though it occurs after the particle has fallen behind the light ring. It therefore appears plausible that a higher-order SPA can be pushed to later times, past the waveform's peak. On the other hand, the onset of the QNM regime is very sudden in Fig.~\ref{fig:wfphase_phasespace}, marking a clear dissociation of the waveform from the orbit and indicating the difficulty of pushing the QNM sum to earlier times.

We expect these types of analyses will further sharpen our understanding of the merger process, particularly when our model is extended to 1PG order, where nonlinear effects arise. Such analyses might additionally have utility in improving EOB and Phenom IMR waveform models.

Separate from the broad, fundamental questions about the merger regime, we have also qualitatively assessed the accuracy of the 0PG waveform, using NR waveforms from the SXS catalog as a benchmark. For sufficiently small mass ratios, our 0PG waveform's error is necessarily dominated by its omission of 1PG terms, corresponding to ${\cal O}(\nu^2)$ terms in the waveform amplitude and ${\cal O}(\nu)$ terms in the phase. Our comparisons with NR are consistent with that expectation, and they suggest that a complete 1PG model could be highly accurate at mass ratios $\varepsilon\approx 1/10$ (and perhaps even closer to $\varepsilon=1$). We have also singled out the most significant 1PG corrections: those associated with the nonzero spin of the final, remnant black hole. Such corrections will appear automatically in our framework at 1PG order, and they are potentially easier to calculate than the full litany of 1PG effects. Hence, as an intermediate step toward 1PG waveforms, there would be value in building an incomplete 1PG model, including only the 1PG terms associated with the remnant spin.  

Our next immediate goal, however, is to build an IMR model that enables the production of fast and accurate waveforms for asymmetric binaries. We aim to achieve this by combining the framework presented in this paper, together with our earlier work on the transition to plunge~\cite{Kuchler:2024esj}, with the post-adiabatic inspiral~\cite{Miller:2020bft,Wardell:2021fyy}. In work presented in Refs.~\cite{KuchlerCapra27,KuchlerLISA,KuchlerAEI}, we have built a fast IMR waveform model using the 1PA inspiral, the transition to plunge through 2PLT order and the geodesic plunge, with switching points between the three regimes. That model will be detailed in a followup paper. Future work will focus on improving the connection between the three regimes and carrying the transition-to-plunge and plunge expansions to higher order, with the aim of extending the sub-radian accuracy of post-adiabatic inspiral waveforms through merger and ringdown.

Further future work includes the generalization to plunges (both prograde and retrograde) into a Kerr black hole. The general framework described in this paper also applies when considering a spinning primary. For a particle plunging from the Kerr ISCO~\cite{Mummery:2022ana,Dyson:2023fws}, obtaining the first-order plunge waveform would then be achieved using the Teukolsky equation~\cite{Teukolsky:1972my} rather than the RWZ equations employed in this paper. The generalization to generic equatorial plunging orbits, which are reached when the inspiral admits eccentricity, is also under investigation.  

The offline/online split in our approach also lends itself to modelling mergers beyond general relativity or including matter effects. Waveform ingredients (beyond-GR or matter corrections to amplitudes and forcing functions) can be computed offline, in a modular way, and then added to the online waveform generation, just as has been done in the inspiral (see, e.g., Refs.~\cite{Spiers:2023cva,Speri:2024qak,Dyson:2025dlj}). We will exploit this fact to explore black hole mergers in alternative theories of gravity~\cite{AyushRoy:InPreparation} or in the presence of dark matter halos~\cite{GeoSumanta:InPreparation}.

Finally, we note that, very recently, during the preparation of our paper, Ref.~\cite{Strusberg:2025qfv} also presented a fast, leading-order IMR waveform model that includes the same 0PG waveform we have detailed here. Their IMR model involves stitching the time-domain 0PA inspiral waveform to the time-domain 0PG plunge waveform, skipping the transition regime in between. Since the transition regime has a frequency width of order $\varepsilon^{2/5}$, omitting it should have a small effect for sufficiently small mass ratios; however, we expect that this omission will incur significant errors for moderate mass ratios. Such stitching might also become impractical beyond leading order, as the post-adiabatic and post-geodesic expansions become increasingly singular at the ISCO. Since the 0PA and 0PG waveforms never have a common frequency, stitching them together in this way would additionally seem to require a discontinuity in the waveform frequency. However, the approach has a clear advantage in its simplicity, and it should suffice for many systems, such as EMRIs, where the merger and ringdown carry little SNR.

\section*{Acknowledgements}

We would like to thank Ben Leather for helpful discussions and collaboration on a closely related project, as well as for providing a Mathematica template for drawing Penrose diagrams. We are also grateful to Ayush Roy and Rodrigo Panosso Macedo for insights gleaned from another related project. A.P. thanks Jeremy Miller for valuable discussions and collaboration in the preliminary stages of this work, as well as Gregorio Carullo, Marc Casals, and Laura Sberna for conversations about QNM sums. L.K. and G.C. thank Mohamed Ould El Hadj for several interesting conversations about the implementation of Ref.~\cite{Folacci:2018cic}. Computational resources have been provided by the Consortium des Équipements de Calcul Intensif (CÉCI), funded by the Fonds de la Recherche Scientifique de Belgique (F.R.S.-FNRS) under Grant No. 2.5020.11 and by the Walloon Region. A.P. and L.K. acknowledge the support of a UKRI Frontier Research Grant (as selected by the ERC) under the Horizon Europe Guarantee scheme [grant number EP/Y008251/1]. A.P. additionally acknowledges the support of a Royal Society University Research Fellowship. G.C. is Research Director of the FNRS. He acknowledges the support of the Win4Project grant ETLOG of the Walloon Region for the Einstein Telescope. This work makes use of the Black Hole Perturbation Toolkit~\cite{BHPToolkit}.

\appendix

\section{Asymptotic match between transition to plunge and plunge}

\subsection{Coefficients of the asymptotic late-time transition-to-plunge solutions}\label{app:earlylate_transition}

Up to 7PLT order, the first coefficients in the late-time solution of the orbital radius~\eqref{g late expansion - DOmega} are given by
\begin{align}
    r_{[0]}^{(2,1)} =& -24\sqrt{6}M^2, \qquad r_{[0]}^{(2,1-5i/2)} =\, 0,\,\, \forall \, i\geq1,
    \\[1ex]
    r_{[1]}^{(3,3/2-5i/2)} =&\; 0,\,\, \forall \, i\geq0,
    \\[1ex]
    r_{[2]}^{(4,2)} =&\; 576M^3, \qquad r_{[2]}^{(4,-1/2)} = -36\,2^{1/4}3^{3/4} M^{3/2} f^t_{[5]},
    \\[1ex]
    r_{[3]}^{(5,5/2)} =&\; 0 \qquad r_{[3]}^{(5,0)} = -54M^2 f^r_{[5]}, \qquad r_{[3]}^{(5,-5i/2)} = 0,\,\, \forall \, i\geq1,
    \\[1ex]
    r_{[4]}^{(6,3)} =& -3072\sqrt{6}M^4, \qquad r_{[4]}^{(6,1/2)} = -36\,6^{1/4}M^{3/2}\left(48\sqrt{2}M f^t_{[5]} - \sqrt{3} f^t_{[7]A}\right),
    \\[1ex]
    \begin{split}
    r_{[5]}^{(7,7/2)} =&\; 0, \qquad r_{[5]}^{(7,1)} = 6M \left(f^t_{[8]A} + 90\sqrt{6}M^2 f^r_{[5]}\right),
    \end{split}
    \\[1ex]
    \begin{split}
    r_{[6]}^{(8,4)} =&\; 92160M^5,
    \\[1ex]
    r_{[6]}^{(8,3/2)} =&\; \left(\frac{2}{3}\right)^{3/4} M^{1/2} f^{t}_{[9]B}\\
    &+ 12\,6^{1/4}M^{3/2}\left[\sqrt{3} f^{t}_{[9]A} + 24M\left(65\sqrt{3}M f^t_{[5]} - 3\sqrt{2} f^t_{[7]A}\right)\right],
    \end{split}
    \\[1ex]
    \begin{split}
    r_{[7]}^{(9,9/2)} =&\; 0
    \\[1ex]
    r_{[7]}^{(9,2)} =&\; \frac{1}{72} \left[\sqrt{6} f^{t}_{[10]B} + 3 \sqrt{6} f^{t}_{[10]C} + 216 M f^{t}_{[10]D}\right.\\
    &\left.- 432 M^2 \left(11 \sqrt{6} f^t_{[8]A} - 3 \sqrt{6} f^{t}_{[10]E} - 36 \sqrt{6}M f^r_{[7]A} + 4752M^2 f^r_{[5]}\right)\right].
    \end{split}
\end{align}
The first coefficients in the late-time solution of $d\Omega/dt$~\eqref{g late expansion - DOmega} up to 7PLT order read
\begin{align}\label{F_[0]^(i,j/2)}
    F_{[0]}^{(3,3/2)} =& \frac{1}{2^{1/4}3^{7/4}M^{1/2}} , \quad F_{[0]}^{(3,-1)} = \frac{f^t_{[5]}}{4M^2}, 
    \\[1ex]
    F_{[1]}^{(4,2-5i/2)} =&\; 0, \,\, \forall \, i\geq0,
    \\[1ex]
    F_{[2]}^{(5,5/2)} =&\; 2\left(\frac{2}{3}\right)^{1/4} M^{1/2}, \qquad F_{[2]}^{(5,0)} = \frac{1}{12M^2}\left(36\sqrt{6}Mf^t_{[5]} - f^t_{[7]A}\right),
    \\[1ex]
    \begin{split}
    F_{[3]}^{(6,3)} =&\; 0,
    \\[1ex]
    F_{[3]}^{(6,1/2)} =&\; \frac{1}{96\,6^{3/4}M^{5/2}}\left[9M\left(8\sqrt{3}Mf^r_{[5]} - \sqrt{2}f^r_{[7]A}\right) - \sqrt{2}f^t_{[8]A}\right],
    \end{split}
    \\[1ex]
    \begin{split}\label{F_[4]^(i,j/2)}
    F_{[4]}^{(7,7/2)} =& -8\left(\frac{2}{3}\right)^{3/4}M^{3/2},
    \\[1ex]
    F_{[4]}^{(7,1)} =&\; \frac{1}{6480M^3}\left[108M\left(1032M^2f^t_{[5]} - 19\sqrt{6}M f^t_{[7]A} - f^t_{[9]A}\right)\right.\\ 
    &\left.- 9\sqrt{6}M f^r_{[8]A} - \sqrt{6}f^t_{[9]B}\right],
    \end{split}
    \\[1ex]
    \begin{split}
    F_{[5]}^{(8,4)} =&\; 0,
    \\[1ex]
    F_{[5]}^{(8,3/2)} =& -\frac{1}{31104\,6^{3/4} M^{7/2}}\left[\sqrt{3} f^t_{[10]B} + 3 \sqrt{3} f^t_{[10]C}\right.\\
    &+ 36 M \left(\sqrt{3} f^r_{[9]B} + 3 \sqrt{2} f^t_{[10]D} + 6 M \left(9 \sqrt{2} f^r_{[9]A} + 4 \sqrt{3} f^t_{[8]A}\right.\right.\\
    &\left.\left.\left.+ 6 \sqrt{3} f^t_{[10]E} + 72 \sqrt{3}M f^r_{[7]A} - 1512 \sqrt{2} M^2 f^r_{[5]}\right)\right)\right],
    \end{split}
    \\[1ex]
    \begin{split}
    F_{[6]}^{(9,9/2)} =& - 32\,2^{1/4} 3^{3/4} M^{5/2},
    \\[1ex]
    F_{[6]}^{(9,2)} =& -\frac{1}{181440 M^3} \left[5 f^r_{[10]B} + 15 f^r_{[10]C} + 8 \sqrt{6} f^t_{[11]D} + 12 \sqrt{6} f^t_{[11]E}\right.\\
    &+ 36 M \left(5 \sqrt{6} f^r_{[10]D} - 4 \left(f^t_{[9]B} - 3 f^t_{[11]B} - 5 f^t_{[11]C} - 9 f^t_{[11]F} - 21 Mf^r_{[8]A}\right.\right.\\
    &\left.\left.\left. - 45 M f^r_{[10]E} + 18 \sqrt{6} M f^t_{[9]A} + 9324 M^2 f^t_{[7]A} - 379008 \sqrt{6} M^3 f^t_{[5]}\right)\right)\right],
    \end{split}
    \\[1ex]
    \begin{split}
    F_{[7]}^{(10,5)} =&\; 0,
    \\[1ex]
    F_{[7]}^{(10,5/2)} =& -\frac{1}{62208\,6^{3/4} M^{7/2}}\left[2 \sqrt{3} f^t_{[12]D} + 3 \sqrt{3} f^t_{[12]L} + \sqrt{3} f^t_{[12]M} + 36 \sqrt{3}M f^r_{[11]D}\right.\\
    &+ 54 \sqrt{3}M f^r_{[11]E} - 18 \sqrt{2}M f^t_{[10]B} - 54 \sqrt{2}M f^t_{[10]C} + 54 \sqrt{2}M f^t_{[12]E}\\
    & + 18 \sqrt{2}M f^t_{[12]F} + 90 \sqrt{2}M f^t_{[12]G} + 90 \sqrt{2}M f^t_{[12]H} + 108 \sqrt{2}M f^t_{[12]J}\\
    &+ 324 \sqrt{2}M^2 f^r_{[9]B} + 972 \sqrt{2}M^2 f^r_{[11]B} + 1620 \sqrt{2}M^2 f^r_{[11]C}\\
    &+ 2916 \sqrt{2}M^2 f^r_{[11]F} - 1296 \sqrt{3}M^2 f^t_{[10]D} + 1296 \sqrt{3}M^2 f^t_{[12]K}\\
    &+ 11664 \sqrt{3}M^3 f^r_{[9]A} - 44064 \sqrt{2}M^3 f^t_{[8]A} - 23328 \sqrt{2}M^3 f^t_{[10]E}\\
    &\left.- 5184 \sqrt{2}M^3 f^t_{[12]A} - 1003104 \sqrt{2}M^4 f^r_{[7]A} + 8491392 \sqrt{3}M^5 f^r_{[5]}\right].
    \end{split}
\end{align}

\subsection{Coefficients of the asymptotic near-ISCO plunge solutions}\label{app:nearISCO_plunge}

Up to 1PG order, the first coefficients in the near-ISCO solutions of the orbital radius (Eqs.~\eqref{g near ISCO} and \eqref{r1nearISCO}) are given by
\begin{align}
    \begin{split}
    r_{\{0\}}^{(2,1)} =& -24\sqrt{6}M^2, \qquad r_{\{0\}}^{(2i+1,i+1/2)} = 0,\,\, \forall \, i\geq1
    \\[1ex]
    r_{\{0\}}^{(4,2)} =&\; 576 M^3, \qquad r_{\{0\}}^{(6,3)} = -3072 \sqrt{6} M^4, \qquad r_{\{0\}}^{(8,4)} = 92160 M^5,
    \end{split}
    \\[1ex]
    \begin{split}
    r_{\{1\}}^{(2,-3/2)} =& r_{\{1\}}^{(3,-1)} = 0,
    \\[1ex]
    r_{\{1\}}^{(4,-1/2)} =& - 36\,2^{1/4}3^{3/4}M^{3/2}f^t_{\{1\},0},
    \\[1ex]
    r_{\{1\}}^{(5,0)} =& -54 M^2 f^r_{\{1\},0},
    \\[1ex]
    r_{\{1\}}^{(6,1/2)} =&\; 36\,2^{1/4} 3^{3/4} M^{3/2} f^t_{\{1\},2} - 1728\,2^{3/4} 3^{1/4} M^{5/2} f^t_{\{1\},0},
    \\[1ex]
    r_{\{1\}}^{(7,1)} =&\; 540 \sqrt{6} M^3 f^r_{\{1\},0} + 18\,2^{1/4} 3^{3/4} M^{3/2} f^t_{\{1\},3},
    \\[1ex]
    \begin{split}r_{\{1\}}^{(8,3/2)} =&\; 18720\,2^{1/4} 3^{3/4} M^{7/2} f^t_{\{1\},0} -  864\,2^{3/4} 3^{1/4} M^{5/2}f^t_{\{1\},2}\\
    &+ 12\,2^{1/4} 3^{3/4} M^{3/2}f^t_{\{1\},4},\end{split}
    \\[1ex]
    \begin{split}r_{\{1\}}^{(9,2)} =&\; 216 \sqrt{6}M^3 f^r_{\{1\},2} - 28512 M^4 f^r_{\{1\},0} - 594\,2^{3/4} 3^{1/4} M^{5/2} f^t_{\{1\},3}\\
    &+ 9\,2^{1/4} 3^{3/4} M^{3/2} f^t_{\{1\},5}.\end{split}
    \end{split}
\end{align}
The first coefficients in the near-ISCO solutions of the forcing terms~\eqref{g near ISCO} and \eqref{F1nearISCO} read%
\begingroup\allowdisplaybreaks%
\begin{align}
    \begin{split}
    F_{\{0\}}^{(3,3/2)} =&\; \frac{1}{2^{1/4} 3^{7/4}M^{1/2}}, \qquad F_{\{0\}}^{(2i,i)} = 0,\,\, \forall \, i\geq2,
    \\[1ex]
    F_{\{0\}}^{(5,5/2)} =& 2 \left(\frac{2}{3}\right)^{1/4}M^{1/2}, \qquad F_{\{0\}}^{(7,7/2)} = -8 \left(\frac{2}{3}\right)^{3/4} M^{3/2},
    \\[1ex]
    F_{\{0\}}^{(9,9/2)} =& - 32\,2^{1/4} 3^{3/4} M^{5/2},
    \end{split}
    \\[1ex]
    \begin{split}
    F_{\{1\}}^{(3,-1)} =&\; \frac{f^t_{\{1\},0}}{4 M^2},
    \\[1ex]
    F_{\{1\}}^{(4,-1/2)} =&\; 0,
    \\[1ex]
    F_{\{1\}}^{(5,0)} =&\; \frac{1}{12 M^2}\left(36 \sqrt{6} M f^t_{\{1\},0} - f^t_{\{1\},2}\right),
    \\[1ex]
    F_{\{1\}}^{(6,1/2)} =&\; \frac{1}{64 M^2}\left[6^{1/4} M^{1/2} \left(8\sqrt{3} M f^r_{\{1\},0} - \sqrt{2} f^r_{\{1\},2}\right) - 2 f^t_{\{1\},3}\right],
    \\[1ex]
    \begin{split}F_{\{1\}}^{(7,1)} =&\, -\frac{1}{240 M^2}\left(4 f^t_{\{1\},4} + 3\,2^{3/4} 3^{1/4} M^{1/2} f^r_{\{1\},3}\right.\\
    &\left. + 76 \sqrt{6} M f^t_{\{1\},2} - 4128 M^2 f^t_{\{1\},0}\right),\end{split}
    \\[1ex]
    \begin{split}F_{\{1\}}^{(8,3/2)} =&\; \frac{1}{96 M^2}\left[6^{1/4} M^{1/2} \left(168 \sqrt{2} M^2 f^r_{\{1\},0} - 8 \sqrt{3} M f^r_{\{1\},2}\right.\right.\\
    &\left.\left.- 4\,6^{1/4} M^{1/2} f^t_{\{1\},3} - \sqrt{2} f^r_{\{1\},4}\right) - f^t_{\{1\},5}\right],\end{split}
    \\[1ex]
    \begin{split}F_{\{1\}}^{(9,2)} =&\, -\frac{1}{560 M^2}\left[4 f^t_{\{1\},6} + 5\,2^{3/4} 3^{1/4} M^{1/2} f^r_{\{1\},5} + 4 M \left(7\,2^{1/4} 3^{3/4} M^{1/2} f^r_{\{1\},3}\right.\right.\\
    &\left.\left. -2 \sqrt{6} f^t_{\{1\},4} + 28 M \left(1504 \sqrt{6} M f^t_{\{1\},0} - 37 f^t_{\{1\},2}\right)\right)\right],\end{split}
    \\[1ex]
    F_{\{1\}}^{(10,5/2)} =&\, -\frac{1}{384 M^2}\left[2 f^t_{\{1\},7} + 3 M^{1/2} \left(2^{3/4} 3^{1/4} f^r_{\{1\},6} - 4 \left(\sqrt{6} M^{1/2} f^t_{\{1\},5}\right.\right.\right.\\
    &- 2^{1/4} 3^{3/4} M f^r_{\{1\},4} + 68 M^{3/2} f^t_{\{1\},3} + 86\,2^{3/4} 3^{1/4} M^2 f^r_{\{1\},2}\\
    &\left.\left.\left.- 728\,2^{1/4} 3^{3/4} M^3 f^r_{\{1\},0}\right)\right)\right].
    \end{split}
\end{align}
\endgroup

\subsection{Self-force matching conditions}\label{app:selfforce_matchingcond}

In the transition-to-plunge regime, the self-force admits an expansion in integer powers of $\lambda$ at fixed phase-space coordinates $\Delta\Omega$ and $\delta M^\pm$, $f^\mu=\sum_{n=5}^\infty \lambda^n f^\mu_{[n]}(\Delta\Omega, \delta M^\pm)$. The coefficients of $\lambda^n$ in this expansion can be written as a sum of terms factored into a $\Delta\Omega$-dependent piece and a $\Delta\Omega$-independent piece (labeled with capital Latin letters); see Ref.~\cite{Kuchler:2024esj}. Using Eq.~\eqref{g late expansion - DOmega} to expand the $\Delta\Omega$-dependent pieces, we find that the late-time behaviour of the transition-to-plunge self-force reads
\begin{equation}\label{sf_tr_latefull}
\begin{split}
    f^\mu(\lambda, \Delta\Omega\to+\infty, \delta M^\pm) =&\, \lambda^5 f^\mu_{[5]} + \lambda^7 \Delta\Omega f^\mu_{[7]A} + \lambda^8\left[F^{(3,3/2)}_{[0]}\Delta\Omega^{3/2} + {\cal O}(\Delta\Omega^{-1})\right] f^\mu_{[8]A}
    \\[1ex]
    &+ \lambda^9\left[\Delta\Omega^2 f^\mu_{[9]A} + \frac{3}{2}\left(F^{(3,3/2)}_{[0]}\right)^2\Delta\Omega^2 f^\mu_{[9]B} + {\cal O}\left(\Delta\Omega^{-1/2}\right)\right]
    \\[1ex]
    &+ {\cal O}_{\Delta\Omega}(\lambda^{10}).
\end{split}
\end{equation}
Here we use the symbol ${\cal O}_{\Delta\Omega}$ to indicate subleading terms in the transition-to-plunge expansion and the symbol ${\cal O}$ to indicate subleading terms in the late-time expansion of a given PLT order. 

The near-ISCO solution of the self-force in the plunge regime follows from Eq.~\eqref{plungesf_early_Omega}:
\begin{equation}\label{sf_plunge_nearISCOfull}
\begin{split}
    f^\mu(\varepsilon, \Omega\to\Omega_*, \delta M^\pm) =&\, \lambda^5 f^{\mu}_{\{1\},0} + \lambda^7 \Delta\Omega f^{\mu}_{\{1\},2} + \lambda^8 \Delta\Omega^{3/2} f^{\mu}_{\{1\},3}
    \\[1ex]
    &+ \lambda^9 \Delta\Omega^2 f^{\mu}_{\{1\},4} + {\cal O}(\lambda^{10}\Delta\Omega^{5/2}) + {\cal O}_\Omega(\varepsilon^2).
\end{split}
\end{equation}
Here we use the symbol ${\cal O}_\Omega$ to indicate subleading terms in the plunge expansion and the symbol ${\cal O}$ to indicate subleading terms in the near-ISCO expansion of a given post-geodesic order. 

Equating the coefficients of equal powers of $\lambda$ and $\Delta\Omega$ in Eqs.~\eqref{sf_tr_latefull} and \eqref{sf_plunge_nearISCOfull}, we obtain the following matching conditions for the self-force:
\begin{subequations}\label{matchingsf}
\begin{align}
    f^\mu_{\{1\},0} &= f^\mu_{[5]},
    \\[1ex]
    f^\mu_{\{1\},2} &= f^\mu_{[7]A},
    \\[1ex]
     f^\mu_{\{1\},3} &= F^{(3,3/2)}_{[0]} f^\mu_{[8]A},
    \\[1ex]
    f^\mu_{\{1\},4} &= f^\mu_{[9]A} + \frac{3}{2}\left(F^{(3,3/2)}_{[0]}\right)^2 f^\mu_{[9]B}.
\end{align}
\end{subequations}
The subleading terms in the ${\Delta\Omega\to+\infty}$ expansions at each order in $\lambda$ in Eq.~\eqref{sf_tr_latefull}  will match with terms that originate from subleading orders in the post-geodesic expansion in Eq.~\eqref{sf_plunge_nearISCOfull}. The coefficients in Eq.~\eqref{matchingsf} can be further connected to inspiral quantities through the asymptotic match between the inspiral and transition-to-plunge regimes~\cite{Kuchler:2024esj}. In particular, the terms $f^\mu_{\{1\},0}$ and $f^\mu_{\{1\},2}$ are given by the inspiral's first-order self-force and its $\Omega$ derivative evaluated at the ISCO, respectively.

The asymptotic match presented in Table~\ref{tab:matchstructureTP} between transition to plunge (up to 7PLT order) and plunge (up to 1PG order) requires the self-force matching conditions in Eqs.~\eqref{matchingsf} in addition to the higher-order ones listed below:
\begin{align}
    f^\mu_{\{1\},5} =&\, \frac{1}{324\,2^{3/4}3^{1/4}M^{3/2}}\left[f^\mu_{[10]B} + 3f^\mu_{[10]C} + 36M\left(\sqrt{6}f^\mu_{[10]D} + 36Mf^\mu_{[10]E}\right)\right],
    \\[1ex]
    f^\mu_{\{1\},6} =&\, \frac{f^\mu_{[11]B}}{3} + \frac{5}{9}f^\mu_{[11]C} + \frac{f^\mu_{[11]D}}{27\sqrt{6}M} + \frac{f^\mu_{[11]E}}{18\sqrt{6}M} + f^\mu_{[11]F},
    \\[1ex]
    \begin{split}
    f^\mu_{\{1\},7} =& -\frac{1}{648\,2^{1/2}3^{3/4}M^{3/2}}\left[10368 M^3 f^\mu_{[12]A} - 2\sqrt{6}f^\mu_{[12]D} - 108 M f^\mu_{[12]E}\right.\\
    & - 36M f^\mu_{[12]F} - 180 M f^\mu_{[12]G} - 180 M f^\mu_{[12]H} - 216M f^\mu_{[12]J}\\
    &\left. - 1296\sqrt{6}M^2 f^\mu_{[12]K} - 3\sqrt{6}f^\mu_{[12]L} - \sqrt{6}f^\mu_{[12]M}\right].
    \end{split}
\end{align}

\section{Transformation between the fixed-\texorpdfstring{$\Omega$}{} and fixed-\texorpdfstring{$r_p$}{} expansions}\label{app:equivalence}

In this appendix we derive the transformation between the fixed-$\Omega$ formulation presented in Sec.~\ref{sec:expfixedOmega} and the fixed-$r_p$ formulation presented in Sec.~\ref{sec:expfixedrp}. As stressed in the body of the paper, the two formulations are related by a transformation on phase space, which is an inherent gauge freedom in the phase-space formalism.

We first recall the expansions of $r_p$ (at fixed $\Omega$) and $\Omega$ (at fixed $r_p$):
\begin{subequations}\label{dict}
\begin{align}\label{dict1}
    r_p &= r_{\{0\}}(\Omega)+ \sum_{n=1}^\infty \varepsilon^n r_{\{ n\}}(\Omega, \delta M^\pm),
    \\[1ex]\label{dict2}
    \Omega &= \Omega_{\{0\}}(r_p) + \sum_{n=1}^\infty  \varepsilon^n \Omega_{\{ n\}}(r_p, \delta M^\pm).
\end{align}    
\end{subequations}
Substituting Eq.~\eqref{dict1} into Eq.~\eqref{dict2} we derive the dictionary between the two expansions. 

At zeroth order in $\varepsilon$, we obtain
\begin{equation}\label{equivalence0th}
    \Omega=\Omega_{\{0\}}(r_{\{0\}}(\Omega)),  
\end{equation}
which is automatically obeyed since $\Omega_{\{0\}}(\cdot)$ and $r_{\{0\}}(\cdot)$ are inverse functions of one another, after choosing the correct branches for $r_{\{0\}}$; see Sec.~\ref{sec:expfixedOmega}. Taking a first and a second $\Omega$ derivative of Eq.~\eqref{equivalence0th}, we obtain 
\begin{subequations}\label{id0}
\begin{align}
    1 &= r_{\{0\}}'(\Omega)\Omega_{\{0\}}'(r_{\{0\}}(\Omega)), 
    \\[1ex]
    0 &= r_{\{0\}}''(\Omega)\Omega_{\{0\}}'(r_{\{0\}}(\Omega))+(r_{\{0\}}'(\Omega))^2\Omega_{\{0\}}''(r_{\{0\}}(\Omega)), 
\end{align} 
\end{subequations}
where a prime denotes differentiation with respect to the argument. 

At order $\varepsilon$, after using Eq.~\eqref{id0}, we find from Eq.~\eqref{dict} that
\begin{align}\label{id1}
   r_{\{1\}}(\Omega,\delta M^\pm)&=-r_{\{0\}}'(\Omega)\Omega_{\{1\}}(r_{\{0\}}(\Omega),\delta M^\pm). 
\end{align}
Taking a first and a second $\Omega$ derivative of this equation, we furthermore obtain 
\begin{subequations}\label{id2}
\begin{align}\label{eq:98}
    r_{\{1 \}}'(\Omega,\delta M^\pm) =& -(r_{\{0\}}'(\Omega))^2\Omega_{\{1\}}'(r_{\{0\}}(\Omega),\delta M^\pm) - r_{\{0\}}''(\Omega_{\{0\}})\Omega_{\{1\}}(r_{\{0\}}(\Omega),\delta M^\pm),
    \\[1ex] 
    \begin{split}r_{\{1 \}}''(\Omega,\delta M^\pm) =&\, -(r_{\{0\}}'(\Omega))^3\Omega_{\{1\}}''(r_{\{0\}}(\Omega),\delta M^\pm)\\
    &-3 r_{\{0\}}'(\Omega)r_{\{0\}}''(\Omega)\Omega_{\{1\}}'(r_{\{0\}}(\Omega),\delta M^\pm)\\ 
    &- r_{\{0\}}'''(\Omega_{\{0\}})\Omega_{\{1\}}(r_{\{0\}}(\Omega),\delta M^\pm),\end{split}
\end{align}
\end{subequations}
where a prime denotes differentiation with respect to the first argument. 

The self-forces in either formulation are related as
\begin{equation}\label{feq}
    \sum_{n=1}^\infty \varepsilon^n \left.f^\mu_{\{ n\}}\right\vert_\text{fixed $\Omega$}(\Omega, \delta M^\pm) = \sum_{n=1}^\infty \varepsilon^n \left.f^\mu_{\{ n\}}\right\vert_\text{fixed $r_p$}(r_p, \delta M^\pm).
\end{equation}
The relationship between the two self-forces can be obtained by plugging Eq.~\eqref{dict} into Eq.~\eqref{feq}. We can express the resulting equations in terms of either $r_p$ or $\Omega$. We choose to express all matching equations in terms of the orbital radius $r_p$. At leading order we simply have 
\begin{equation}\label{id3}
   \left.f^\mu_{\{1\}}\right\vert_\text{fixed $r_p$}(r_p, \delta M^\pm) = \left.f^\mu_{\{1\}}\right\vert_\text{fixed $\Omega$}(\Omega_{\{0\}}(r_p), \delta M^\pm).
\end{equation}

In order to obtain the dictionary between the evolution equations, we start by considering the rate of change of the orbital frequency. In the fixed-$r_p$ formulation, from Eqs.~\eqref{dict2} and \eqref{dOmegafixedrp} we have
\begin{equation}\label{dOmegadtfixedrp_comp}
    \frac{d\Omega}{dt} = \frac{d\Omega}{dr_p}\frac{dr_p}{dt} = \left[\Omega_{\{0\}}'(r_p) + \sum_{n=1}^\infty \varepsilon^n \Omega_{\{ n\}}'(r_p, \delta M^\pm)\right] \left[F_{\{0\}}^{r_p}(r_p) + \sum_{n=1}^\infty \varepsilon^n F^{r_p}_{\{ n\}}(r_p, \delta M^\pm)\right].
\end{equation}
We then use Eq.~\eqref{dict2} to re-expand the evolution equation~\eqref{dOmegafixedOmega} in the fixed-$\Omega$ formulation,
\begin{equation}\label{dOmegadtfixedOmega_comp}
    \frac{d\Omega}{dt} = F_{\{0\}}^\Omega(\Omega_{\{0\}}(r_p)) + \varepsilon \left[F^\Omega_{\{1 \}}(\Omega_{\{0\}}(r_p),\delta M^\pm) + \partial_\Omega F_{\{0\}}^\Omega (\Omega_{\{0\}}(r_p)) \Omega_{\{ 1\}}(r_p,\delta M^\pm)\right] + {\cal O}(\varepsilon^2).
\end{equation}
Comparing Eqs.~\eqref{dOmegadtfixedrp_comp} and \eqref{dOmegadtfixedOmega_comp} order by order in $\varepsilon$, we obtain, up to 1PG order,
\begin{subequations}\label{id4}
\begin{align}\label{eq3a}
   F_{\{0\}}^{r_p}(r_p) =&\, \frac{F^\Omega_{\{0\}}(\Omega_{\{0\}}(r_p))}{\Omega_{\{0\}}'(r_p)},
   \\[1ex]\label{eq3b}
   \begin{split}F^{r_p}_{\{1\}}(r_p,\delta M^\pm) =&\, \frac{1}{\Omega_{\{0\}}'(r_p)}\left[F^\Omega_{\{1\}}(\Omega_{\{0\}}(r_p),\delta M^\pm) + \partial_\Omega F_{\{0\}}^\Omega (\Omega_{\{0\}}(r_p)) \Omega_{\{1\}}(r_p,\delta M^\pm)\right.\\
   &\left. - F_{\{0\}}^{r_p}(r_p) \Omega_{\{1\}}'(r_p,\delta M^\pm) \right].\end{split}
\end{align}
\end{subequations}

At geodesic order, the two formulations are mathematically equivalent, as can now be easily verified from Eq.~\eqref{eq3a} using Eqs.~\eqref{FOmegaG}, \eqref{OmegarG} and \eqref{rGdot}. At 1PG order, the 1PG set of differential equations~\eqref{1PGeqsfixedOmega} and \eqref{1PGeqsfixedrp} can also be recovered one from the other; this is easily verified using the relations~\eqref{id0}, \eqref{id1}, \eqref{id2}, \eqref{id3} and \eqref{id4}.

\section{Vector and tensor spherical harmonics}\label{app:vecten_sphericalharm}

We list the components of the vector and tensor spherical harmonics as defined in Appendix~A of Ref.~\cite{Martel:2005ir}. The even- and odd-parity vector spherical harmonics read
\begin{gather}
    Y^{\ell m}_\theta \coloneqq \partial_\theta Y^{\ell m}, \qquad Y^{\ell m}_\phi \coloneqq \partial_\phi Y^{\ell m},
    \\[1ex]
    X^{\ell m}_\theta \coloneqq -\frac{1}{\sin\theta}\partial_\phi Y^{\ell m}, \qquad X^{\ell m}_\phi \coloneqq \sin\theta \, \partial_\theta Y^{\ell m},
\end{gather}
where $Y^{\ell m}=Y^{\ell m}(\theta, \phi)$ are the standard scalar spherical harmonics. The even- and odd-parity tensor spherical harmonics are given by
\begin{align}
    Y^{\ell m}_{\theta\theta} &\coloneqq \left[\partial_\theta^2 + \frac{1}{2}\ell(\ell+1)\right]Y^{\ell m},
    \\[1ex]
    Y^{\ell m}_{\theta\phi} &\coloneqq \left[\partial_\theta \partial_\phi - \frac{\cos\theta}{\sin\theta} \partial_\phi\right]Y^{\ell m},
    \\[1ex]
    Y^{\ell m}_{\phi\phi} &\coloneqq \left[\partial_\phi^2  + \sin\theta \cos\theta \, \partial_\theta + \frac{1}{2}\ell(\ell+1)\sin^2\theta\right]Y^{\ell m},
    \\[1ex]
    X^{\ell m}_{\theta\theta} &\coloneqq -\frac{1}{\sin\theta}\left[\partial_\theta \partial_\phi - \frac{\cos\theta}{\sin\theta}\partial_\phi\right]Y^{\ell m},
    \\[1ex]
    X^{\ell m}_{\theta\phi} &\coloneqq \frac{1}{2}\left[\sin\theta\, \partial_\theta^2 - \frac{1}{\sin\theta}\partial_\phi^2 - \cos\theta \, \partial_\theta\right]Y^{\ell m},
    \\[1ex]
    X^{\ell m}_{\phi\phi} &\coloneqq \left[\sin\theta \, \partial_\theta \partial_\phi - \cos\theta \, \partial_\phi\right]Y^{\ell m}.
\end{align}

\section{First-order Regge-Wheeler-Zerilli equations in generic slicing}\label{sec:RWZ s slicing}

In this appendix we extend the main steps in Secs.~\ref{sec:RWZt}--\ref{sec:inhomogsol} to generic hyperboloidal slicing $s$. Here we require $s$ to approach retarded time $u$ at future null infinity and advanced time $v$ at the horizon.

We highlight how this slicing leads to well-behaved solutions in our puncture scheme, and we explain how those solutions can be used to obtain the correct (though singular) solutions in $t$ slicing.

\subsection{Punctured RWZ equations}

We start from the RWZ equations on phase space for a generic choice of time slicing $s$, as given in Eq.~\eqref{RWZs}. Following the same procedure as in the main text we can split the field $R^\text{e/o}_{\ell m}$ into a puncture $R^{\text{e/o}\,\mathcal{P}}_{\ell m}$ and a residual field $R^{\text{e/o}\,\mathcal{R}}_{\ell m}$ as
\begin{equation}
    R^\text{e/o}_{\ell m}(r_p,r) = R^{\text{e/o}\,\mathcal{R}}_{\ell m}(r_p,r) + R^{\text{e/o}\,\mathcal{P}}_{\ell m}(r_p,r),
\end{equation}
where
\begin{equation}
    R^{\text{e/o}\,\mathcal{P}}_{\ell m}(r_p,r) \coloneqq R^{\text{e/o}\,(1)}_{\ell m}(6M,r)\theta(r_p - r_\mathcal{P}).
\end{equation}
The inspiral field $R^{\text{e/o}\,(1)}_{\ell m}(6M,r)$, which hereafter we also denote with the short notation $R^{\text{e/o}\,(1)}_{\ell m}\vert_*$, satisfies Eq.~\eqref{RWZs} evaluated at $r_p=6M$:
\begin{equation}
\begin{split}
    &\left(\partial_x^2 -  V^{\text{e/o}}_\ell(r)\right) \left.R^{\text{e/o}\,(1)}_{\ell m}\right\vert_* + im\left.\Omega_{\{0\}}\right\vert_*\left.\frac{dt_p}{ds}\right\vert_*\left(2H\partial_x + \frac{dH}{dx}\right) \left.R^{\text{e/o}\,(1)}_{\ell m}\right\vert_*
    \\[1ex]
    &+\left(1 - H^2\right)\left[im\left.\Omega_{\{0\}}\right\vert_*\left.\frac{d^2t_p}{ds^2}\right\vert_* + m^2\left.\Omega_{\{0\}}^2\right\vert_*\left(\left.\frac{dt_p}{ds}\right\vert_*\right)^2 \right] \left.R^{\text{e/o}\,(1)}_{\ell m}\right\vert_* = S^\text{e/o}_{\ell m}(6M, r) \coloneqq \left.S^\text{e/o}_{\ell m}\right\vert_*.
\end{split}
\end{equation}
The residual field then solves an equation structurally analogous to Eq.~\eqref{RWZs},
\begin{equation}\label{RWZs app}
\begin{split}
    &\left(\partial_x^2 -  V_\ell^\text{e/o}\right) R^{\text{e/o}\,\mathcal{R}}_{\ell m} - 2H\frac{dt_p}{ds}\left(\dot r_{\{0\}} \partial_x\partial_{r_p} - im\Omega_{\{0\}}\partial_x\right) R^{\text{e/o}\,\mathcal{R}}_{\ell m} - \frac{dH}{dx}\frac{dt_p}{ds}\left(\dot r_{\{0\}} \partial_{r_p} - im\Omega_{\{0\}}\right)R^{\text{e/o}\,\mathcal{R}}_{\ell m}
    \\[1ex]
    &+ \left(1 - H^2\right)\left[im\Omega_{\{0\}}\frac{d^2t_p}{ds^2} + \left(m^2\Omega_{\{0\}}^2 + im \partial_{r_p}\Omega_{\{0\}} \dot r_{\{0\}}\right)\left(\frac{dt_p}{ds}\right)^2 \right.
    \\[1ex]
    &\left. - \dot r_{\{0\}}\frac{d^2t_p}{ds^2}\partial_{r_p} - \dot r_{\{0\}}^2\left(\frac{dt_p}{ds}\right)^2\partial_{r_p}^2 - \left(\ddot r_{\{0\}} - 2im\Omega_{\{0\}}\dot r_{\{0\}}\right)\left(\frac{dt_p}{ds}\right)^2\partial_{r_p}\right]R^{\text{e/o}\,\mathcal{R}}_{\ell m} = S^\text{e/o eff}_{\ell m},
\end{split}
\end{equation}
where $S^\text{e/o eff}_{\ell m}(r_p, r)$ is now the effective source in $s$ slicing. As in the main text, the effective source can be split into the ordinary point-particle source in the region $r_p< r_{\cal P}^-$,
\begin{equation}
    S^{\text{e/o pp}}_{\ell m}(r_p, r) \coloneqq S^{\text{e/o}}_{\ell m}(r_p, r)\theta(r_\mathcal{P}-r_p),
\end{equation}
and the following extended source in the region $6M\geq r_p\geq r_{\cal P}^-$:
\begin{equation}
\begin{split}
    S^\text{e/o ext}_{\ell m}(r_p, r) \coloneqq\,&\; \theta(r_p - r_\mathcal{P})\bigl[S^{\text{e/o}}_{\ell m}(r_p, r) - S^{\text{e/o}}_{\ell m}(6M, r)\bigr]
    \\[1ex]
    &- \theta(r_p - r_{\cal P})\, im \left(\Omega_{\{0\}}\frac{dt_p}{ds} - \left.\Omega_{\{0\}}\right\vert_*\left.\frac{dt_p}{ds}\right\vert_*\right)\left(2H\partial_x + \frac{dH}{dx}\right) \left.R^{\text{e/o}\,(1)}_{\ell m}\right\vert_*\\
    &- \theta(r_p - r_{\cal P})\left(1-H^2\right)\Biggl[im\left(\Omega_{\{0\}}\frac{d^2t_p}{ds^2} - \left.\Omega_{\{0\}}\right\vert_*\left.\frac{d^2t_p}{ds^2}\right\vert_*\right)\\
    &+ \left(\frac{dt_p}{ds}\right)^2\left(m^2\Omega_{\{0\}}^2 + im \partial_{r_p}\Omega_{\{0\}} \dot r_{\{0\}}\right) - \left(\left.\frac{dt_p}{ds}\right\vert_*\right)^2 m^2\left.\Omega_{\{0\}}\right\vert_*^2\Biggr] \left.R^{\text{e/o}\,(1)}_{\ell m}\right\vert_*
    \\[1ex]
    &+\delta(r_p - r_{\cal P})\Biggl[\left.\frac{dt_p}{ds}\right\vert_{\mathcal P} \left.\dot r_{\{0\}}\right\vert_{\mathcal P}\left(2H\partial_x + \frac{dH}{dx}\right) \left.R^{\text{e/o}\,(1)}_{\ell m}\right\vert_*\\
    &- \left(1 - H^2\right)\left(\left.\frac{d^2t_p}{ds^2}\right\vert_{\mathcal P} \left.\dot r_{\{0\}}\right\vert_\mathcal{P} + \left(\left.\frac{dt_p}{ds}\right\vert_{\mathcal P}\right)^2\left(\left.\ddot r_{\{0\}}\right\vert_\mathcal{P} + 2im\left.\Omega_{\{0\}}\right\vert_\mathcal{P}\left.\dot r_{\{0\}}\right\vert_\mathcal{P}\right)\right) \!\!\! \left.R^{\text{e/o}\,(1)}_{\ell m}\right\vert_*\Biggr]
    \\[1ex]
    &+\partial_{r_p}\delta(r_p - r_{\cal P})\left(1-H^2\right)\left(\left.\frac{dt_p}{ds}\right\vert_{\mathcal P}\right)^2\left.\dot r_{\{0\}}^2\right\vert_\mathcal{P} \left.R^{\text{e/o}\,(1)}_{\ell m}\right\vert_*.
\end{split}
\end{equation}

\subsection{Frequency-domain equations}

We next introduce the following forward and inverse transforms:
\begin{subequations}
\begin{align}\label{stransformforw}
    \hat g(\omega, r) &= \int_{2M}^{6M} \frac{dr_p}{\dot r_{\{0\}}(r_p) \, \frac{dt_p}{ds}(r_p)}g(r_p, r)e^{i\omega s_G(r_p) - i m \phi_G(r_p)},
    \\[1ex]\label{stransforminv}
    g(r_p, r) &= -\frac{1}{2\pi}\int^{+\infty}_{-\infty} d\omega \, \hat g(\omega, r)e^{-i\omega s_G(r_p) + i m \phi_G(r_p)},
\end{align}
\end{subequations}
where
\begin{equation}
    s_G(r_p) \coloneqq \int^{r_p} \frac{dr_p'}{\dot r_{\{0\}}(r_p')}\frac{ds}{dt_p}(r_p')
\end{equation}
and $dt_p/ds$ is given in Eq.~\eqref{dtpds}.

The transform of Eq.~\eqref{RWZs app} reads
\begin{equation}\label{RWZs freq domain}
    \left(\partial_x^2 -  V_\ell^\text{e/o}\right) \hat R^{\text{e/o}\,\mathcal{R}}_{\ell m} + i\omega \left(2H\partial_x + \frac{dH}{dx}\right) \hat R^{\text{e/o}\,\mathcal{R}}_{\ell m} + \left(1 - H^2\right)\omega^2 \hat R^{\text{e/o}\,\mathcal{R}}_{\ell m} = \hat S^\text{e/o eff}_{\ell m},
\end{equation}
where, following the same reasoning as in Sec.~\ref{sec:f domain eqs}, no boundary terms appear. The transformed effective source $\hat S^\text{e/o eff}_{\ell m} = \hat S^{\text{e/o pp}}_{\ell m} + \hat S^\text{e/o ext}_{\ell m}$ is given by
\begin{subequations}
\begin{align}
    \hat S^{\text{e/o pp}}_{\ell m} \coloneqq&\, \int_{2M}^{r_{\cal P}}\frac{dr_p\,e^{i \omega s_G-i m \phi_G}}{\dot r_{\{0\}}\,dt_p/ds} S^{\text{e/o}}_{\ell m},
    \\[1ex]
    \begin{split}\label{Seffhat s slicing}
    \hat S^\text{e/o ext}_{\ell m} \coloneqq&\, \int_{r_{\cal P}}^{6M}\frac{dr_p\,e^{i \omega s_G-i m \phi_G}}{\dot r_{\{0\}}\,dt_p/ds} \left(S^{\text{e/o}}_{\ell m} - \left.S^{\text{e/o}}_{\ell m}\right\vert_*\right)
    \\[1ex]
    &- {\cal I}_m^1(r_{\cal P};\omega)\left(2H\partial_x + \frac{dH}{dx}\right)\left.R^{\text{e/o}\,(1)}_{\ell m}\right\vert_* - \left(1-H^2\right){\cal I}_m^2(r_{\cal P};\omega)\left.R^{\text{e/o}\,(1)}_{\ell m}\right\vert_*
    \\[1ex]
    &+ \left[2H \partial_x + \frac{dH}{dx} - i\left(1-H^2\right) \left(\omega + m \left.\Omega_{\{0\}}\right\vert_{\cal P} \left. \frac{dt_p}{ds}\right\vert_{\cal P}\right)\right]\left.R^{\text{e/o}\,(1)}_{\ell m}\right\vert_*e^{i \omega s_G(r_{\cal P})-i m \phi_G(r_{\cal P})},
    \end{split}
\end{align}
\end{subequations}
with
\begin{subequations}
\begin{align}
    {\cal I}_m^1(r_{\cal P};\omega) \coloneqq&\, \int_{r_{\cal P}}^{6M} \frac{dr_p\,e^{i \omega s_G-i m \phi_G}}{\dot r_{\{0\}} \, dt_p/ds} \left[i m \left(\Omega_{\{0\}}\frac{dt_p}{ds} - \left.\Omega_{\{0\}}\right\vert_*\left.\frac{dt_p}{ds}\right\vert_*\right)\right],
    \\[1ex]
    \begin{split}{\cal I}_m^2(r_{\cal P};\omega) \coloneqq&\, \int_{r_{\cal P}}^{6M} \frac{dr_p\,e^{i \omega s_G-i m \phi_G}}{\dot r_{\{0\}} \, dt_p/ds} \Bigg[im\left(\Omega_{\{0\}}\frac{d^2t_p}{ds^2} - \left.\Omega_{\{0\}}\right\vert_*\left.\frac{d^2t_p}{ds^2}\right\vert_*\right)\\
    &+ \left(\frac{dt_p}{ds}\right)^2\left(m^2\Omega_{\{0\}}^2 + im \partial_{r_p}\Omega_{\{0\}} \dot r_{\{0\}}\right) - \left(\left.\frac{dt_p}{ds}\right\vert_*\right)^2 m^2\left.\Omega_{\{0\}}^2\right\vert_*\Bigg].\end{split}
\end{align}
\end{subequations}

\subsection{Inhomogeneous solutions}

The homogeneous solutions to Eq.~\eqref{RWZs freq domain} are related to the homogeneous solutions to Eq.~\eqref{tRWZtransformed} simply by an overall factor of $e^{-i\omega \kappa(x)}$. Recalling the $t$-slicing solutions in Eq.~\eqref{inup}, we readily obtain the ``in'' and ``up'' solutions in $s$ slicing as
\begin{subequations}\label{inup-s}
\begin{equation}
    \hat R_{\ell}^\text{e/o in}(\omega,x) \sim \left\{ \begin{array}{ll}
    A_{\ell}^\text{e/o in}(\omega) e^{-2i\omega x} + A_{\ell}^\text{e/o out}(\omega) & \text{as } r\to+\infty \, (x\to+\infty),\\
    1 & \text{as } r\to2M \, (x\to-\infty),
    \end{array} \right.
\end{equation}
\begin{equation}
    \hat R_{\ell}^\text{e/o up}(\omega,x) \sim \left\{ \begin{array}{ll}
    1 & \text{as } r\to+\infty \, (x\to+\infty),\\
    B_{\ell}^\text{e/o in}(\omega) + B_{\ell}^\text{e/o out}(\omega) e^{+2i\omega x} & \text{as } r\to2M \, (x\to-\infty).
    \end{array} \right.
\end{equation}
\end{subequations}

Using the two independent homogeneous solutions to Eq.~\eqref{RWZs freq domain}, $\hat R_{\ell}^\text{e/o in}$ and $\hat R_{\ell}^\text{e/o up}$, we construct the following Green function:
\begin{equation}
\begin{split}
    \hat G^\text{e/o}_{\ell}(\omega, x, x') = \frac{1}{W_\ell (\omega)}&\left[\theta(x-x')\hat R_{\ell}^\text{e/o in}(\omega, x')\hat R_{\ell}^\text{e/o up}(\omega, x)\right.\\
    &\;\;\left.+ \theta(x'-x)\hat R_{\ell}^\text{e/o in}(\omega, x)\hat R_{\ell}^\text{e/o up}(\omega, x')\right],
\end{split}
\end{equation}
with
\begin{equation}
    W_\ell(\omega) = e^{2i\omega\kappa(x)}\left[\hat R_{\ell}^\text{e/o in}(\omega, x)\partial_x \hat R_{\ell}^\text{e/o up}(\omega, x) - \hat R_{\ell}^\text{e/o up}(\omega, x)\partial_x \hat R_{\ell}^\text{e/o in}(\omega, x)\right].
\end{equation}
The Green function satisfies
\begin{equation}
    \left[\left(\partial_x^2 -  V_{\ell}^{\text{e/o}}\right) + i\omega\left(2H\partial_x + \frac{dH}{dx}\right) + \left(1 - H^2\right)\omega^2 \right] \hat G^\text{e/o}_{\ell}(\omega, x, x') = \frac{\delta(x-x^\prime)}{e^{2i\omega \kappa(x)}}.
\end{equation}
It is straightforward to verify that $W_\ell$ is independent of the field point $r$, $\partial_x W_\ell=0$, and it can therefore be computed from the asymptotic solutions~\eqref{inup-s} at $r\to+\infty$, leading to
\begin{equation}
    W_\ell(\omega) = 2i\omega A_\ell^{\rm in}\left(\omega\right).
\end{equation}
The inhomogeneous solution can then be obtained as
\begin{equation}\label{Rhat res s}
\begin{split}
    \hat R^{\text{e/o}\,\mathcal{R}}_{\ell m}(\omega, r) =&\,
    \int_{-\infty}^{+\infty}dx' \hat G^\text{e/o}_{\ell}(\omega, x, x') e^{2i\omega\kappa(x')} \hat S^\text{e/o eff}_{\ell m}(\omega, x^\prime)
    \\[1ex]
    =&\,\frac{\hat R_{\ell}^\text{e/o up}}{2i\omega A_\ell^{\rm in}\left(\omega\right)}\int_{2M}^r \frac{dr^\prime}{f(r^\prime)} \hat R_{\ell}^\text{e/o in}(\omega, r^\prime)e^{2i\omega\kappa(r^\prime)}\hat S^\text{e/o eff}_{\ell m}(\omega, r^\prime)\\
    &+ \frac{\hat R_{\ell}^\text{e/o in}}{2i\omega A_\ell^{\rm in}\left(\omega\right)}\int_r^{+\infty}\frac{dr^\prime}{f(r^\prime)} \hat R_{\ell}^\text{e/o up}(\omega, r^\prime)e^{2i\omega\kappa(r^\prime)}\hat S^\text{e/o eff}_{\ell m}(\omega, r^\prime).
\end{split}
\end{equation}

\subsection{Regularity, slicing, and punctures at the boundaries}
\label{sec:regularity at boundaries}

To assess regularity near the boundaries, we consider $s=u$ ($\kappa=x$, $H=+1$) at large $r$ and $s=v$ ($\kappa=-x$, $H=-1$) near $r=2M$. In either case, the effective source~\eqref{Seffhat s slicing} in a neighbourhood of each boundary reduces to
\begin{align}
    \hat S^\text{e/o ext}_{\ell m} = \mp\, 2\left[{\cal I}_m^1(r_{\cal P};\omega) 
    - e^{i \omega s_G(r_{\cal P})-i m \phi_G(r_{\cal P})}\right]\left.\partial_x R^{\text{e/o}\,(1)}_{\ell m}\right\vert_*.
\end{align}
At large $r$, $R^{\text{e/o}\,(1)}_{\ell m}\vert_*$ behaves as a regular series in $1/r$, beginning at $r^0$, such that $\hat S^\text{e/o ext}_{\ell m}= {\cal O}(1/r^2)$. At the horizon, $R^{\text{e/o}\,(1)}_{\ell m}\vert_*$ behaves as a regular series in $(r-2M)$, beginning at $(r-2M)^0$, and $\partial_x = f\partial_r$, such that $\hat S^\text{e/o ext}_{\ell m} = {\cal O}(r-2M)$. Referring to the asymptotic behaviour~\eqref{inup-s}, we then see that (i) the integrals in Eq.~\eqref{Rhat res s} converge, (ii) at the horizon, the solution behaves as the regular homogeneous solution $\hat R_{\ell}^\text{e/o in}$, and (iii) at infinity, the solution behaves as the regular homogeneous solution $\hat R_{\ell}^\text{e/o up}$.

If we were using $t$ slicing rather than hyperboloidal slicing, the integrals in Eq.~\eqref{Rhat res s} would not converge, nor would either integral vanish in the limits $r\to2M$ and $r\to\infty$. With $s=t$ ($H=0$), the effective source~\eqref{Seffhat s slicing} behaves like $R^{\text{e/o}\,(1)}_{\ell m}\vert_*$ at the boundaries. In $t$ slicing, $R^{\text{e/o}\,(1)}_{\ell m}\vert_*$ behaves as $e^{\pm i \omega x}$ toward the boundaries, and both integrals in Eq.~\eqref{Rhat res s} then diverge for any value of $r$. 

Such ill behaviour is familiar from infrared divergences that arise in the inspiral multiscale expansion in $t$ slicing~\cite{Pound:2015wva,Miller:2020bft,Miller:2023ers,Cunningham:2024dog}. As in that case, we can obtain the physically correct solution by introducing punctures at the boundaries. The physically correct solution in this case is obtained in hyperboloidal slicing, where the field behaves as a smooth outgoing wave at future null infinity and a smooth ingoing one at the horizon. To obtain the correct boundary conditions in $t$ slicing, we can transform the hyperboloidal-slicing solution $\hat R^{\text{e/o}\,\cal R}_{\ell m}$ to $t$ slicing, expand it near the horizon and infinity, and treat the results (which will be singular at the boundaries) as  punctures. 

To transform the hyperboloidal solution to $t$ slicing, we first transform it back to phase space, using Eq.~\eqref{stransforminv}. At a given point in spacetime, the field in $s$ slicing is then ${}_{[s]}R^{\rm e/o}_{\ell m}(r_p(t_p(s)),r) e^{-im\phi_p(t_p(s))}$, where we add a left-subscript to indicate the slicing; at the same point, the field in $t$ slicing is ${}_{[t]}R^{\rm e/o}_{\ell m}(r_p(t),r) e^{-im\phi_p(t)}$. Since the field at a given point must be independent slicing, we can infer 
\begin{equation}\label{slicing transformation}
{}_{[t]}R^{\text{e/o}\,\cal R}_{\ell m}(r_p(t),r) = {}_{[s]}R^{\text{e/o}\,\cal R}_{\ell m}(r_p(t_p(s)),r) e^{im[\phi_p(t)-\phi_p(t_p(s))]}.
\end{equation}
For this to give us the $t$-slicing field variable, the right-hand side must be expressed entirely in terms of the independent variables $r_p(t)$ and $r$. 
We put it in that form by expanding for $r_p(t)$ near the ISCO; this suffices because we only require the boundary punctures in the window $6M>r_p>r_{\cal P}$. 

Define $\Delta t \coloneqq t_p(s) - t$. Then 
\begin{align}
r_p(t_p(s)) &= r_p(t) + \dot r_p(r_p(t))\Delta t + \frac{1}{2}\ddot r_p(r_p(t))(\Delta t)^2 +\ldots\label{rp(tp(s))}\\ 
\phi_p(t_p(s)) &= \phi_p(t) + \Omega_{\{0\}}(r_p(t))\Delta t + \frac{1}{2}\dot \Omega_{\{0\}}(r_p(t))(\Delta t)^2 +\ldots
\end{align}
Although $\Delta t$ need not be small, these expansions are well behaved because successive terms are smaller in the limit  $r_p\to 6M$. Concretely, Eqs.~\eqref{rp_transition_late} and \eqref{FDO_transition_late} imply $\dot r_p={\cal O}[(6M-r_p)^{3/2}]=\dot \Omega_{\{0\}}$, and each additional derivative introduces $d/dt = {\cal O}[(6M-r_p)^{1/2}]$.

Given that $s=t-\kappa(r)$, we have $t_p(s)=s+\kappa(r_p(t_p(s)))$ and so $\Delta t = \kappa(r_p(t_p(s)))-\kappa(r)$. Formally expanding for small time derivatives and appealing to Eq.~\eqref{rp(tp(s))} then yields
\begin{equation}
    \Delta t = \Delta t_0(r_p,r) +\Delta t_1(r_p,r) + \ldots,
\end{equation}
where $\Delta t_0\coloneqq[\kappa(r_p) - \kappa(r)]$, $\Delta t_1\coloneqq\kappa'(r_p)\dot r_p(r_p)\Delta t_0(r_p,r)$, $\kappa'(r_p)=\partial_{r_p}\kappa(r_p),$ and here it is understood that $r_p=r_p(t)$.  Equation~\eqref{slicing transformation} then becomes
\begin{multline}\label{slicing transformation expanded}
\hspace{-10pt}{}_{[t]}R^{\text{e/o}\,\cal R}_{\ell m}(r_p,r) = e^{-im\Omega_{\{0\}} \Delta t_0}\biggl\{{}_{[s]}R^{\text{e/o}\,\cal R}_{\ell m}(r_p,r) \\
+ \left[\dot r_p\Delta t_0\partial_{r_p}-im\left(\Omega_{\{0\}} \Delta t_1+\frac{1}{2}\dot\Omega_{\{0\}}(\Delta t_0)^2\right)\right]{}_{[s]}R^{\text{e/o}\,\cal R}_{\ell m}(r_p,r) + \ldots \biggr\},
\end{multline}
where again it is understood that $r_p=r_p(t)$. The first subleading term, in square brackets, scales as $(r_p-6M)^{3/2}$, and the ellipses denote terms of order $(r_p-6M)^{2}$ and smaller. 

The transformation~\eqref{slicing transformation expanded} can be compared to Eq.~(161) in Ref.~\cite{Miller:2020bft}, which represents the analogous result in the inspiral. Like in the inspiral, the $t$-slicing field ${}_{[t]}R^{\text{e/o}\,\cal R}_{\ell m}$ is singular at the boundaries. This can be seen from the factors of $\kappa(r)$ in $\Delta t_n$, which reduce to $\pm x$ near the boundaries, diverging logarithmically when $r\to 2M$ and linearly when $r\to\infty$.

Finally, to obtain the behaviour of the frequency-domain, $t$-slicing solution near the boundaries, we expand Eq.~\eqref{slicing transformation expanded} in the limits $r\to2M$ and $r\to \infty$ and then apply the Fourier transform~\eqref{transformforw}. This defines for us our punctures at the boundaries, which we move to the right-hand side of the $t$-slicing field equation~\eqref{tRWZtransformed}, defining a new effective source. Such procedures are detailed extensively in Ref.~\cite{Miller:2023ers}. Since the punctures are necessarily approximate particular solutions, they cancel the old effective source near the boundaries. If the near-horizon and large-$r$ expansions of the punctures are carried to sufficiently high order, the new effective source will behave sufficiently well toward the boundaries, the Green-function integrals over $r'$ will converge, and the sum of the new residual field plus the new punctures will be precisely equivalent to the hyperboloidal solution, simply expressed in terms of a Fourier transform based on the singular $t$ coordinate.

\section{Contribution from the early-time effective source}\label{app:Sext}

In this appendix we consider the solution to Eq.~\eqref{RWZpunctured} with the extended source~\eqref{Seffext}. Our aim is to show that its contribution can be neglected if we only require the solution sufficiently far to the future of the puncture window (i.e., for values of $r_p$ not too close to $6M$).

We start by obtaining the source's  transform~\eqref{transformforw}:
\begin{equation}\label{hatSeffext}
\begin{split}
    \hat S^\text{e/o ext}_{\ell m}(\omega, r) =&\; \int_{r_\mathcal{P}}^{6M} \frac{dr_p}{\dot r_{\{0\}}}e^{i\omega t_G - i m \phi_G} \left[S^{\text{e/o}}_{\ell m}(r_p, r) - S^{\text{e/o}}_{\ell m}(6M, r)\right]
    \\[1ex]
    &\;-R^{\text{e/o}\,(1)}_{\ell m}(6M,r)\int_{r_\mathcal{P}}^{6M} \frac{dr_p}{\dot r_{\{0\}}}e^{i\omega t_G - i m \phi_G}\left(m^2\Omega_{\{0\}}^2 - m^2\left.\Omega_{\{0\}}^2\right\vert_* + i m \partial_{r_p}\Omega_{\{0\}} \, \dot r_{\{0\}}\right)
    \\[1ex]
    &\;-i\left(\omega + m\left.\Omega_{\{0\}}\right\vert_\mathcal{P}\right) e^{i\omega t_G(r_\mathcal{P}) - i m \phi_G(r_\mathcal{P})}R^{\text{e/o}\,(1)}_{\ell m}(6M,r).
\end{split}
\end{equation}
We anticipate that we are interested in the limit $r_\mathcal{P}\to6M$. In this limit, the $r_p$ integrals in Eq.~\eqref{hatSeffext} vanish. We therefore only focus on the solution at infinity sourced by the third term in Eq.~\eqref{hatSeffext}. 

Using the same Green-function technique we have adopted in Sec.~\ref{sec:inhomogsol}, we obtain
\begin{equation}
\begin{split}
    \left.R^{\text{e/o}\,\mathcal{R}}_{\ell m}\right\vert_{{\rm ext}\,\infty}(r_p) = -\frac{1}{2\pi}&\int_{-\infty}^{+\infty} d\omega\,e^{-i\omega t_G(r_p) + im\phi_G(r_p)}\\
    &\times\frac{e^{i\omega x}}{2i\omega A^{\rm in}_\ell(\omega)}\int_{2M}^{\infty}\frac{dr'}{f(r')} \hat R^\text{e/o in}_{\ell}(\omega, r')\hat S^\text{e/o ext}_{\ell m}(\omega, r').
\end{split}
\end{equation}
As discussed in Appendix~\ref{sec:regularity at boundaries}, the radial integral here does not converge if we use the effective source described in the body of the paper, due to the poor behaviour of the source in $t$ slicing. We adopt the prescription in Appendix~\ref{sec:regularity at boundaries} to avoid that divergence. We can then write the equation above as
\begin{equation}\label{R res ext}
    \left.R^{\text{e/o}\,\mathcal{R}}_{\ell m}\right\vert_{{\rm ext}\,\infty}(r_p) = -\frac{1}{2\pi}\int_{-\infty}^{+\infty} d\omega\,F^\text{e/o}_{\ell m}(\omega; r_p, r_\mathcal{P}) e^{i\omega [t_G(r_\mathcal{P})-t_G(r_p)+x]},
\end{equation}
where we have defined
\begin{equation}
\begin{split}
    F^{\text{e/o}}_{\ell m}(\omega; r_p, r_\mathcal{P}) \coloneqq& -  \frac{e^{i m [\phi_G(r_p) - \phi_G(r_\mathcal{P})]}}{2 A_\ell^{\rm in}\left(\omega\right)}\left(1+\frac{m\left.\Omega_{\{0\}}\right\vert_\mathcal{P}}{\omega}\right)
    \\
    &\times \int_{2M}^{\infty}\frac{dr'}{f(r')} \hat R^\text{e/o in}_\ell(\omega, r') R^{\text{e/o}(1)}_{\ell m}(6M, r'). 
\end{split}
\end{equation}

The integral \eqref{R res ext} contains a rapidly oscillating integrand when $r_{\cal P}$ is near $6M$ (and $r_p$ is not too close to $6M$). Integrating by parts $N+1$ times, we find
\begin{equation}\label{R res ext integration by parts}
    \left.R^{\text{e/o}\,\mathcal{R}}_{\ell m}\right\vert_{{\rm ext}\,\infty}(r_p) = \frac{1}{2\pi}\sum_{n=0}^N \left[i^{n+1}\frac{d^n F^{\text{e/o}}_{\ell m}(\omega; r_p, r_\mathcal{P})}{d\omega^n}\frac{e^{i\omega [t_G(r_\mathcal{P})-t_G(r_p)+x]}}{[t_G(r_\mathcal{P})-t_G(r_p)+x]^{n+1}}\right]_{-\infty}^{+\infty} +{\cal O}(N+1),
\end{equation}
where ${\cal O}(N+1)$ denotes the remaining integral, which is proportional to $1/[t_G(r_\mathcal{P})-t_G(r_p)+x]^{N+1}$. Assuming the solution~\eqref{R res ext} exists, $F^{\text{e/o}}_{\ell m}$ must decay to zero as $\omega\to\pm\infty$, implying its derivatives do likewise. Hence, the sum in Eq.~\eqref{R res ext integration by parts} vanishes, leaving the ${\cal O}(N+1)$ remainder. If $r_{\cal P}$ is taken toward $6M$, then $t_G(r_\mathcal{P})\to-\infty$, causing the ${\cal O}(N+1)$ remainder to go to zero. Since $N$ is arbitrary, it follows that $R^{\text{e/o}\,\mathcal{R}}_{\ell m}\vert_{{\rm ext}\,\infty}(r_p)$ goes to zero faster than any power of $1/t_G(r_\mathcal{P})$ as $r_{\cal P}\to6M$ (at fixed values of $r_p$ and $x$).

\section{Regularization procedure for the excitation coefficients}\label{app:regularization}

In this appendix we justify the regularization procedure introduced in Sec.~\ref{sec:QNMs}. We start by considering Eq.~\eqref{waveformF},
\begin{equation}\label{waveformF_app}
    \left.R^{\text{e/o}\,\mathcal{R}}_{\ell m}\right\vert_{{\rm pp}\,\infty}(r_p)e^{-i m \phi_G(r_p)} = -\frac{1}{2\pi}\int_{-\infty}^{+\infty} d\omega\, e^{-i\omega[t_G(r_p) - x]} \frac{C^\text{e/o}_{\ell m}(\omega)}{2i\omega A_\ell^{\rm in}(\omega)},
\end{equation}
where the integration along the real $\omega$ axis gives a finite result. The numerator of the integrand, $C^\text{e/o}_{\ell m}(\omega)$, is given by the radial integral~\eqref{Clm}, which we rewrite here for convenience:
\begin{equation}
    C^\text{e/o}_{\ell m}(\omega) \coloneqq \int_{2M}^{r_{\cal P}}\frac{dr'}{\dot r_{\{0\}}(r')} e^{i\omega t_G(r') - im\phi_G(r')} K^\text{e/o}_{\ell m}(\omega, r').
\end{equation}
Now, analogously to the procedure described at the end of Sec.~\ref{sec:QNMs},  we can rewrite $C^\text{e/o}_{\ell m}(\omega)$ by subtracting the near-horizon behaviour $q^\text{e/o}_{\ell m}$, given in Eq.~\eqref{qlm}, from the integrand and adding back its antiderivative $Q^\text{e/o}_{\ell m}$:
\begin{equation}
    C^\text{e/o}_{\ell m}(\omega) = \left.C^\text{e/o}_{\ell m}\right\vert_\text{reg}(\omega)-Q^\text{e/o}_{\ell m}(\omega,2M).
\end{equation}
Integrating Eq.~\eqref{qlm} gives $Q^\text{e/o}_{\ell m}(\omega,r)\sim (r-2M)^{1-4iM\omega}$, which vanishes at the horizon. Therefore, Eq.~\eqref{waveformF} is equivalent to 
\begin{equation}
    \left.R^{\text{e/o}\,\mathcal{R}}_{\ell m}\right\vert_{{\rm pp}\,\infty}(r_p)e^{-i m \phi_G(r_p)} = -\frac{1}{2\pi}\int_{-\infty}^{+\infty} d\omega\, e^{-i\omega[t_G(r_p) - x]} \frac{\left.C^\text{e/o}_{\ell m}\right\vert_\text{reg}(\omega)}{2i\omega A_\ell^{\rm in}(\omega)}.
\end{equation}
After closing the contour in the complex plane and using the residue theorem, this then justifies computing the excitation factors appearing in Eq.~\eqref{QNMsum} by using $\left.C^\text{e/o}_{\ell m}\right\vert_\text{reg}$ rather than simply $C^\text{e/o}_{\ell m}$ in Eq.~\eqref{Bl_Dlm}.

\bibliographystyle{utphys} 
\bibliography{refs}

\providecommand{\href}[2]{#2}\begingroup\raggedright\begin{thebibliography}{100}

\bibitem{LIGOScientific:2016aoc}
{\bf LIGO Scientific, Virgo} Collaboration, B.~P. Abbott {\em et al.}, ``{Observation of Gravitational Waves from a Binary Black Hole Merger},'' {\em Phys. Rev. Lett.} {\bf 116} (2016), no.~6, 061102, \href{http://www.arXiv.org/abs/1602.03837}{{\tt 1602.03837}}.

\bibitem{LIGOScientific:2018mvr}
{\bf LIGO Scientific, Virgo} Collaboration, B.~P. Abbott {\em et al.}, ``{GWTC-1: A Gravitational-Wave Transient Catalog of Compact Binary Mergers Observed by LIGO and Virgo during the First and Second Observing Runs},'' {\em Phys. Rev. X} {\bf 9} (2019), no.~3, 031040, \href{http://www.arXiv.org/abs/1811.12907}{{\tt 1811.12907}}.

\bibitem{LIGOScientific:2020ibl}
{\bf LIGO Scientific, Virgo} Collaboration, R.~Abbott {\em et al.}, ``{GWTC-2: Compact Binary Coalescences Observed by LIGO and Virgo During the First Half of the Third Observing Run},'' {\em Phys. Rev. X} {\bf 11} (2021) 021053, \href{http://www.arXiv.org/abs/2010.14527}{{\tt 2010.14527}}.

\bibitem{KAGRA:2021vkt}
{\bf KAGRA, VIRGO, LIGO Scientific} Collaboration, R.~Abbott {\em et al.}, ``{GWTC-3: Compact Binary Coalescences Observed by LIGO and Virgo during the Second Part of the Third Observing Run},'' {\em Phys. Rev. X} {\bf 13} (2023), no.~4, 041039, \href{http://www.arXiv.org/abs/2111.03606}{{\tt 2111.03606}}.

\bibitem{ET:2019dnz}
{\bf ET} Collaboration, M.~Maggiore {\em et al.}, ``{Science Case for the Einstein Telescope},'' {\em JCAP} {\bf 03} (2020) 050, \href{http://www.arXiv.org/abs/1912.02622}{{\tt 1912.02622}}.

\bibitem{Evans:2021gyd}
M.~Evans {\em et al.}, ``{A Horizon Study for Cosmic Explorer: Science, Observatories, and Community},'' \href{http://www.arXiv.org/abs/2109.09882}{{\tt 2109.09882}}.

\bibitem{Baibhav:2019gxm}
V.~Baibhav, E.~Berti, D.~Gerosa, M.~Mapelli, N.~Giacobbo, Y.~Bouffanais, and U.~N. Di~Carlo, ``{Gravitational-wave detection rates for compact binaries formed in isolation: LIGO/Virgo O3 and beyond},'' {\em Phys. Rev. D} {\bf 100} (2019), no.~6, 064060, \href{http://www.arXiv.org/abs/1906.04197}{{\tt 1906.04197}}.

\bibitem{Kalogera:2021bya}
V.~Kalogera {\em et al.}, ``{The Next Generation Global Gravitational Wave Observatory: The Science Book},'' \href{http://www.arXiv.org/abs/2111.06990}{{\tt 2111.06990}}.

\bibitem{amaroseoane2017laser}
P.~Amaro-Seoane {\em et al.}, ``Laser {I}nterferometer {S}pace {A}ntenna,'' \href{http://www.arXiv.org/abs/1702.00786}{{\tt 1702.00786}}.

\bibitem{Babak:2017tow}
S.~Babak, J.~Gair, A.~Sesana, E.~Barausse, C.~F. Sopuerta, C.~P.~L. Berry, E.~Berti, P.~Amaro-Seoane, A.~Petiteau, and A.~Klein, ``{Science with the space-based interferometer LISA. V: Extreme mass-ratio inspirals},'' {\em Phys. Rev. D} {\bf 95} (2017), no.~10, 103012, \href{http://www.arXiv.org/abs/1703.09722}{{\tt 1703.09722}}.

\bibitem{LISA:2022yao}
{\bf LISA} Collaboration, P.~A. Seoane {\em et al.}, ``{Astrophysics with the Laser Interferometer Space Antenna},'' {\em Living Rev. Rel.} {\bf 26} (2023), no.~1, 2, \href{http://www.arXiv.org/abs/2203.06016}{{\tt 2203.06016}}.

\bibitem{Barausse:2020mdt}
E.~Barausse, I.~Dvorkin, M.~Tremmel, M.~Volonteri, and M.~Bonetti, ``{Massive Black Hole Merger Rates: The Effect of Kiloparsec Separation Wandering and Supernova Feedback},'' {\em Astrophys. J.} {\bf 904} (2020), no.~1, 16, \href{http://www.arXiv.org/abs/2006.03065}{{\tt 2006.03065}}.

\bibitem{Mangiagli:2022niy}
A.~Mangiagli, C.~Caprini, M.~Volonteri, S.~Marsat, S.~Vergani, N.~Tamanini, and H.~Inchausp\'e, ``{Massive black hole binaries in LISA: Multimessenger prospects and electromagnetic counterparts},'' {\em Phys. Rev. D} {\bf 106} (2022), no.~10, 103017, \href{http://www.arXiv.org/abs/2207.10678}{{\tt 2207.10678}}.

\bibitem{Bellovary:2024akp}
J.~Bellovary, Y.~Luo, T.~Quinn, F.~Munshi, M.~Tremmel, and J.~Wadsley, ``{Intermediate Mass Ratio Inspirals in Milky Way Galaxies},'' \href{http://www.arXiv.org/abs/2411.12117}{{\tt 2411.12117}}.

\bibitem{Jani:2019ffg}
K.~Jani, D.~Shoemaker, and C.~Cutler, ``{Detectability of Intermediate-Mass Black Holes in Multiband Gravitational Wave Astronomy},'' {\em Nature Astron.} {\bf 4} (2019), no.~3, 260--265, \href{http://www.arXiv.org/abs/1908.04985}{{\tt 1908.04985}}.

\bibitem{LISAConsortiumWaveformWorkingGroup:2023arg}
{\bf LISA Consortium Waveform Working Group} Collaboration, N.~Afshordi {\em et al.}, ``{Waveform Modelling for the Laser Interferometer Space Antenna},'' \href{http://www.arXiv.org/abs/2311.01300}{{\tt 2311.01300}}.

\bibitem{Damour:2009sm}
T.~Damour, ``{Gravitational Self Force in a Schwarzschild Background and the Effective One Body Formalism},'' {\em Phys. Rev. D} {\bf 81} (2010) 024017, \href{http://www.arXiv.org/abs/0910.5533}{{\tt 0910.5533}}.

\bibitem{Nagar:2022fep}
A.~Nagar and S.~Albanesi, ``{Toward a gravitational self-force-informed effective-one-body waveform model for nonprecessing, eccentric, large-mass-ratio inspirals},'' {\em Phys. Rev. D} {\bf 106} (2022), no.~6, 064049, \href{http://www.arXiv.org/abs/2207.14002}{{\tt 2207.14002}}.

\bibitem{vandeMeent:2023ols}
M.~van~de Meent, A.~Buonanno, D.~P. Mihaylov, S.~Ossokine, L.~Pompili, N.~Warburton, A.~Pound, B.~Wardell, L.~Durkan, and J.~Miller, ``{Enhancing the SEOBNRv5 effective-one-body waveform model with second-order gravitational self-force fluxes},'' {\em Phys. Rev. D} {\bf 108} (2023), no.~12, 124038, \href{http://www.arXiv.org/abs/2303.18026}{{\tt 2303.18026}}.

\bibitem{Leather:2025nhu}
B.~Leather, A.~Buonanno, and M.~van~de Meent, ``{Inspiral-merger-ringdown waveforms with gravitational self-force results within the effective-one-body formalism},'' \href{http://www.arXiv.org/abs/2505.11242}{{\tt 2505.11242}}.

\bibitem{Wardell:2021fyy}
B.~Wardell, A.~Pound, N.~Warburton, J.~Miller, L.~Durkan, and A.~Le~Tiec, ``{Gravitational Waveforms for Compact Binaries from Second-Order Self-Force Theory},'' {\em Phys. Rev. Lett.} {\bf 130} (2023), no.~24, 241402, \href{http://www.arXiv.org/abs/2112.12265}{{\tt 2112.12265}}.

\bibitem{Albertini:2022rfe}
A.~Albertini, A.~Nagar, A.~Pound, N.~Warburton, B.~Wardell, L.~Durkan, and J.~Miller, ``{Comparing second-order gravitational self-force, numerical relativity, and effective one body waveforms from inspiralling, quasicircular, and nonspinning black hole binaries},'' {\em Phys. Rev. D} {\bf 106} (2022), no.~8, 084061, \href{http://www.arXiv.org/abs/2208.01049}{{\tt 2208.01049}}.

\bibitem{Katz:2021yft}
M.~L. Katz, A.~J.~K. Chua, L.~Speri, N.~Warburton, and S.~A. Hughes, ``{Fast extreme-mass-ratio-inspiral waveforms: New tools for millihertz gravitational-wave data analysis},'' {\em Phys. Rev. D} {\bf 104} (2021), no.~6, 064047, \href{http://www.arXiv.org/abs/2104.04582}{{\tt 2104.04582}}.

\bibitem{Amaro-Seoane:2012lgq}
P.~Amaro-Seoane, ``{Relativistic dynamics and extreme mass ratio inspirals},'' {\em Living Rev. Rel.} {\bf 21} (2018), no.~1, 4, \href{http://www.arXiv.org/abs/1205.5240}{{\tt 1205.5240}}.

\bibitem{Abac:2025saz}
A.~Abac {\em et al.}, ``{The Science of the Einstein Telescope},'' \href{http://www.arXiv.org/abs/2503.12263}{{\tt 2503.12263}}.

\bibitem{Blanchet:2013haa}
L.~Blanchet, ``{Post-Newtonian Theory for Gravitational Waves},'' {\em Living Rev. Rel.} {\bf 17} (2014) 2, \href{http://www.arXiv.org/abs/1310.1528}{{\tt 1310.1528}}.

\bibitem{Barack:2018yvs}
L.~Barack and A.~Pound, ``{Self-force and radiation reaction in general relativity},'' {\em Rept. Prog. Phys.} {\bf 82} (2019), no.~1, 016904,
\href{http://www.arXiv.org/abs/1805.10385}{{\tt 1805.10385}}.

\bibitem{Pound:2021qin}
A.~{Pound} and B.~{Wardell}, ``{Black Hole Perturbation Theory and Gravitational Self-Force},'' in {\em Handbook of Gravitational Wave Astronomy}, p.~38.
\newblock 2022.
\newblock \href{http://www.arXiv.org/abs/2101.04592}{{\tt 2101.04592}}.

\bibitem{Duez:2018jaf}
M.~D. Duez and Y.~Zlochower, ``{Numerical Relativity of Compact Binaries in the 21st Century},'' {\em Rept. Prog. Phys.} {\bf 82} (2019), no.~1, 016902, \href{http://www.arXiv.org/abs/1808.06011}{{\tt 1808.06011}}.

\bibitem{London:2014cma}
L.~London, D.~Shoemaker, and J.~Healy, ``{Modeling ringdown: Beyond the fundamental quasinormal modes},'' {\em Phys. Rev. D} {\bf 90} (2014), no.~12, 124032, \href{http://www.arXiv.org/abs/1404.3197}{{\tt 1404.3197}}. [Erratum: Phys.Rev.D 94, 069902 (2016)].

\bibitem{Berti:2025hly}
E.~Berti {\em et al.}, ``{Black hole spectroscopy: from theory to experiment},'' \href{http://www.arXiv.org/abs/2505.23895}{{\tt 2505.23895}}.

\bibitem{Buonanno:1998gg}
A.~Buonanno and T.~Damour, ``{Effective one-body approach to general relativistic two-body dynamics},'' {\em Phys. Rev.} {\bf D59} (1999) 084006,
\href{http://www.arXiv.org/abs/gr-qc/9811091}{{\tt gr-qc/9811091}}.

\bibitem{Buonanno:2000ef}
A.~Buonanno and T.~Damour, ``{Transition from inspiral to plunge in binary black hole coalescences},'' {\em Phys. Rev.} {\bf D62} (2000) 064015,
\href{http://www.arXiv.org/abs/gr-qc/0001013}{{\tt gr-qc/0001013}}.

\bibitem{Damour:2002vi}
T.~Damour, B.~R. Iyer, P.~Jaranowski, and B.~S. Sathyaprakash, ``{Gravitational waves from black hole binary inspiral and merger: The Span of third postNewtonian effective one-body templates},'' {\em Phys. Rev. D} {\bf 67} (2003) 064028, \href{http://www.arXiv.org/abs/gr-qc/0211041}{{\tt gr-qc/0211041}}.

\bibitem{Damour:2007xr}
T.~Damour and A.~Nagar, ``{Faithful effective-one-body waveforms of small-mass-ratio coalescing black-hole binaries},'' {\em Phys. Rev.} {\bf D76} (2007) 064028,
\href{http://www.arXiv.org/abs/0705.2519}{{\tt 0705.2519}}.

\bibitem{Buonanno:2009qa}
A.~Buonanno, Y.~Pan, H.~P. Pfeiffer, M.~A. Scheel, L.~T. Buchman, and L.~E. Kidder, ``{Effective-one-body waveforms calibrated to numerical relativity simulations: Coalescence of non-spinning, equal-mass black holes},'' {\em Phys. Rev. D} {\bf 79} (2009) 124028, \href{http://www.arXiv.org/abs/0902.0790}{{\tt 0902.0790}}.

\bibitem{Taracchini:2014zpa}
A.~Taracchini, A.~Buonanno, G.~Khanna, and S.~A. Hughes, ``{Small mass plunging into a Kerr black hole: Anatomy of the inspiral-merger-ringdown waveforms},'' {\em Phys. Rev. D} {\bf 90} (2014), no.~8, 084025, \href{http://www.arXiv.org/abs/1404.1819}{{\tt 1404.1819}}.

\bibitem{Riemenschneider:2021ppj}
G.~Riemenschneider, P.~Rettegno, M.~Breschi, A.~Albertini, R.~Gamba, S.~Bernuzzi, and A.~Nagar, ``{Assessment of consistent next-to-quasicircular corrections and postadiabatic approximation in effective-one-body multipolar waveforms for binary black hole coalescences},'' {\em Phys. Rev. D} {\bf 104} (2021), no.~10, 104045, \href{http://www.arXiv.org/abs/2104.07533}{{\tt 2104.07533}}.

\bibitem{Ajith:2007qp}
P.~Ajith {\em et al.}, ``{Phenomenological template family for black-hole coalescence waveforms},'' {\em Class. Quant. Grav.} {\bf 24} (2007) S689--S700, \href{http://www.arXiv.org/abs/0704.3764}{{\tt 0704.3764}}.

\bibitem{Ajith:2007kx}
P.~Ajith {\em et al.}, ``{A Template bank for gravitational waveforms from coalescing binary black holes. I. Non-spinning binaries},'' {\em Phys. Rev. D} {\bf 77} (2008) 104017, \href{http://www.arXiv.org/abs/0710.2335}{{\tt 0710.2335}}. [Erratum: Phys.Rev.D 79, 129901 (2009)].

\bibitem{Ajith:2009bn}
P.~Ajith {\em et al.}, ``{Inspiral-merger-ringdown waveforms for black-hole binaries with non-precessing spins},'' {\em Phys. Rev. Lett.} {\bf 106} (2011) 241101, \href{http://www.arXiv.org/abs/0909.2867}{{\tt 0909.2867}}.

\bibitem{Santamaria:2010yb}
L.~Santamaria {\em et al.}, ``{Matching post-Newtonian and numerical relativity waveforms: systematic errors and a new phenomenological model for non-precessing black hole binaries},'' {\em Phys. Rev. D} {\bf 82} (2010) 064016, \href{http://www.arXiv.org/abs/1005.3306}{{\tt 1005.3306}}.

\bibitem{Miller:2020bft}
J.~Miller and A.~Pound, ``{Two-timescale evolution of extreme-mass-ratio inspirals: waveform generation scheme for quasicircular orbits in Schwarzschild spacetime},'' {\em Phys. Rev. D} {\bf 103} (2021), no.~6, 064048, \href{http://www.arXiv.org/abs/2006.11263}{{\tt 2006.11263}}.

\bibitem{Mathews:2025nyb}
J.~Mathews and A.~Pound, ``{Post-adiabatic waveform-generation framework for asymmetric precessing binaries},'' \href{http://www.arXiv.org/abs/2501.01413}{{\tt 2501.01413}}.

\bibitem{Hinderer:2008dm}
T.~Hinderer and E.~E. Flanagan, ``{Two timescale analysis of extreme mass ratio inspirals in Kerr. I. Orbital Motion},'' {\em Phys. Rev.} {\bf D78} (2008) 064028,
\href{http://www.arXiv.org/abs/0805.3337}{{\tt 0805.3337}}.

\bibitem{Stein:2019buj}
L.~C. Stein and N.~Warburton, ``{Location of the last stable orbit in Kerr spacetime},'' {\em Phys. Rev. D} {\bf 101} (2020), no.~6, 064007, \href{http://www.arXiv.org/abs/1912.07609}{{\tt 1912.07609}}.

\bibitem{Ori:2000zn}
A.~Ori and K.~S. Thorne, ``{The Transition from inspiral to plunge for a compact body in a circular equatorial orbit around a massive, spinning black hole},'' {\em Phys. Rev.} {\bf D62} (2000) 124022,
\href{http://www.arXiv.org/abs/gr-qc/0003032}{{\tt gr-qc/0003032}}.

\bibitem{Compere:2021iwh}
G.~Comp\`ere and L.~K\"uchler, ``{Self-consistent adiabatic inspiral and transition motion},'' {\em Phys. Rev. Lett.} {\bf 126} (2021), no.~24, 241106, \href{http://www.arXiv.org/abs/2102.12747}{{\tt 2102.12747}}.

\bibitem{Compere:2021ERR}
G.~Comp\`ere and L.~K\"uchler, ``Erratum: Self-consistent adiabatic inspiral and transition motion [phys. rev. lett. 126, 241106 (2021)],'' {\em Phys. Rev. Lett.} {\bf 128} (Jan, 2022) 029901.

\bibitem{Compere:2021zfj}
G.~Comp\`ere and L.~K\"uchler, ``{Asymptotically matched quasi-circular inspiral and transition-to-plunge in the small mass ratio expansion},'' {\em SciPost Phys.} {\bf 13} (2022), no.~2, 043, \href{http://www.arXiv.org/abs/2112.02114}{{\tt 2112.02114}}.

\bibitem{Kuchler:2024esj}
L.~K\"uchler, G.~Comp\`ere, L.~Durkan, and A.~Pound, ``{Self-force framework for transition-to-plunge waveforms},'' {\em SciPost Phys.} {\bf 17} (2024) 056, \href{http://www.arXiv.org/abs/2405.00170}{{\tt 2405.00170}}.

\bibitem{Sundararajan:2007jg}
P.~A. Sundararajan, G.~Khanna, and S.~A. Hughes, ``{Towards adiabatic waveforms for inspiral into Kerr black holes. I. A New model of the source for the time domain perturbation equation},'' {\em Phys. Rev. D} {\bf 76} (2007) 104005, \href{http://www.arXiv.org/abs/gr-qc/0703028}{{\tt gr-qc/0703028}}.

\bibitem{Sundararajan:2008zm}
P.~A. Sundararajan, G.~Khanna, S.~A. Hughes, and S.~Drasco, ``{Towards adiabatic waveforms for inspiral into Kerr black holes: II. Dynamical sources and generic orbits},'' {\em Phys. Rev. D} {\bf 78} (2008) 024022, \href{http://www.arXiv.org/abs/0803.0317}{{\tt 0803.0317}}.

\bibitem{Sundararajan:2010sr}
P.~A. Sundararajan, G.~Khanna, and S.~A. Hughes, ``{Binary black hole merger gravitational waves and recoil in the large mass ratio limit},'' {\em Phys. Rev. D} {\bf 81} (2010) 104009, \href{http://www.arXiv.org/abs/1003.0485}{{\tt 1003.0485}}.

\bibitem{Rifat:2019ltp}
N.~E.~M. Rifat, S.~E. Field, G.~Khanna, and V.~Varma, ``{Surrogate model for gravitational wave signals from comparable and large-mass-ratio black hole binaries},'' {\em Phys. Rev. D} {\bf 101} (2020), no.~8, 081502, \href{http://www.arXiv.org/abs/1910.10473}{{\tt 1910.10473}}.

\bibitem{Islam:2022laz}
T.~Islam, S.~E. Field, S.~A. Hughes, G.~Khanna, V.~Varma, M.~Giesler, M.~A. Scheel, L.~E. Kidder, and H.~P. Pfeiffer, ``{Surrogate model for gravitational wave signals from nonspinning, comparable-to large-mass-ratio black hole binaries built on black hole perturbation theory waveforms calibrated to numerical relativity},'' {\em Phys. Rev. D} {\bf 106} (2022), no.~10, 104025, \href{http://www.arXiv.org/abs/2204.01972}{{\tt 2204.01972}}.

\bibitem{Hadar:2009ip}
S.~Hadar and B.~Kol, ``{Post-ISCO Ringdown Amplitudes in Extreme Mass Ratio Inspiral},'' {\em Phys. Rev. D} {\bf 84} (2011) 044019, \href{http://www.arXiv.org/abs/0911.3899}{{\tt 0911.3899}}.

\bibitem{Folacci:2018cic}
A.~Folacci and M.~Ould El~Hadj, ``{Multipolar gravitational waveforms and ringdowns generated during the plunge from the innermost stable circular orbit into a Schwarzschild black hole},'' {\em Phys. Rev. D} {\bf 98} (2018), no.~8, 084008, \href{http://www.arXiv.org/abs/1806.01577}{{\tt 1806.01577}}.

\bibitem{Cheung:2023vki}
M.~H.-Y. Cheung, E.~Berti, V.~Baibhav, and R.~Cotesta, ``{Extracting linear and nonlinear quasinormal modes from black hole merger simulations},'' {\em Phys. Rev. D} {\bf 109} (2024), no.~4, 044069, \href{http://www.arXiv.org/abs/2310.04489}{{\tt 2310.04489}}. [Erratum: Phys.Rev.D 110, 049902 (2024)].

\bibitem{Lim:2022veo}
H.~Lim, S.~A. Hughes, and G.~Khanna, ``{Measuring quasinormal mode amplitudes with misaligned binary black hole ringdowns},'' {\em Phys. Rev. D} {\bf 105} (2022), no.~12, 124030, \href{http://www.arXiv.org/abs/2204.06007}{{\tt 2204.06007}}.

\bibitem{Carullo:2024smg}
G.~Carullo, ``{Ringdown amplitudes of nonspinning eccentric binaries},'' {\em JCAP} {\bf 10} (2024) 061, \href{http://www.arXiv.org/abs/2406.19442}{{\tt 2406.19442}}.

\bibitem{Mitman:2025hgy}
K.~Mitman {\em et al.}, ``{Probing the ringdown perturbation in binary black hole coalescences with an improved quasi-normal mode extraction algorithm},'' \href{http://www.arXiv.org/abs/2503.09678}{{\tt 2503.09678}}.

\bibitem{AncillaryFiles}
``Self-force framework for merger-ringdown waveforms (ancillary files).'' (\url{https://github.com/gcompere/Self-force-framework-for-merger-ringdown-waveforms}).

\bibitem{Pound:2012nt}
A.~Pound, ``{Second-order gravitational self-force},'' {\em Phys. Rev. Lett.} {\bf 109} (2012) 051101, \href{http://www.arXiv.org/abs/1201.5089}{{\tt 1201.5089}}.

\bibitem{Pound:2012dk}
A.~Pound, ``{Nonlinear gravitational self-force. I. Field outside a small body},'' {\em Phys. Rev. D} {\bf 86} (2012) 084019, \href{http://www.arXiv.org/abs/1206.6538}{{\tt 1206.6538}}.

\bibitem{Pound:2017psq}
A.~Pound, ``{Nonlinear gravitational self-force: second-order equation of motion},'' {\em Phys. Rev. D} {\bf 95} (2017), no.~10, 104056, \href{http://www.arXiv.org/abs/1703.02836}{{\tt 1703.02836}}.

\bibitem{Upton:2021oxf}
S.~D. Upton and A.~Pound, ``{Second-order gravitational self-force in a highly regular gauge},'' {\em Phys. Rev. D} {\bf 103} (2021), no.~12, 124016, \href{http://www.arXiv.org/abs/2101.11409}{{\tt 2101.11409}}.

\bibitem{Detweiler:2011tt}
S.~Detweiler, ``{Gravitational radiation reaction and second order perturbation theory},'' {\em Phys. Rev. D} {\bf 85} (2012) 044048, \href{http://www.arXiv.org/abs/1107.2098}{{\tt 1107.2098}}.

\bibitem{Pound:2014xva}
A.~Pound and J.~Miller, ``{Practical, covariant puncture for second-order self-force calculations},'' {\em Phys. Rev. D} {\bf 89} (2014), no.~10, 104020, \href{http://www.arXiv.org/abs/1403.1843}{{\tt 1403.1843}}.

\bibitem{Lewis:InPrep}
J.~Lewis, T.~Kakehi, A.~Pound, and T.~Tanaka, ``Post-adiabatic dynamics and waveform generation in gravitational self-force theory: a pseudo-hamiltonian framework.'' in preparation.

\bibitem{Miller:2023ers}
J.~Miller, B.~Leather, A.~Pound, and N.~Warburton, ``{Worldtube puncture scheme for first- and second-order self-force calculations in the Fourier domain},'' {\em Phys. Rev. D} {\bf 109} (2024), no.~10, 104010, \href{http://www.arXiv.org/abs/2401.00455}{{\tt 2401.00455}}.

\bibitem{Pound:2015wva}
A.~Pound, ``{Second-order perturbation theory: problems on large scales},'' {\em Phys. Rev. D} {\bf 92} (2015), no.~10, 104047, \href{http://www.arXiv.org/abs/1510.05172}{{\tt 1510.05172}}.

\bibitem{Cunningham:2024dog}
K.~Cunningham, C.~Kavanagh, A.~Pound, D.~Trestini, N.~Warburton, and J.~Neef, ``{Gravitational memory: new results from post-Newtonian and self-force theory},'' \href{http://www.arXiv.org/abs/2410.23950}{{\tt 2410.23950}}.

\bibitem{MTW}
C.~W. Misner, K.~S. Thorne, and J.~A. Wheeler, {\em Gravitation}.
\newblock W.H.Freeman \& Co Ltd, 1973.

\bibitem{Kuchler:2023jbu}
{K\"uchler, Lorenzo}, {\em {Inspiral, transition and plunge: a framework for complete waveforms in the small-mass-ratio expansion}}.
\newblock PhD thesis, U. Brussels, 2023.

\bibitem{Gralla:2008fg}
S.~E. Gralla and R.~M. Wald, ``{A Rigorous Derivation of Gravitational Self-force},'' {\em Class. Quant. Grav.} {\bf 25} (2008) 205009, \href{http://www.arXiv.org/abs/0806.3293}{{\tt 0806.3293}}. [Erratum: Class.Quant.Grav. 28, 159501 (2011)].

\bibitem{Pound:2015fma}
A.~Pound, ``{Gauge and motion in perturbation theory},'' {\em Phys. Rev. D} {\bf 92} (2015), no.~4, 044021, \href{http://www.arXiv.org/abs/1506.02894}{{\tt 1506.02894}}.

\bibitem{Warburton:2013yj}
N.~Warburton, L.~Barack, and N.~Sago, ``{Isofrequency pairing of geodesic orbits in Kerr geometry},'' {\em Phys. Rev. D} {\bf 87} (2013), no.~8, 084012, \href{http://www.arXiv.org/abs/1301.3918}{{\tt 1301.3918}}.

\bibitem{Folacci:2018vtf}
A.~Folacci and M.~Ould El~Hadj, ``{Electromagnetic radiation generated by a charged particle plunging into a Schwarzschild black hole: Multipolar waveforms and ringdowns},'' {\em Phys. Rev. D} {\bf 98} (2018), no.~2, 024021, \href{http://www.arXiv.org/abs/1805.11950}{{\tt 1805.11950}}.

\bibitem{Rom:2022uvv}
B.~Rom and R.~Sari, ``{Extreme mass-ratio binary black hole merger: Characteristics of the test-particle limit},'' {\em Phys. Rev. D} {\bf 106} (2022), no.~10, 104040, \href{http://www.arXiv.org/abs/2204.11738}{{\tt 2204.11738}}.

\bibitem{Strusberg:2025qfv}
I.~Strusberg, B.~Rom, and R.~Sari, ``{Universal Waveforms for Extreme Mass-Ratio Inspiral},'' \href{http://www.arXiv.org/abs/2505.07941}{{\tt 2505.07941}}.

\bibitem{Chandrasekhar:579245}
S.~Chandrasekhar, {\em {The mathematical theory of black holes}}.
\newblock The International Series of Monographs on Physics. Oxford Univ. Press, New York, 1983.

\bibitem{Martel:2005ir}
K.~Martel and E.~Poisson, ``{Gravitational perturbations of the Schwarzschild spacetime: A Practical covariant and gauge-invariant formalism},'' {\em Phys. Rev. D} {\bf 71} (2005) 104003, \href{http://www.arXiv.org/abs/gr-qc/0502028}{{\tt gr-qc/0502028}}.

\bibitem{Spiers:2023mor}
A.~Spiers, A.~Pound, and B.~Wardell, ``{Second-order perturbations of the Schwarzschild spacetime: Practical, covariant, and gauge-invariant formalisms},'' {\em Phys. Rev. D} {\bf 110} (2024), no.~6, 064030, \href{http://www.arXiv.org/abs/2306.17847}{{\tt 2306.17847}}.

\bibitem{PhysRev.108.1063}
T.~Regge and J.~A. Wheeler, ``Stability of a schwarzschild singularity,'' {\em Phys. Rev.} {\bf 108} (Nov, 1957) 1063--1069.

\bibitem{PhysRevD.2.2141}
F.~J. Zerilli, ``Gravitational field of a particle falling in a schwarzschild geometry analyzed in tensor harmonics,'' {\em Phys. Rev. D} {\bf 2} (Nov, 1970) 2141--2160.

\bibitem{Hopper:2010uv}
S.~Hopper and C.~R. Evans, ``{Gravitational perturbations and metric reconstruction: Method of extended homogeneous solutions applied to eccentric orbits on a Schwarzschild black hole},'' {\em Phys. Rev. D} {\bf 82} (2010) 084010, \href{http://www.arXiv.org/abs/1006.4907}{{\tt 1006.4907}}.

\bibitem{Cunningham:1978zfa}
C.~T. Cunningham, R.~H. Price, and V.~Moncrief, ``{Radiation from collapsing relativistic stars. I - Linearized odd-parity radiation},'' {\em Astrophys. J.} {\bf 224} (1978) 643.

\bibitem{Moncrief:1974am}
V.~Moncrief, ``{Gravitational perturbations of spherically symmetric systems. I. The exterior problem.},'' {\em Annals Phys.} {\bf 88} (1974) 323--342.

\bibitem{Cunningham:1979px}
C.~T. Cunningham, R.~H. Price, and V.~Moncrief, ``{Radiation from collapsing relativistic stars. II. Linearized even-parity radiation},'' {\em Astrophys. J.} {\bf 230} (1979) 870--892.

\bibitem{1975RSPSA.344..441C}
S.~{Chandrasekhar} and S.~{Detweiler}, ``{The Quasi-Normal Modes of the Schwarzschild Black Hole},'' {\em Proceedings of the Royal Society of London Series A} {\bf 344} (Aug., 1975) 441--452.

\bibitem{Berti:2009kk}
E.~Berti, V.~Cardoso, and A.~O. Starinets, ``{Quasinormal modes of black holes and black branes},'' {\em Class. Quant. Grav.} {\bf 26} (2009) 163001, \href{http://www.arXiv.org/abs/0905.2975}{{\tt 0905.2975}}.

\bibitem{Berti:2006wq}
E.~Berti and V.~Cardoso, ``{Quasinormal ringing of Kerr black holes. I. The Excitation factors},'' {\em Phys. Rev. D} {\bf 74} (2006) 104020, \href{http://www.arXiv.org/abs/gr-qc/0605118}{{\tt gr-qc/0605118}}.

\bibitem{Zhang:2013ksa}
Z.~Zhang, E.~Berti, and V.~Cardoso, ``{Quasinormal ringing of Kerr black holes. II. Excitation by particles falling radially with arbitrary energy},'' {\em Phys. Rev. D} {\bf 88} (2013) 044018, \href{http://www.arXiv.org/abs/1305.4306}{{\tt 1305.4306}}.

\bibitem{Oshita:2021iyn}
N.~Oshita, ``{Ease of excitation of black hole ringing: Quantifying the importance of overtones by the excitation factors},'' {\em Phys. Rev. D} {\bf 104} (2021), no.~12, 124032, \href{http://www.arXiv.org/abs/2109.09757}{{\tt 2109.09757}}.

\bibitem{Leaver:1986gd}
E.~W. Leaver, ``{Spectral decomposition of the perturbation response of the Schwarzschild geometry},'' {\em Phys. Rev. D} {\bf 34} (1986) 384--408.

\bibitem{BHPToolkit}
``{Black Hole Perturbation Toolkit}.'' (\href{http://bhptoolkit.org/}{bhptoolkit.org}).

\bibitem{Vega:2009qb}
I.~Vega, P.~Diener, W.~Tichy, and S.~L. Detweiler, ``{Self-force with (3+1) codes: A Primer for numerical relativists},'' {\em Phys. Rev. D} {\bf 80} (2009) 084021, \href{http://www.arXiv.org/abs/0908.2138}{{\tt 0908.2138}}.

\bibitem{Bertiwebsite}
E.~Berti, ``Ringdown.'' https://pages.jh.edu/eberti2/ringdown/.

\bibitem{Berti:2005ys}
E.~Berti, V.~Cardoso, and C.~M. Will, ``{On gravitational-wave spectroscopy of massive black holes with the space interferometer LISA},'' {\em Phys. Rev. D} {\bf 73} (2006) 064030, \href{http://www.arXiv.org/abs/gr-qc/0512160}{{\tt gr-qc/0512160}}.

\bibitem{Leaver:1985ax}
E.~W. Leaver, ``{An Analytic representation for the quasi normal modes of Kerr black holes},'' {\em Proc. Roy. Soc. Lond. A} {\bf 402} (1985) 285--298.

\bibitem{sxscatalog}
``{SXS Gravitational Waveform Database}.'' https://data.black-holes.org/waveforms/index.html.

\bibitem{Boyle:2019kee}
M.~Boyle {\em et al.}, ``{The SXS Collaboration catalog of binary black hole simulations},'' {\em Class. Quant. Grav.} {\bf 36} (2019), no.~19, 195006, \href{http://www.arXiv.org/abs/1904.04831}{{\tt 1904.04831}}.

\bibitem{sxspackage}
M.~Boyle and M.~Scheel, ``The sxs package.'' 10.5281/zenodo.4034006.

\bibitem{LeTiec:2011bk}
A.~Le~Tiec, A.~H. Mroue, L.~Barack, A.~Buonanno, H.~P. Pfeiffer, N.~Sago, and A.~Taracchini, ``{Periastron Advance in Black Hole Binaries},'' {\em Phys. Rev. Lett.} {\bf 107} (2011) 141101, \href{http://www.arXiv.org/abs/1106.3278}{{\tt 1106.3278}}.

\bibitem{LeTiec:2011dp}
A.~Le~Tiec, E.~Barausse, and A.~Buonanno, ``{Gravitational Self-Force Correction to the Binding Energy of Compact Binary Systems},'' {\em Phys. Rev. Lett.} {\bf 108} (2012) 131103, \href{http://www.arXiv.org/abs/1111.5609}{{\tt 1111.5609}}.

\bibitem{LeTiec:2013uey}
A.~Le~Tiec {\em et al.}, ``{Periastron Advance in Spinning Black Hole Binaries: Gravitational Self-Force from Numerical Relativity},'' {\em Phys. Rev. D} {\bf 88} (2013), no.~12, 124027, \href{http://www.arXiv.org/abs/1309.0541}{{\tt 1309.0541}}.

\bibitem{Nagar:2013sga}
A.~Nagar, ``{Gravitational recoil in nonspinning black hole binaries: the span of test-mass results},'' {\em Phys. Rev. D} {\bf 88} (2013), no.~12, 121501, \href{http://www.arXiv.org/abs/1306.6299}{{\tt 1306.6299}}.

\bibitem{LeTiec:2017ebm}
A.~Le~Tiec and P.~Grandcl\'ement, ``{Horizon Surface Gravity in Corotating Black Hole Binaries},'' {\em Class. Quant. Grav.} {\bf 35} (2018), no.~14, 144002, \href{http://www.arXiv.org/abs/1710.03673}{{\tt 1710.03673}}.

\bibitem{vandeMeent:2020xgc}
M.~van~de Meent and H.~P. Pfeiffer, ``{Intermediate mass-ratio black hole binaries: Applicability of small mass-ratio perturbation theory},'' {\em Phys. Rev. Lett.} {\bf 125} (2020), no.~18, 181101, \href{http://www.arXiv.org/abs/2006.12036}{{\tt 2006.12036}}.

\bibitem{jonathan_blackman_2024_13159569}
J.~Blackman and S.~Collaboration, ``{Binary black-hole simulation SXS:BBH:1220},'' Aug., 2024.

\bibitem{sxs_collaboration_2024_13147606}
S.~Collaboration, ``{Binary black-hole simulation SXS:BBH:2477},'' Aug., 2024.

\bibitem{sxs_collaboration_2019_3273103}
S.~Collaboration, ``{Binary black-hole simulation SXS:BBH:1931},'' July, 2019.

\bibitem{sxs_collaboration_2024_13147573}
S.~Collaboration, ``{Binary black-hole simulation SXS:BBH:2471},'' Aug., 2024.

\bibitem{Giesler:2019uxc}
M.~Giesler, M.~Isi, M.~A. Scheel, and S.~Teukolsky, ``{Black Hole Ringdown: The Importance of Overtones},'' {\em Phys. Rev. X} {\bf 9} (2019), no.~4, 041060, \href{http://www.arXiv.org/abs/1903.08284}{{\tt 1903.08284}}.

\bibitem{Bhagwat:2019dtm}
S.~Bhagwat, X.~J. Forteza, P.~Pani, and V.~Ferrari, ``{Ringdown overtones, black hole spectroscopy, and no-hair theorem tests},'' {\em Phys. Rev. D} {\bf 101} (2020), no.~4, 044033, \href{http://www.arXiv.org/abs/1910.08708}{{\tt 1910.08708}}.

\bibitem{Okounkova:2020vwu}
M.~Okounkova, ``{Revisiting non-linearity in binary black hole mergers},'' \href{http://www.arXiv.org/abs/2004.00671}{{\tt 2004.00671}}.

\bibitem{Cook:2020otn}
G.~B. Cook, ``{Aspects of multimode Kerr ringdown fitting},'' {\em Phys. Rev. D} {\bf 102} (2020), no.~2, 024027, \href{http://www.arXiv.org/abs/2004.08347}{{\tt 2004.08347}}.

\bibitem{Dhani:2020nik}
A.~Dhani, ``{Importance of mirror modes in binary black hole ringdown waveform},'' {\em Phys. Rev. D} {\bf 103} (2021), no.~10, 104048, \href{http://www.arXiv.org/abs/2010.08602}{{\tt 2010.08602}}.

\bibitem{Mourier:2020mwa}
P.~Mourier, X.~Jim\'enez~Forteza, D.~Pook-Kolb, B.~Krishnan, and E.~Schnetter, ``{Quasinormal modes and their overtones at the common horizon in a binary black hole merger},'' {\em Phys. Rev. D} {\bf 103} (2021), no.~4, 044054, \href{http://www.arXiv.org/abs/2010.15186}{{\tt 2010.15186}}.

\bibitem{Finch:2021iip}
E.~Finch and C.~J. Moore, ``{Modeling the ringdown from precessing black hole binaries},'' {\em Phys. Rev. D} {\bf 103} (2021), no.~8, 084048, \href{http://www.arXiv.org/abs/2102.07794}{{\tt 2102.07794}}.

\bibitem{Forteza:2021wfq}
X.~J. Forteza and P.~Mourier, ``{High-overtone fits to numerical relativity ringdowns: Beyond the dismissed n=8 special tone},'' {\em Phys. Rev. D} {\bf 104} (2021), no.~12, 124072, \href{http://www.arXiv.org/abs/2107.11829}{{\tt 2107.11829}}.

\bibitem{Baibhav:2023clw}
V.~Baibhav, M.~H.-Y. Cheung, E.~Berti, V.~Cardoso, G.~Carullo, R.~Cotesta, W.~Del~Pozzo, and F.~Duque, ``{Agnostic black hole spectroscopy: Quasinormal mode content of numerical relativity waveforms and limits of validity of linear perturbation theory},'' {\em Phys. Rev. D} {\bf 108} (2023), no.~10, 104020, \href{http://www.arXiv.org/abs/2302.03050}{{\tt 2302.03050}}.

\bibitem{Nee:2023osy}
P.~J. Nee, S.~H. V\"olkel, and H.~P. Pfeiffer, ``{Role of black hole quasinormal mode overtones for ringdown analysis},'' {\em Phys. Rev. D} {\bf 108} (2023), no.~4, 044032, \href{http://www.arXiv.org/abs/2302.06634}{{\tt 2302.06634}}.

\bibitem{Zhu:2023mzv}
H.~Zhu, J.~L. Ripley, A.~C\'ardenas-Avenda\~no, and F.~Pretorius, ``{Challenges in quasinormal mode extraction: Perspectives from numerical solutions to the Teukolsky equation},'' {\em Phys. Rev. D} {\bf 109} (2024), no.~4, 044010, \href{http://www.arXiv.org/abs/2309.13204}{{\tt 2309.13204}}.

\bibitem{Qiu:2023lwo}
Y.~Qiu, X.~J. Forteza, and P.~Mourier, ``{Linear versus nonlinear modeling of black hole ringdowns},'' {\em Phys. Rev. D} {\bf 109} (2024), no.~6, 064075, \href{http://www.arXiv.org/abs/2312.15904}{{\tt 2312.15904}}.

\bibitem{Giesler:2024hcr}
M.~Giesler {\em et al.}, ``{Overtones and Nonlinearities in Binary Black Hole Ringdowns},'' \href{http://www.arXiv.org/abs/2411.11269}{{\tt 2411.11269}}.

\bibitem{Buonanno:2006ui}
A.~Buonanno, G.~B. Cook, and F.~Pretorius, ``{Inspiral, merger and ring-down of equal-mass black-hole binaries},'' {\em Phys. Rev. D} {\bf 75} (2007) 124018, \href{http://www.arXiv.org/abs/gr-qc/0610122}{{\tt gr-qc/0610122}}.

\bibitem{Oshita:2024wgt}
N.~Oshita and V.~Cardoso, ``{Reconstruction of ringdown with excitation factors},'' {\em Phys. Rev. D} {\bf 111} (2025), no.~10, 104043, \href{http://www.arXiv.org/abs/2407.02563}{{\tt 2407.02563}}.

\bibitem{DeAmicis:2024not}
M.~De~Amicis, S.~Albanesi, and G.~Carullo, ``{Inspiral-inherited ringdown tails},'' {\em Phys. Rev. D} {\bf 110} (2024), no.~10, 104005, \href{http://www.arXiv.org/abs/2406.17018}{{\tt 2406.17018}}.

\bibitem{KuchlerCapra27}
L.~K\"uchler, ``Progress towards merger-ringdown waveforms from self-force theory.'' Talk given at 27th Capra Meeting on Radiation Reaction in General Relativity (National University of Singapore), https://www.caprameeting.org/capra-meetings/capra-27/abstracts, 2024.

\bibitem{KuchlerLISA}
L.~K\"uchler, ``Self-force framework for transition-to-plunge waveforms.'' Talk given at 15th International LISA Symposium (University College Dublin), https://www.lisasymposium2024.ie/programme/, 2024.

\bibitem{KuchlerAEI}
L.~K\"uchler, ``Self-force framework for transition-to-plunge waveforms.'' Talk given at Fundamental Physics Meets Waveforms with LISA Workshop (Max Planck Institute for Gravitational Physics, Potsdam), https://workshops.aei.mpg.de/fpmeetswavelisa/program/, 2024.

\bibitem{Mummery:2022ana}
A.~Mummery and S.~Balbus, ``{Inspirals from the Innermost Stable Circular Orbit of Kerr Black Holes: Exact Solutions and Universal Radial Flow},'' {\em Phys. Rev. Lett.} {\bf 129} (2022), no.~16, 161101, \href{http://www.arXiv.org/abs/2209.03579}{{\tt 2209.03579}}.

\bibitem{Dyson:2023fws}
C.~Dyson and M.~van~de Meent, ``{Kerr-fully diving into the abyss: analytic solutions to plunging geodesics in Kerr},'' {\em Class. Quant. Grav.} {\bf 40} (2023), no.~19, 195026, \href{http://www.arXiv.org/abs/2302.03704}{{\tt 2302.03704}}.

\bibitem{Teukolsky:1972my}
S.~A. Teukolsky, ``{Rotating black holes - separable wave equations for gravitational and electromagnetic perturbations},'' {\em Phys. Rev. Lett.} {\bf 29} (1972) 1114--1118.

\bibitem{Spiers:2023cva}
A.~Spiers, A.~Maselli, and T.~P. Sotiriou, ``{Measuring scalar charge with compact binaries: High accuracy modeling with self-force},'' {\em Phys. Rev. D} {\bf 109} (2024), no.~6, 064022, \href{http://www.arXiv.org/abs/2310.02315}{{\tt 2310.02315}}.

\bibitem{Speri:2024qak}
L.~Speri, S.~Barsanti, A.~Maselli, T.~P. Sotiriou, N.~Warburton, M.~van~de Meent, A.~J.~K. Chua, O.~Burke, and J.~Gair, ``{Probing fundamental physics with Extreme Mass Ratio Inspirals: a full Bayesian inference for scalar charge},'' \href{http://www.arXiv.org/abs/2406.07607}{{\tt 2406.07607}}.

\bibitem{Dyson:2025dlj}
C.~Dyson, T.~F.~M. Spieksma, R.~Brito, M.~van~de Meent, and S.~Dolan, ``{Environmental effects in extreme mass ratio inspirals: perturbations to the environment in Kerr},'' \href{http://www.arXiv.org/abs/2501.09806}{{\tt 2501.09806}}.

\bibitem{AyushRoy:InPreparation}
A.~Roy, L.~K\"uchler, A.~Pound, and R.~P. Macedo, ``Black hole mergers beyond general relativity: the self-force approach.'' in preparation.

\bibitem{GeoSumanta:InPreparation}
G.~Comp\`ere, S.~Chakraborty, and L.~Machet, ``Plunging in black hole surrounded by a dark matter halo.'' in preparation.

\end{thebibliography}\endgroup

\end{document}